\documentclass[twocolumn,twocolappendix]{aastex63}
\shorttitle{X-ray Constraint on AGN Torus Location}
\shortauthors{Uematsu et al.}
\graphicspath{{./}{figures/}}

\begin{document}


\title{X-ray Constraint on Location of AGN Torus in Circinus Galaxy}

\correspondingauthor{Ryosuke Uematsu}
\email{uematsu@kusastro.kyoto-u.ac.jp}

\author{Ryosuke Uematsu}
\affiliation{Department of Astronomy, Kyoto University, Kyoto 606-8502, Japan}

\author[0000-0001-7821-6715]{Yoshihiro Ueda}
\affiliation{Department of Astronomy, Kyoto University, Kyoto 606-8502, Japan}

\author[0000-0002-0114-5581]{Atsushi Tanimoto}
\affiliation{Department of Physics, The University of Tokyo, Tokyo 113-0033, Japan}

\author[0000-0002-6808-2052]{Taiki Kawamuro}
\affiliation{National Astronomical Observatory of Japan, Tokyo 181-8588, Japan}
\affiliation{Nu\'{c}leo de Astronom\'{i}a de la Facultad de Ingenier\'{i}a, Universidad Diego Portales, Av. Ej\'{e}ercito Libertador 441, Santiago, Chile}

\author[0000-0001-5353-7635]{Kenta Setoguchi}
\affiliation{Department of Astronomy, Kyoto University, Kyoto 606-8502, Japan}

\author[0000-0002-5701-0811]{Shoji Ogawa}
\affiliation{Department of Astronomy, Kyoto University, Kyoto 606-8502, Japan}

\author[0000-0002-9754-3081]{Satoshi Yamada}
\affiliation{Department of Astronomy, Kyoto University, Kyoto 606-8502, Japan}

\author{Hirokazu Odaka}
\affiliation{Department of Physics, The University of Tokyo, Tokyo 113-0033, Japan}
\affiliation{Kavil IPMU, The University of Tokyo, Chiba 277-8583, Japan}



\begin{abstract}

The location of the 
obscuring ``torus'' in an active galactic nucleus
(AGN) is still an unresolved issue. 
The line widths of X-ray fluorescence
lines originated from the torus, particularly Fe K$\alpha$, carry key
information on the radii of line emitting regions. 
Utilizing XCLUMPY \citep{2019ApJ...877...95T}, an X-ray clumpy torus
model, we develop a realistic model of emission line profiles from an
AGN torus where we take into account line broadening due to the
Keplerian motion around the black hole. Then, we apply the updated
model to the best available broadband spectra (3--100 keV) of the
Circinus galaxy observed with Suzaku, XMM-Newton, Nuclear
Spectroscopic Telescope Array (NuSTAR), and Chandra, 
including 0.62 Ms Chandra/HETG data. We confirm that the torus is Compton-thick
(hydrogen column-density along the equatorial plane is
$N_\mathrm{H}^\mathrm{Equ}=2.16^{+0.24}_{-0.16}\times 10^{25}\ \mathrm{cm}^{-2}$)
geometrically thin 
(torus angular width
$\sigma=10.3^{+0.7}_{-0.3}\ \mathrm{degrees}$),
viewed edge-on 
(inclination 
$i=78.3^{+0.4}_{-0.9}\ \mathrm{degrees}$), and has super-solar abundance
($1.52^{+0.04}_{-0.06}$ times solar).
Simultaneously analyzing the Chandra/HETG first, second, and
third order spectra with 
consideration of the spatial extent of the Fe K$\alpha$ line emitting region, 
we constrain the inner radius of the torus to be $1.9^{+3.1}_{-0.8}\times 10^5$
times the gravitational radius, or $1.6^{+1.5}_{-0.9}\times 10^{-2}\ \mathrm{pc}$
for a black hole mass of $(1.7\pm 0.3)\times 10^6\ M_{\odot}$. This is
about 3 times smaller than that estimated from the dust sublimation
radius, suggesting that the inner side of the dusty region of the torus
is composed of dust-free gas.

\end{abstract}



\section{Introduction} \label{section:intro}

Formation mechanisms of supermassive black holes (SMBHs) in galactic
centers are an important topic in modern astronomy, which are related
to the understanding of galaxy-SMBH coevolution (see e.g.,
\citealt{2013ARA&A..51..511K} for a review). An active galactic nucleus (AGN)
is a key object to tackle this problem because it represents the
process of SMBH growth by mass accretion. One of the major remaining
questions in AGN phenomena is mass feeding processes onto SMBHs. 
Past studies have built a consensus that the central engine of an AGN
is surrounded by obscuring material called a ``torus''\footnote
{In this paper, we use the term ``torus'' as the whole structure 
of X-ray obscuring material including possible dust-free gas 
at the innermost region.
In this context, our ``torus'' is not the same as ``dust (or dusty)
torus'' often referred to in optical and infrared astronomy.} 
(see e.g., \citealt{2017NatAs...1..679R} for a
recent review). This torus is considered to be a link between 
the inner accretion disk and host-galaxy scale gas
playing as a reservoir
supplying mass toward the central SMBH. Hence, elucidating the nature of
AGN tori is indispensable to understand the physical origin of AGN
feeding.

X-ray observations are powerful tool to study AGN tori because X-rays
have strong penetrating power and serve as an unbiased tracer of
material in various physical conditions, including gas and dust. Several
groups developed X-ray spectral models that reproduce reflection
components (including Compton reflection continuum and fluorescence
lines) from a torus with a simplified geometry (e.g., \citealt{2009MNRAS.397.1549M},
, \citealt{2009ApJ...692..608I}, \citealt{2011MNRAS.413.1206B},
\citealt{2018ApJ...854...42B}). Many studies show that an AGN torus
is not smooth but is composed of clumpy media. Accordingly,
\citet{2019ApJ...877...95T} have constructed a new X-ray clumpy torus
model called XCLUMPY, utilizing the Monte Carlo simulation for
astrophysics and cosmology framework (MONACO:
\citealt{2011ApJ...740..103O, 2016MNRAS.462.2366O})(see also \citealt{2019A&A...629A..16B}
for a similar X-ray clumpy torus model). XCLUMPY assumes the same
distribution of clumps as that in the infrared clumpy torus model CLUMPY
constructed by \citet{2008ApJ...685..147N,2008ApJ...685..160N}. By
applying XCLUMPY to broadband X-ray spectra of nearby AGNs, \citet{2020ApJ...897....2T},
\citet{2020ApJ...897..107Y}, and \citet{2021ApJ...906...84O} constrained
their basic torus parameters, such as the
torus angular width (torus scale height relative to radius) and column
density. These works 
are consistent with 
the existence of a relation between torus covering factor and 
Eddington ratio, as reported by \citet{2017Natur.549..488R}.

A key parameter to understand the whole structure of an AGN is the inner
radius (i.e., the distance from the SMBH) of the torus. Optical and
infrared reverberation observations of nearby AGNs have determined the
innermost radii of ``dusty'' regions in their tori (e.g., \citealt{2009ApJ...700L.109K}),
which are found to be proportional to the square root of AGN luminosity
as theoretically expected \citep{2010ApJ...724L.183K}. However, it remains unclear
whether or not there is a significant amount of dust-free gas that
cannot be probed by infrared observations inside the dust sublimation
radius. Unfortunately, most of previous X-ray studies were not able to
constrain it because the torus models used are ``scale-free'', that is,
tori of a similar geometrical form with different absolute scales
produce exactly the same reflection component including the continuum
and lines. 

High resolution X-ray spectroscopy has a great potential to constrain
the true innermost radius of an AGN torus including dust-free gas.
In AGN X-ray spectra, most of the narrow fluorescence lines from cold material
are believed to come from the torus, among which Fe K$\alpha$ at
6.4 keV is the strongest 
(see e.g, \citealt{2014arXiv1412.1177R}). Assuming that the motion of
torus matter basically obeys Keplerian rotation around the SMBH, it is
possible to estimate the radius
by measuring the line width (or ideally, line profile) due
to Doppler broadening. 
\citet{2015ApJ...802...98M} and \citet{2015ApJ...812..113G} estimated
the radii of Fe K$\alpha$ line emitting regions in nearby AGNs. They referred
to the Fe K$\alpha$ line widths compiled by
\citet{2010ApJS..187..581S,2011ApJ...738..147S} using the
Chandra/HETG data. However, it has been pointed out that
these measurements might be subject to systematic uncertainties.
Firstly, 
\citet{2010ApJS..187..581S,2011ApJ...738..147S} approximately fitted the
Fe K$\alpha$ line by a single-Gaussian model. In reality, Fe K$\alpha$
line consists of two lines (K$\alpha$1, K$\alpha$2) often accompanied 
by Compton shoulders. 
Secondly, the grating spectra may be 
affected by spatial extent of the emitting region.
\citet{2016MNRAS.463L.108L} reported that the 
Chandra/HETG first order spectra of nearby AGNs systematically showed
broader line widths than the higher order ones.
This can be explained by the spatial extent, to which the first order 
spectra are the most sensitive.\footnote{\texttt{https://cxc.harvard.edu/newsletters/news\_24/}} 

In this paper, we update the fluorescence line component of the XCLUMPY
model by including line broadening effects due to the Keplerian motion
around the SMBH. Then, we apply it to the best
available X-ray data observed with Suzaku, XMM-Newton, NuSTAR,
and Chandra (including Chandra/HETG data of total 0.62 Ms) of the Circinus
galaxy ($z=0.0014$). 
To reliably estimate the torus inner radius, we simultaneously analyze
the first, second, and third order spectra of the Chandra/HETG
by taking into account the source
extent of the Fe K$\alpha$ line emitting region.
The structure of this paper is as follows. In
Section~\ref{section:model}, we describe the update of the XCLUMPY
model. Sections~\ref{observation} and ~\ref{results} present the
observations and the results of the spectral analysis, respectively.
Section~\ref{feline} focuses on a detailed analysis of the Fe K$\alpha$
line. In Section~\ref{discussion}, we compare our result with previous
works and discuss the implication.
For the Circinus galaxy, 
we assume a distance of
4.2 Mpc \citep{1977A&A....55..445F} when calculating luminosities and spatial scales. 
The Solar abundances of \citet{1989GeCoA..53..197A} are adopted. 
An attached error corresponds to the 90\% confidence region for a single parameter.

\section{XCLUMPY Model Update} \label{section:model}
For our study, we update the original XCLUMPY model \citep{2019ApJ...877...95T}
by taking into account (1) Doppler line broadening by the Keplerian motion of each
clump and (2) variable metal abundance. We explain the first part below
(Section~\ref{subsection:geometry}, \ref{subsection:doppler}). 
\citet{2018ApJ...867...80H} suggested that the metal abundance of the
Circinus galaxy were super-solar ($\approx$1.7) with Compton-shoulder
diagnostics of the Fe K$\alpha$ line in a fixed torus geometry, whereas
\citet{2019ApJ...877...95T} showed that the fraction of the Compton shoulder also depended
on torus geometry and column density. To solve the degeneracy, we newly
introduce the metal abundance parameter, defined as the ratios to the
Solar values, which ranges from 1.0 to 1.8 with a step of 0.2.

\subsection{Geometry of XCLUMPY Model} \label{subsection:geometry}
Here we briefly review the torus geometry of the XCLUMPY model.
XCLUMPY assumes a power-law distribution of spherical clumps 
with a fixed radius of $R_{\rm clump}$ in
the radial direction within the inner and outer radii ($r_{\rm in}$ and
$r_{\rm out}$), and normal distribution in the elevation direction.
The number density of clumps is expressed in
Equation~\ref{equation:xclumpy} (where we use the spherical coordinate
system; $r$ is a radius, $\theta$ is a polar angle, $\phi$ is an
azimuthal angle):
\begin{equation}
 \label{equation:xclumpy}
 d(r,\theta,\phi) \propto \left(\frac{r}{r_\mathrm{in}}\right)^{-0.5}\exp{\left(\frac{-\left(\theta-\pi/2\right)^2}{\sigma^2}\right)}
\end{equation}
This model has three variable torus parameters: equatorial hydrogen
column density ($N_\mathrm{H}^\mathrm{Equ}$), torus angular width
($\sigma$), and inclination angle ($i$). 
The number of clumps along the equatorial plane is set to be 10. \citet{2019ApJ...877...95T}
assume $r_{\rm in}=0.05$ pc, $r_{\rm out}=1$ pc, and $R_{\rm
clump}=0.002$ pc, although only their scale ratios are meaningful because
of the geometric similarity (see Section~\ref{section:intro}).

\subsection{Doppler Effects} \label{subsection:doppler}
\begin{figure}[htbp] \label{figure:velocity}
 \plotone{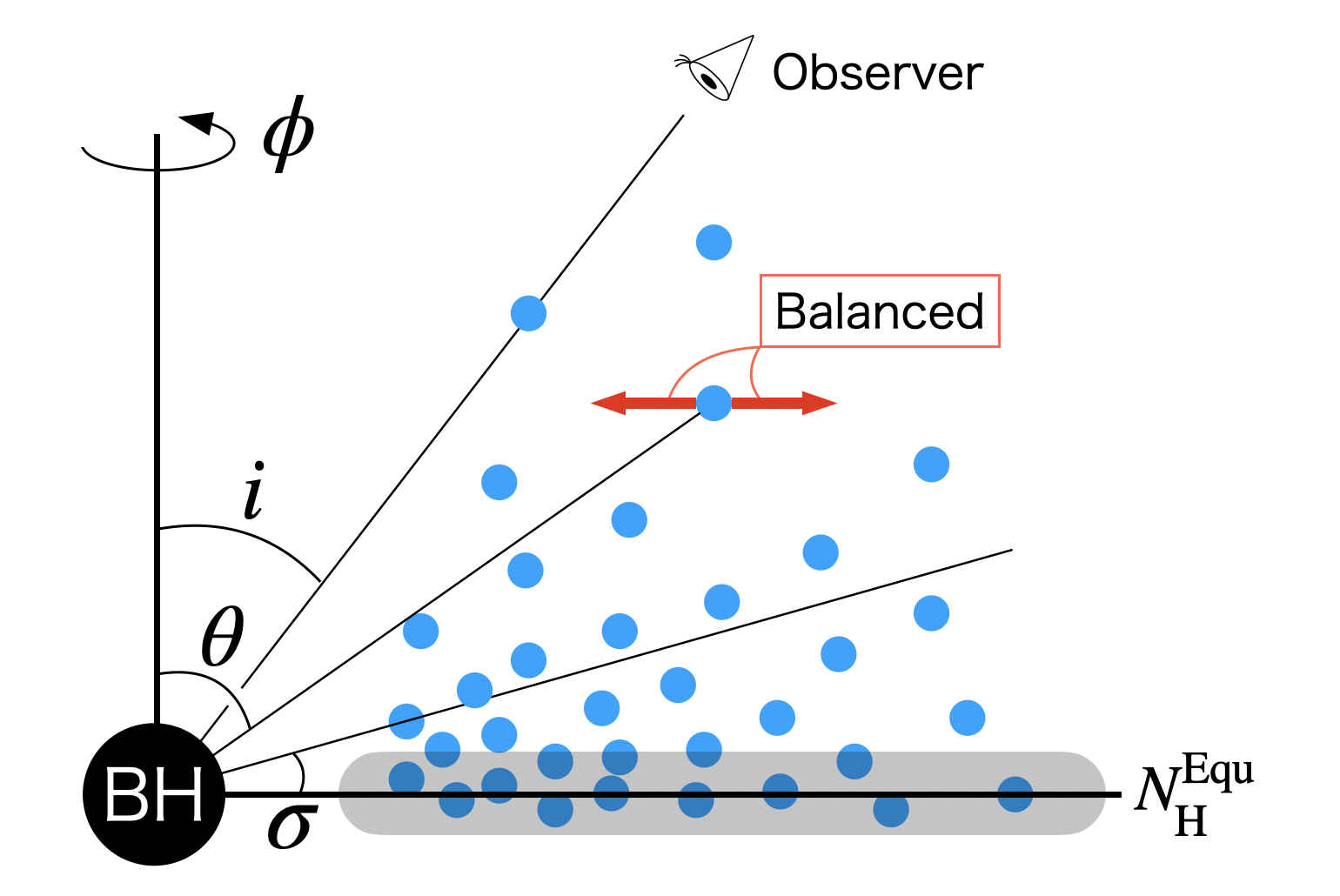}
 \caption{Geometry of the XCLUMPY model. $N_\mathrm{H}^\mathrm{Equ}$ is the equatorial hydrogen column density, $\sigma$ the torus angular width, and $i$ the inclination angle. The red arrows represent the force balance of a clump.}
\end{figure}

In the original XCLUMPY model, it was assumed that all clumps were
completely at rest.
In our updated model, we consider 
the azimuthal velocity of a clump located at $r$ and $\theta$:
\begin{equation} \label{verocity}
  v_{\phi}=\sqrt{\frac{GM}{r}}\sin\theta=c\sqrt{\frac{r_{\rm g}}{r}}\sin\theta,
\end{equation}
where $G$ is the gravitational constant, $M$ the black hole mass, $c$ the light speed, 
and $r_{\rm g} \equiv GM/c^2$ the gravitational radius. 
It assumes the balance between the centrifugal force and gravity force
projected onto the equatorial plane (see Figure~\ref{figure:velocity}). 
For simplicity, we ignore any velocity field in the elevation direction, 
which is not important for the spectra of edge-on sources (like the 
Circinus galaxy).

To save computation time, we do not repeat Monte Carlo simulations by
fully taking into account the motions of clumps in all interactions 
(e.g., Compton scattering and photoelectric absorption of continuum
photons). Instead, as an approximation, we only consider the motion of a
clump that emits or scatters a fluorescence-line photon at its last
interaction. This enables us to perform a post analysis of the photon
event files that were used to make the table models of XCLUMPY.
Although Doppler broadening of absorption edge features in the
reflection continuum cannot be reproduced with this approximation, it is
sufficient for studies focusing on the profile of emission lines.
In the simulation of MONACO, the last interaction point and 
the last direction of each photon are recorded. Then, we correct the
energy and weight of each photon for the Doppler and relativistic
beaming effects, respectively, to produce a new table model of
the fluorescence lines.
Since a Keplerian velocity is determined by the ratio of the radius
to the gravitational radius, we introduce the parameter 
$\log(r_{\mathrm{in}}/r_{\mathrm{g}})$, which 
ranges from 4.0 to 7.0 with a step of 0.5.

In Appendix~\ref{appendix:feline}, we present examples of the Fe K$\alpha$ line profiles, 
together with the spatial distribution of photon emitting positions, 
for 4 different sets of the torus parameters.

\section{Observations and Data Reduction} \label{observation}
We have selected the Circinus galaxy ($z=0.0014$), one of the closest
type 2 AGN, as our target. The Circinus galaxy is ideal for our study
because it shows the brightest narrow Fe K$\alpha$ line among all AGNs
\citep{2011ApJ...727...19F} and because very deep Chandra/HETG
data of 681 ks are available (\citealt{2014ApJ...791...81A}).
Since it is a heavily obscured, Compton-thick
AGN, one can fairly assume that its Fe K$\alpha$ line is dominated by the
emission from the torus and that the contribution from
the accretion disk is ignorable. In this paper, we analyze the
X-ray data obtained with Suzaku, XMM-Newton, NuSTAR, and Chandra, 
which are summarized in Table~\ref{table:observation}.

\begin{deluxetable}{lllc}
 \tablecaption{Summary of Observations \label{table:observation}}
 \tablewidth{0pt}
 \tablehead{
 \colhead{Observatory} & \colhead{ObsID} & \colhead{Date} & \colhead{Exposure}
 }
 \colnumbers
 \startdata
 Chandra HETG & 374 & 2000 June 15 & 7.26 \\
 Chandra HETG & 62877 & 2000 June 16 & 61.4 \\
 Chandra HETG & 4770 & 2004 June 2 & 56.1 \\
 Chandra HETG & 4771 & 2004 Nov 28 & 60.2 \\
 Suzaku & 701036010 & 2006 Jul 21 & 108\tablenotemark{a} \\
 Chandra HETG & 10226 & 2008 Dec 8 & 20.1 \\
 Chandra HETG & 10223 & 2008 Dec 15 & 105 \\
 Chandra HETG & 10832 & 2008 Dec 19 & 21.1 \\
 Chandra HETG & 10833 & 2008 Dec 22 & 29.0 \\
 Chandra HETG & 10224 & 2008 Dec 23 & 78.7 \\
 Chandra HETG & 10844 & 2008 Dec 24 & 27.7 \\
 Chandra HETG & 10225 & 2008 Dec 26 & 69.3 \\
 Chandra HETG & 10842 & 2008 Dec 27 & 37.5 \\
 Chandra HETG & 10843 & 2008 Dec 29 & 58.2 \\
 Chandra ACIS-S\tablenotemark{b} & 12823 & 2010 Dec 17 & 154 \\
 XMM-Newton & 0701981001 & 2013 Feb 3 & 58.9 \\
 NuSTAR & 60002039002 & 2013 Jan 25 & 53.9 \\
 \enddata
 \tablecomments{
 Column (1): observatory. Column (2): observation ID. Column (3): observation start date. Column (4): exposure time in units of ks.
 }
 \tablenotetext{a}{
 Based on the good time interval of an XIS camera with the
 shortest exposure among all XISs.}
 \tablenotetext{b}{Non grating observation.}
\end{deluxetable}

\subsection{Suzaku}

Suzaku \citep{2007PASJ...59S...1M}, the fifth Japanese X-ray
astronomical satellite, observed the Circinus galaxy in 2006 July.
Analysis results utilizing these data were published in \citet{2009ApJ...691..131Y},
\citet{2014ApJ...791...81A}, and \citet{2019ApJ...877...95T}, although \citet{2014ApJ...791...81A} did not include them
in the spectral fitting.
Suzaku carried four X-ray CCD cameras (X-ray Imaging
Spectrometer, XIS: \citealt{2007PASJ...59S..23K}) and a collimated hard
X-ray instrument (Hard X-ray Detector, HXD:
\citealt{2007PASJ...59S..35T}). XIS0, XIS2, and XIS3 are
frontside-illuminated CCDs (XIS-FI: 0.4--12.0~keV) and XIS1 is a
backside-illuminated one (XIS-BI: 0.2--12.0~keV). The HXD consists of two
types of detectors: a GSO/BGO phoswich counter and PIN silicon diodes
\citep{2007PASJ...59S..53K}. The GSO/BGO counter is sensitive to 40--600
keV and PIN diodes to 10--70 keV. We analyzed the data
using HEAsoft v6.27 and the calibration database (CALDB) released on 2018
October 23 (XIS) and 2011 September 15 (HXD).

We reprocessed the unfiltered XIS data with the \textsc{aepipeline}
script. Events were extracted from a circular region with a radius of
100\arcsec\ centered at the source peak. 
We took the background from a source-free circular region with a radius
of 100\arcsec. We generated the XIS redistribution matrix files (RMF) and
ancillary response files (ARF) by using \textsc{xisrmfgen} and \textsc{xissimarfgen},
respectively. To improve the statistics, we combined the spectra of
XIS-FI detectors (XIS0, XIS2, and XIS3) with \textsc{addascaspec}. We binned the XIS spectra
to contain at least 100 counts per bin.

The unfiltered HXD data were also reprocessed with the \textsc{aepipeline}
script. We generated the PIN and GSO spectra by using \textsc{hxdpinxbpi}
and \textsc{hxdgsoxbpi}, respectively. We utilized the tuned background
files \citep{2009PASJ...61S..17F} to obtain the non-X-ray background
(NXB). The cosmic X-ray background (CXB) was added to the NXB of the PIN
spectrum. The CXB in the GSO spectrum was ignored.

\subsection{XMM-Newton}

XMM-Newton \citep{2001A&A...365L...1J}, the second ESA X-ray
astronomical satellite, observed the Circinus galaxy six times between
2001 and 2018 (\citealt{2003MNRAS.343L...1M}, \citealt{2006A&A...455..153M}, \citealt{2014ApJ...791...81A}, \citealt{2019ApJ...877...95T}). 
In this paper, we utilize the data in 2013
February, which were taken simultaneously with NuSTAR (see
below).
XMM-Newton carries three X-ray CCD cameras:
EPIC/pn \citep{2001A&A...365L..18S} and two EPIC/MOSs
\citep{2001A&A...365L..27T}. We analyzed only the data of pn, which has
the largest effective area, using the science analysis software (SAS)
v18.0.0 and CALDB released on 2019 November 28.

We reprocessed the unfiltered pn data with the \textsc{epproc}
script. Source events were extracted from a circular region with a radius of
100\arcsec\ centered at the source peak, for consistency with the Suzaku spectra.
We subtracted the background taken
from a source-free circular region with a radius of 100\arcsec. We generated
the ARF and RMF by using \textsc{arfgen} and \textsc{rmfgen}, respectively. We binned the
pn spectrum to contain at least 100 counts per bin.

\subsection{NuSTAR}

NuSTAR \citep{2013ApJ...770..103H}, the first X-ray
astronomical satellite capable of focusing hard X-rays above 10 keV, 
observed the Circinus galaxy four times in 2013.
The results were published in 
\citet{2014ApJ...791...81A}, \citet{2015ApJ...805...41B}, and \citet{2019ApJ...877...95T}.
NuSTAR carries two
focal plane modules (FPMs), which are sensitive to 3--79 keV. We
utilized the data observed in 2013 January, 
which are the only data observed on-axis. 
We used HEAsoft v6.27 and CALDB released on 2020 May 14 for data reduction.

We reprocessed the FPMs data by using \textsc{nupipeline} and
\textsc{nuproducts}. Events were extracted from a circular region with a radius
of 100\arcsec\ centered at the source peak, and we subtracted the background
taken from a source-free circular region with a radius of 100\arcsec. We
binned the FPMs spectra to contain at least 100 counts per bin.

\subsection{Chandra}

Chandra \citep{2002PASP..114....1W}
observed the Circinus galaxy several times since 2000 with the Advanced
CCD Imaging Spectrometer (ACIS: \citealt{2003SPIE.4851...28G}) with or
without the High Energy Transmission Grating (HETG:
\citealt{2005PASP..117.1144C}). 
Many authors published papers using a part of these data 
(with HETG: \citealt{2001ApJ...546L..13S}; \citealt{2002A&A...396..793B}; \citealt{2010ApJS..187..581S}; \citealt{2011ApJ...738..147S}; \citealt{2014ApJ...791...81A}; \citealt{2016MNRAS.463L.108L,2016MNRAS.459L.105L}; \citealt{2018ApJ...867...80H}; \citealt{2019PASJ...71...68K};, without HETG: \citealt{2013MNRAS.436.2500M}; \citealt{2014ApJ...791...81A}, \citealt{2019PASJ...71...68K}).
Its excellent angular resolution (${<}1$\arcsec)
enables us to separate the AGN from nearby X-ray
sources by utilizing the ACIS imaging data. We analyzed 14 observations 
(13 with HETG and 1 without HETG). 
We followed the standard analysis procedures with the Chandra
interactive analysis of observations (CIAO v4.12) software and
calibration files (CALDB v4.9.1).

\subsubsection{HETG Data}

Since our main purpose is to measure
the width of the Fe K$\alpha$ line, 
it is essential to fully utilize the available Chandra/HETG
data, which achieve the best energy resolution in the iron-K band among
existing missions. We analyzed the data of the whole 13 observations,
whose summed net exposure amounted to 0.62~Ms. Source events were
extracted from 4.8\arcsec\ full-width HEG regions in the cross-dispersion
direction centered at the AGN. The background was extracted from both
sides of the source extraction region by carefully selecting areas not
containing the two bright sources: CGX1 and CGX2 (see Section~\ref{subsubsection:ACIS}). We combined
the spectra from all the observations. To perform $\chi^2$ minimization
simultaneously with the other spectra, the first order HEG spectrum was
binned to contain at least 50 counts per bin.

\subsubsection{ACIS Imaging Data} \label{subsubsection:ACIS}

The X-ray image of the Circinus galaxy is complex and is extended over a
few arcmins, consisting of the AGN core, point-like sources, and
diffuse emission (\citealt{2009ApJ...691..131Y}, \citealt{2013MNRAS.436.2500M}, \citealt{2014ApJ...791...81A},
\citealt{2019PASJ...71...68K}). Hence, the Suzaku, XMM-Newton, and NuSTAR spectra taken
with a large aperture (100\arcsec) contain all these components. To estimate
the contribution from the contaminating sources near the AGN in these
spectra, we re-analyzed the imaging data observed with ACIS-S without
HETG in 2010 December. We extracted the spectra of CGX1 (ultraluminous
X-ray source; ULX) and CGX2 (supernova remnant; SNR) from circular
regions with radii of 3\arcsec\ centered at the source peaks. We subtracted
the backgrounds taken from annuli with inner and outer radii of 3\arcsec\ and
4\arcsec\ surrounding the sources. There are also dimmer point sources and
diffuse emission around the AGN. We extracted their integrated spectrum
(we hereafter refer to it as ``diffuse spectrum'') from a circular
region with a radius of 100\arcsec\ centered at the AGN by masking the three
regions with radii of 3\arcsec\ centered at the AGN, CGX1, and CGX2. We binned
the CGX1, CGX2, and diffuse spectra to contain at least 50, 30 and 100
counts per bin, respectively.

\section{Spectral Analysis and Results} \label{results}

We first evaluate the contribution of the contaminating sources with the
Chandra imaging data, basically following \citet{2014ApJ...791...81A} with
revisions. Then, we perform a simultaneous fit to the 3--100 keV
spectra of the Circinus galaxy observed with Suzaku, XMM-Newton, NuSTAR,
and Chandra/HETG, 
in order to best determine the broadband spectral model of the AGN. 
Finally, we perform a detailed analysis of the Fe
K$\alpha$ line profile with the Chandra/HETG spectra of 
different grating orders, 
where we take into account the effect of spatial extent.
In the spectral fit, we always multiply the Galactic absorption, 
modelled by \textsf{phabs} in XSPEC \citep{1996ASPC..101...17A}, to the
spectral models of the AGN and galaxy emission. Its hydrogen column
density is fixed at $7.02\times 10^{21}\ \mathrm{cm}^{-2}$, a value
estimated by the method of \citet{2013MNRAS.431..394W}.

\subsection{Contamination} \label{subsection:contamination}
\subsubsection{CGX1}

\begin{figure*}[t!]
 \plottwo{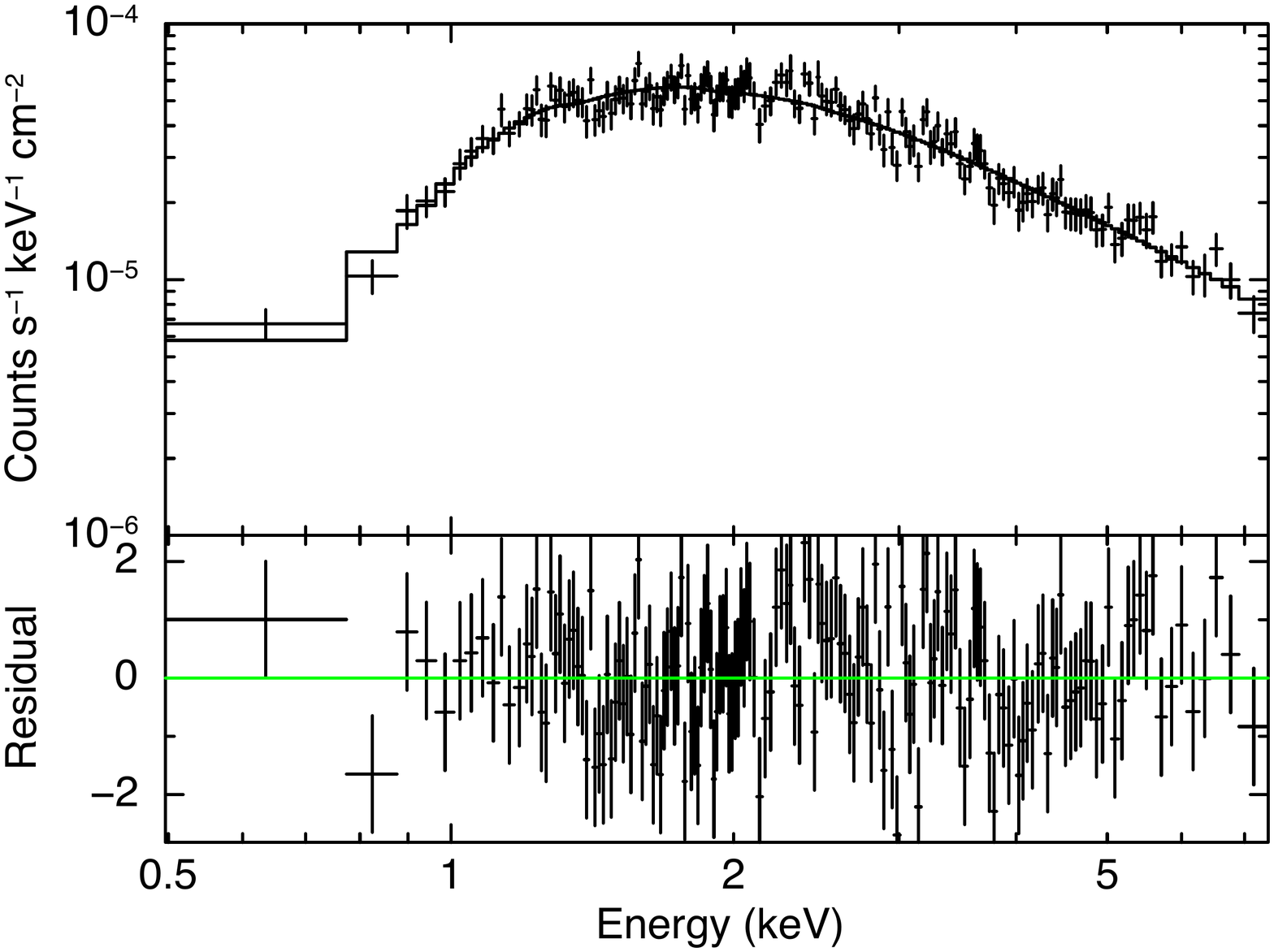}{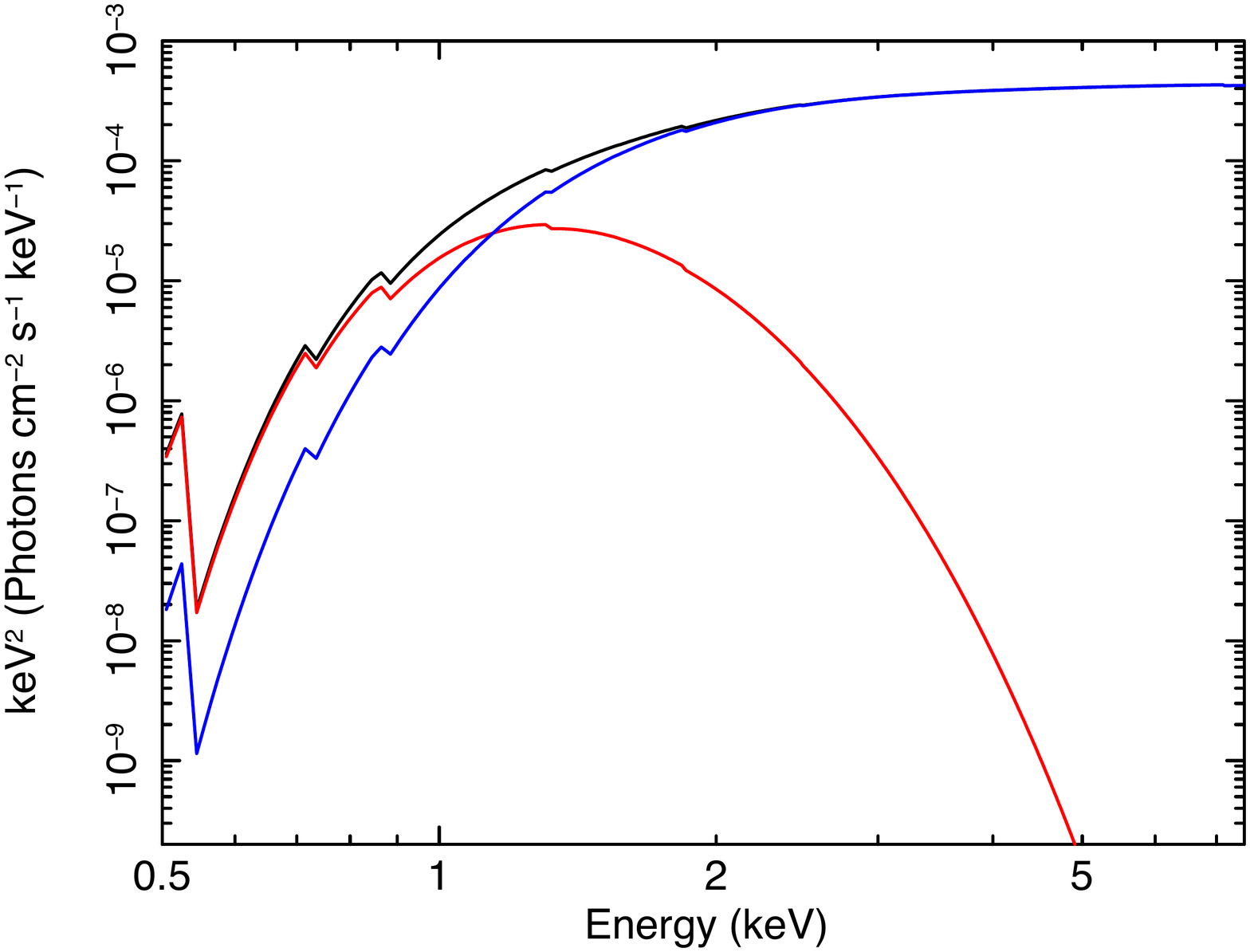}
 \caption{(Left) The folded spectrum of CGX1 obtained with Chandra/ACIS.
 The lower panel shows the fitting residuals. (Right) The best-fit
 models. 
 The black, blue, and red lines represent the total component, the Comptonized component, and the disk black body component, respectively. \label{figure:cgx1}}
\end{figure*}

\citet{2019ApJ...877...57Q} showed that CGX1 was an ULX. 
Here we adopt a spectral model developed by \citet{2017ApJ...839...46S},
a physically motivated model that well reproduces broadband spectra
of a ULX covering the hard X-ray band above 10 keV.
This model consists of two components: disk blackbody radiation from the
accretion disk and Comptonization in the optically thick corona.
The seed photon temperature of the Comptonized component ($T_0$) is 
set to be twice that of the disk innermost temperature ($T_\mathrm{in}$)
(see \citealt{2017ApJ...839...46S} for details). 
The model is
represented as follows in the XSPEC 
terminology:
\begin{equation}
 \label{equation:cgx1}
 \mathrm{CGX1}=\textsf{phabs*(diskbb+mscomptt)},
\end{equation}
where \textsf{mscomptt} is a local model by \citet{2017ApJ...839...46S}.
We obtain a good fit with the model ($\chi^2_{\mathrm{red}}=175/161$).
Table~\ref{table:cgx1} summarizes the best-fit
parameters. 
Figure~\ref{figure:cgx1} shows the X-ray spectrum folded
with the energy response and the best-fit model.

\begin{deluxetable*}{CCCCCCCCCC}
 \tablecaption{Best Fit Parameters of CGX1 \label{table:cgx1}}
 \tablewidth{0pt}
 \tablehead{
 \colhead{$N_{\mathrm{H}}$} & \colhead{$kT_{\mathrm{in}}$} & \colhead{$N_{\mathrm{dbb}}$} & z & \colhead{$kT_{\mathrm{0}}/kT_{\mathrm{in}}$} & \colhead{$kT_{\mathrm{e}}$} & \colhead{$\tau$} & \colhead{$N_{\mathrm{mscomptt}}$} & \colhead{$F_{2-10}$} & \colhead{$\chi^2/\mathrm{dof}$} \\
 (1) & (2) & (3) & (4) & (5) & (6) & (7) & (8) & (9) &  
 }
 \startdata
 1.07^{+0.21}_{-0.19} & 0.21^{+0.04}_{-0.04} & 29.6^{+70.4a}_{-20.3} & 0.0014\ (\mathrm{fixed}) & 2.0\ (\mathrm{fixed}) & 4.0\ (\mathrm{fixed}) & 5.03^{+0.26}_{-0.30} & (1.02^{+0.16}_{-0.11})\times 10^{-4} & 9.7\times 10^{-13} & 175/161  \\
 \enddata
 \tablecomments{
 Column (1): hydrogen column density in units of $10^{22}$ atoms
 $\mathrm{cm}^{-2}$. Column (2): innermost temperature of the
 disk. Column (3): normalization of diskbb. Column (4): reshift fixed at
 that of the Circinus galaxy. Column (5): 
 ratio of the seed-photon temperature ($T_0$) to the disk innermost temperature ($T_{\mathrm{in}}$). Column (6): plasma
 temperature in units of keV. Column (7): Thomson scattering optical depth. Column (8): normalization of mscomptt.
 Column (9): observed X-ray flux in the 2--10 keV band in units of $\mathrm{erg\ cm^{-2}\ s^{-1}}$.
 }
 \tablenotetext{a}{The parameter reaches a limit of its allowed range.}
\end{deluxetable*}

\subsubsection{CGX2}

\begin{figure*}[t!]
 \plottwo{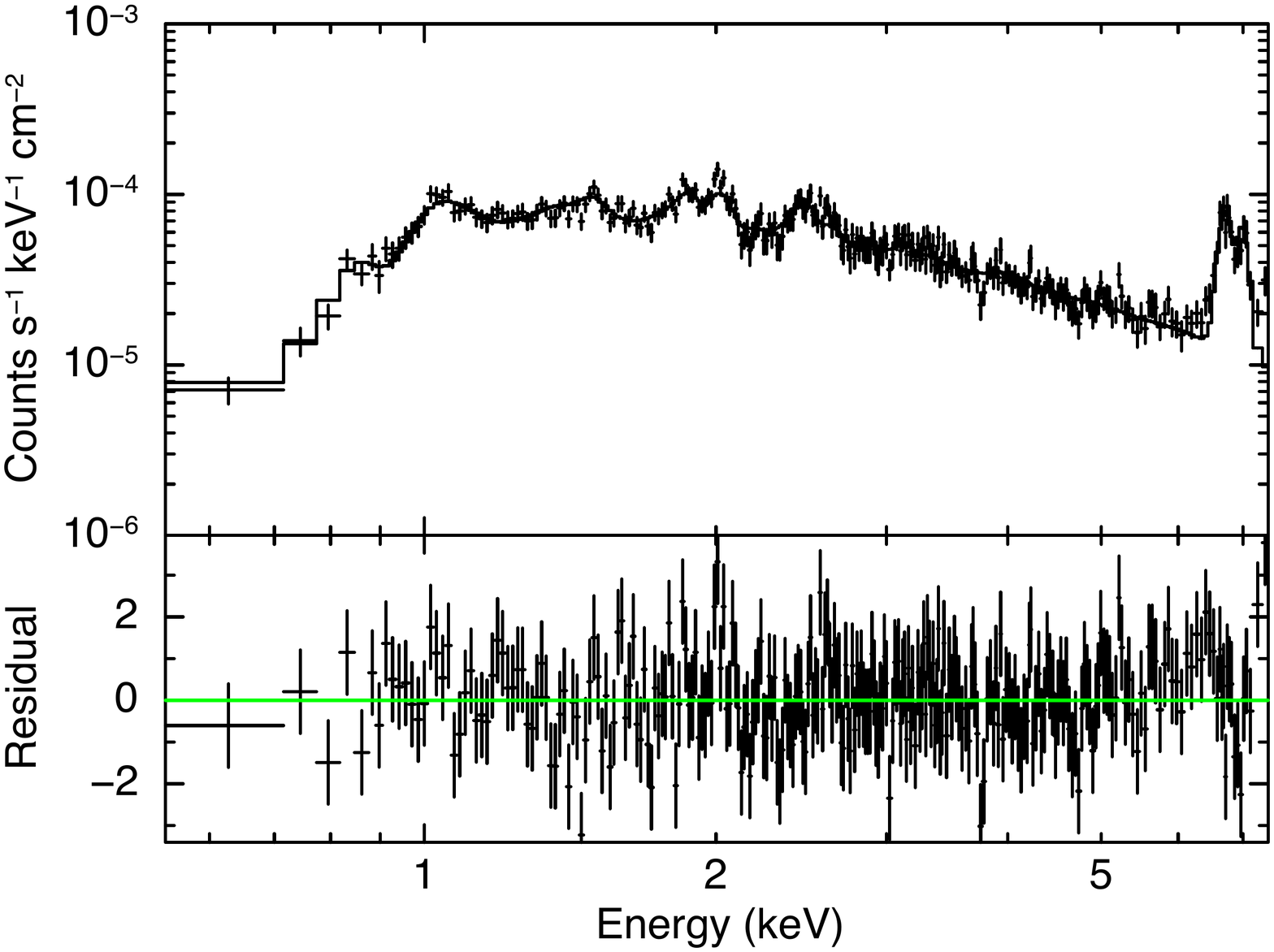}{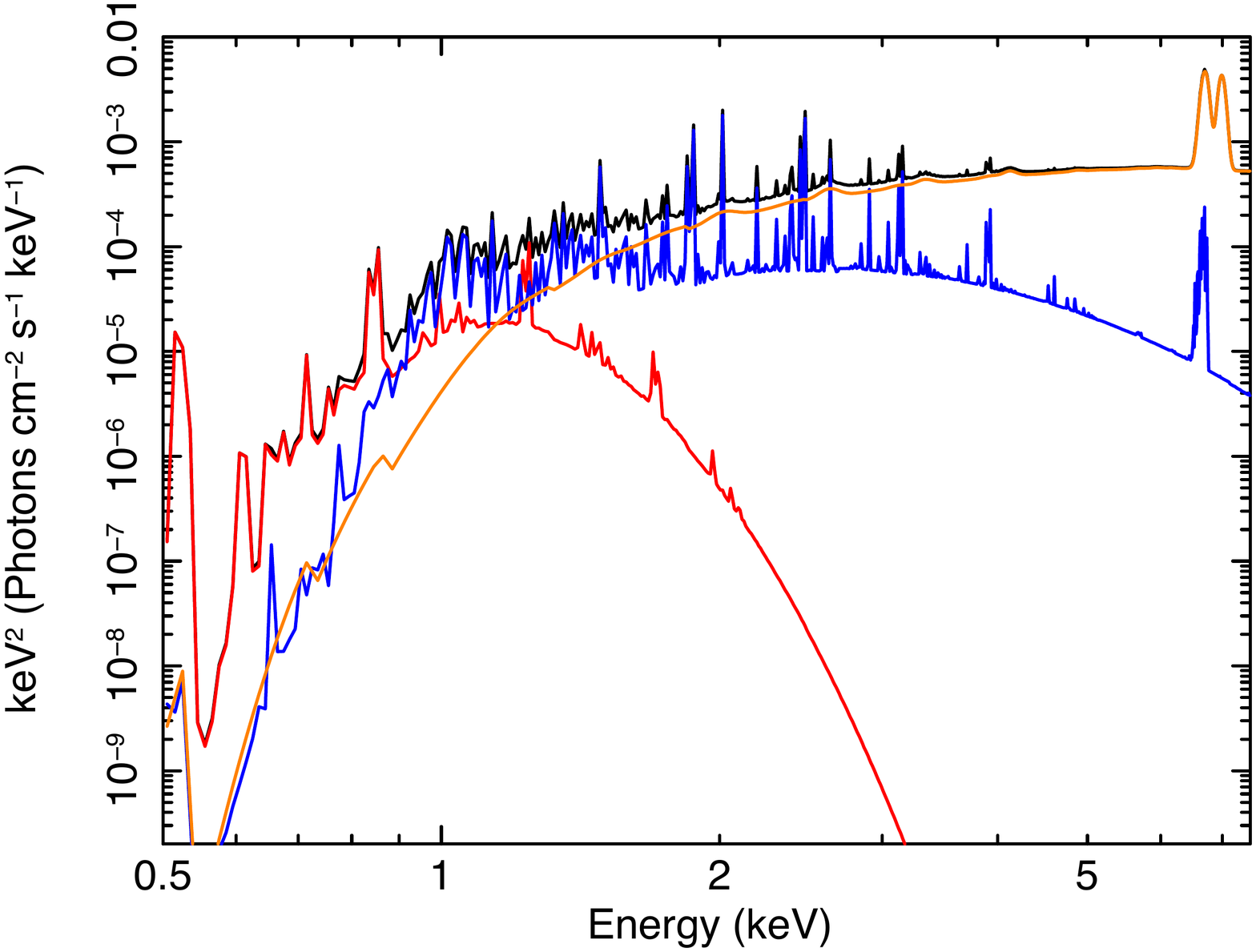}
 \caption{(Left) The folded spectrum of CGX2 observed with
 Chandra/ACIS. The lower panel shows the fitting residuals. (Right)
 The best-fit models. The black line represents the total
 spectrum. The red, blue, and organge lines correspond to the three
 components of optically-thin thermal emission in descending order of
 temperature. \label{figure:cgx2}}

\end{figure*}

To model the spectrum of CGX2, an SNR, we use basically the same model
as that in \citet{2014ApJ...791...81A}, which consists of three
components of optically-thin thermal plasma.
The model is represented as follows in the XSPEC terminology:
\begin{eqnarray}
 \label{equation:cgx2}
 \mathrm{CGX2} &=& \textsf{phabs} \\
 &*& \{\textsf{apec1+apec2+gsmooth(0.065\ keV)*apec3}\} \nonumber
\end{eqnarray}
Here we employ the \textsf{apec} model instead of \textsf{mekal} adopted by \citet{2014ApJ...791...81A}
because the latter does not cover the energy band above ${\sim}50$ keV.  
We confirm that it successfully reproduces the spectrum
($\chi^2_{\mathrm{red}}=308/270$). Table~\ref{table:cgx2} summarizes the
best fit parameters. Figure~\ref{figure:cgx2} shows the folded X-ray
spectrum and the best-fit models.

\begin{deluxetable*}{lCCCCC}
 \tablecaption{Best Fit Parameters of CGX2 \label{table:cgx2}}
 \tablewidth{0pt}
 \tablehead{
 \colhead{Model} & \colhead{$N_{\mathrm{H}}$} & \colhead{$kT$} & \colhead{$Z$} & \colhead{$z$} & \colhead{Norm.} \\
 & \colhead{($10^{22}$)} & \colhead{(\textrm{keV})} & & &
 }
 \startdata
 phabs & 1.64^{+0.09}_{-0.05} & \cdots & \cdots & \cdots & \cdots \\
 apec1 & \cdots & 0.143^{+0.003}_{-0.012} & 0.5\ (\mathrm{fixed}) & (8.28^{+0.20}_{-1.25}) \times 10^{-2} & (1.75^{+1.24}_{-0.65}) \times 10^{-1} \\
 apec2 & \cdots & 1.15^{+0.05}_{-0.05} & 5.0^{+0.9}_{-0.6} & (-5.65^{+0.70}_{-8.58}) \times 10^{-3} & (2.16^{+0.39}_{-0.10}) \times 10^{-4} \\
 apec3 & \cdots & 9.88^{+0.71}_{-0.73} & 5.0\ (\mathrm{linked}) & (-2.63^{+0.15}_{-0.44}) \times 10^{-3} & (6.42^{+0.15}_{-0.49}) \times 10^{-4} \\
 \enddata
 \tablecomments{
 The metal abundance of the lowest temperature component is fixed
 at 0.5. Those of the high and medium temperature components are
 linked together.
 }
\end{deluxetable*}

\subsubsection{Diffuse Emission}

The diffuse emission is mainly originated from X-ray
reflection by gas around the nucleus of the intrinsic AGN
emission. Accordingly, we adopt \textsf{pexmon}
\citep{1995MNRAS.273..837M}, a Compton reflection model from optically
thick matter, and a scattered component by optically thin matter (cutoff
power-law), to reproduce the spectrum. We also add emission lines
reported in \citet{2014ApJ...791...81A} that are not included in the
\textsf{pexmon} model (Table~\ref{table:lines}).  The model is represented as
follows in the XSPEC terminology:

\begin{equation}
 \label{equation:diffuse}
 \mathrm{diffuse}=\textsf{phabs*(pexmon+zcutoffpl+zgausses)}
\end{equation}
We link the photon index and cutoff energy
to those of the intrinsic AGN
component in the simultaneous fitting of the broadband spectra (see
Section~\ref{subsubsection:spectral}). 
The inclination angle in the \textsf{pexmon} model is fixed at 60\degr.

\subsection{Broadband Spectral Fitting} \label{subsection:broad}

\subsubsection{Spectral Models} \label{subsubsection:spectral}

Earlier works showed that many high-order ionization lines were present 
in the spectrum of the Circinus galaxy below 3 keV 
(e.g., \citealt{2001ApJ...546L..13S}).
This suggests that the soft band is characterized by significant emission from
photoionized plasma.
To focus on the reflection component from the torus, we ignore the
data below 3 keV. We perform
simultaneous fits to the spectra observed with Chandra/HETG
(HEG first order, 3--8 keV), Chandra/ACIS (imaging, 3--7 keV),
Suzaku/XIS-FI (3--10 keV), Suzaku/XIS-BI (3--10 keV),
Suzaku/PIN (20--40 keV), Suzaku/GSO (50--100 keV),
XMM-Newton/pn (3--10 keV), NuSTAR/FPMA (8--70 keV), 
and NuSTAR/FPMB (8--70 keV).

The X-ray spectrum of an obscured AGN mainly consists of three
components: a transmitted component, a reflection component from the
torus, and an unabsorbed scattered component (e.g.,
\citealt{2016ApJS..225...14K}, \citealt{2018ApJ...853..146T}).
We accordingly model the emission from the AGN (not including
the diffuse component mentioned in Section~\ref{subsection:contamination}) as follows in the XSPEC terminology:
\begin{eqnarray}
 \label{equation:agn}
 \mathrm{AGN} &=& \textsf{phabs} \nonumber\\
 &*& (\textsf{zvphabs*cabs*zcutoffpl + const1*zcutoffpl} \nonumber\\
 &+& \textsf{atable\{xclumpyv\_R.fits\}} \nonumber\\
 &+& \textsf{const2*atable\{xclumpyvd\_L.fits\}} \nonumber\\
 &+& \textsf{zgausses})
\end{eqnarray}
\begin{enumerate}
 \item The first term (\textsf{phabs}) represents the Galactic absorption.
 \item The first term in the parentheses represents the transmitted
	  component through the torus. The \textsf{zvphabs} and \textsf{cabs}
	  models represent photo-electric absorption and Compton
	  scattering by the torus, respectively.
 \item The second term in the parentheses represents the unabsorbed
	  scattered component. The parameters of \textsf{zcutoffpl}
	  (photon index, cutoff energy, normalization) are linked to
	  those of the transmitted component. The \textsf{const1} factor
	  gives the scattering fraction.
 \item The third and fourth terms in the parentheses represent the
	  reflection continuum and the fluorescence lines from the
	  torus. The metal abundance, photon index, cutoff energy, and
	  normalization are set to those of the transmitted
	  component. We introduce a relative normalization factor of the
	  emission lines to the reflection continuum (\textsf{const2}) to
	  take into account systematic uncertainties due to the
	  simplified assumption of the torus geometry. 
 \item We add the emission lines reported by \citet{2014ApJ...791...81A} 
	  not included in the XCLUMPY model (Table \ref{table:lines}).
\end{enumerate}

The whole model used for the simultaneous fitting is 
represented as follows:
\begin{eqnarray}
 \label{equation:model}
 \mathrm{model1} &=& \textsf{const3}*\mathrm{AGN} \nonumber\\
 \mathrm{model2} &=& \textsf{const3}*\mathrm{diffuse} \nonumber\\
 \mathrm{model3} &=& \textsf{const3}*(\mathrm{AGN} + \mathrm{diffuse} + \textsf{const4}*\mathrm{CGX1} \nonumber\\
 &+& \mathrm{CGX2})
\end{eqnarray}
\begin{enumerate}
 \item The model forms of `CGX1', `CGX2', `diffuse' and
 `AGN' are described in Equations~(\ref{equation:cgx1})-(\ref{equation:agn}),
 respectively. The parameters of CGX1 and CGX2 were frozen
 at the best-fit values obtained by the spectral analysis of the
 Chandra/ACIS imaging data (Section~\ref{subsection:contamination}).
 As mentioned above, the photon index and cutoff energy
 of the diffuse component are linked to those of the AGN one. 
\item We apply model1 to the Chandra/HETG spectrum, model2
 to the Chandra/ACIS diffuse spectrum, and model3 to the other spectra.
 \item The \textsf{const3} is a cross-normalization factor to correct for a
 possible difference in the absolute flux calibration among
 different instruments. We fix that of Chandra/HETG at
 unity as a reference. Those of Suzaku/XIS-FI
 ($C_{\mathrm{FI}}$), Suzaku/XIS-BI
 ($C_{\mathrm{BI}}$), XMM-Newton/pn
 ($C_{\mathrm{pn}}$), NuSTAR/FPMA
 ($C_{\mathrm{FPMA}}$) and NuSTAR/FPMB
 ($C_{\mathrm{FPMB}}$) are left as free parameters. Those of
 Suzaku/PIN and Suzaku/GSO are set to be
 $1.16\times C_{\mathrm{FI}}$ based on the calibration with the
 Crab Nebula. The cross-normalization of Chandra/ACIS
 diffuse spectrum is fixed at unity.
 \item The \textsf{const4} accounts for time variability of the ULX
 (CGX1). Those of Suzaku/XIS-FI
 ($T_{\mathrm{FI}}$) and XMM-Newton/pn
 ($T_{\mathrm{pn}}$) are left as free parameters. Those
 of Suzaku/XIS-BI,
 Suzaku/PIN, and Suzaku/GSO are linked to
 $T_{\mathrm{FI}}$, whereas those of 
 NuSTAR/FPMA and NuSTAR/FPMB are to
 $T_{\mathrm{pn}}$ (because the NuSTAR and XMM-Newton observations
 were simultaneous).
 We note that the X-ray spectrum of a ULX changes with luminosity; generally, it becomes softer at higher luminosities (e.g.,
 \citealt{2014ApJ...793...21W}; \citealt{2017ApJ...839...46S}). 
 Applying the same model adopted here to the spectra of IC 342 X-1 observed
 in multiple epochs, \citet{2017ApJ...839...46S} obtained $\sim$20\% smaller
 optical depths ($\tau$) of the Comptonizing corona when the luminosity
 increased by a factor of $2-3$. Accordingly, we repeat our
 analysis by setting $\tau=4$ for the CGX1 spectra in the Suzaku and
 NuSTAR/XMM-Newton data, where its luminosity is estimated to be $\sim2$
 higher than in the Chandra data (Table~4). We confirm that it has little
 effect on the result of our broadband spectral fitting of the Circinus
 galaxy because the luminosity of CGX1 is sufficiently small compared
 with that of the AGN. Thus, we adopt
 the result obtained by considering only the luminosity change of CGX1.
\end{enumerate}

All the redshift parameters in the AGN and diffuse components are linked
together for the spectrum of each instrument. Since the AGN contains a
strong Fe K$\alpha$ line, the redshift determination based on it very
sensitively depends on the precise energy-scale calibration of each
instrument; in fact, we find that adopting a common redshift value for
all the instruments leads to highly unacceptable fits.
(If we fix the redshifts of all instruments to 0.0014, 
we obtain even a much worse fit with $\chi^2/\mathrm{dof}=2468/1996$.)
To take into account the possible calibration uncertainties in an approximated way,
we allow the linked value of the redshift to vary among different
instruments except for the Suzaku/HXD and NuSTAR spectra, for which we
fix $z=0.0014$. We confirm that the resultant redshift parameters are
consistent with $z=0.0014$ within the calibration uncertainties
(Chandra Calibration Status Summary\footnote{\texttt{https://cxc.cfa.harvard.edu/cal/summary/ Calibration\_Status\_Report.html}}, \citealt{2007PASJ...59S..23K}, XMM-Newton Calibration Technical Note 
\footnote{\texttt{https://xmmweb.esac.esa.int/docs/documents/CAL-TN-0018.pdf}}). 
Note that, for simplicity, we have ignored these small
scale uncertainties in modeling the CGX1 and CGX2 spectra because they
do not affect our results on the AGN component.

\subsubsection{Results}

\begin{figure*}[htb]
\plottwo{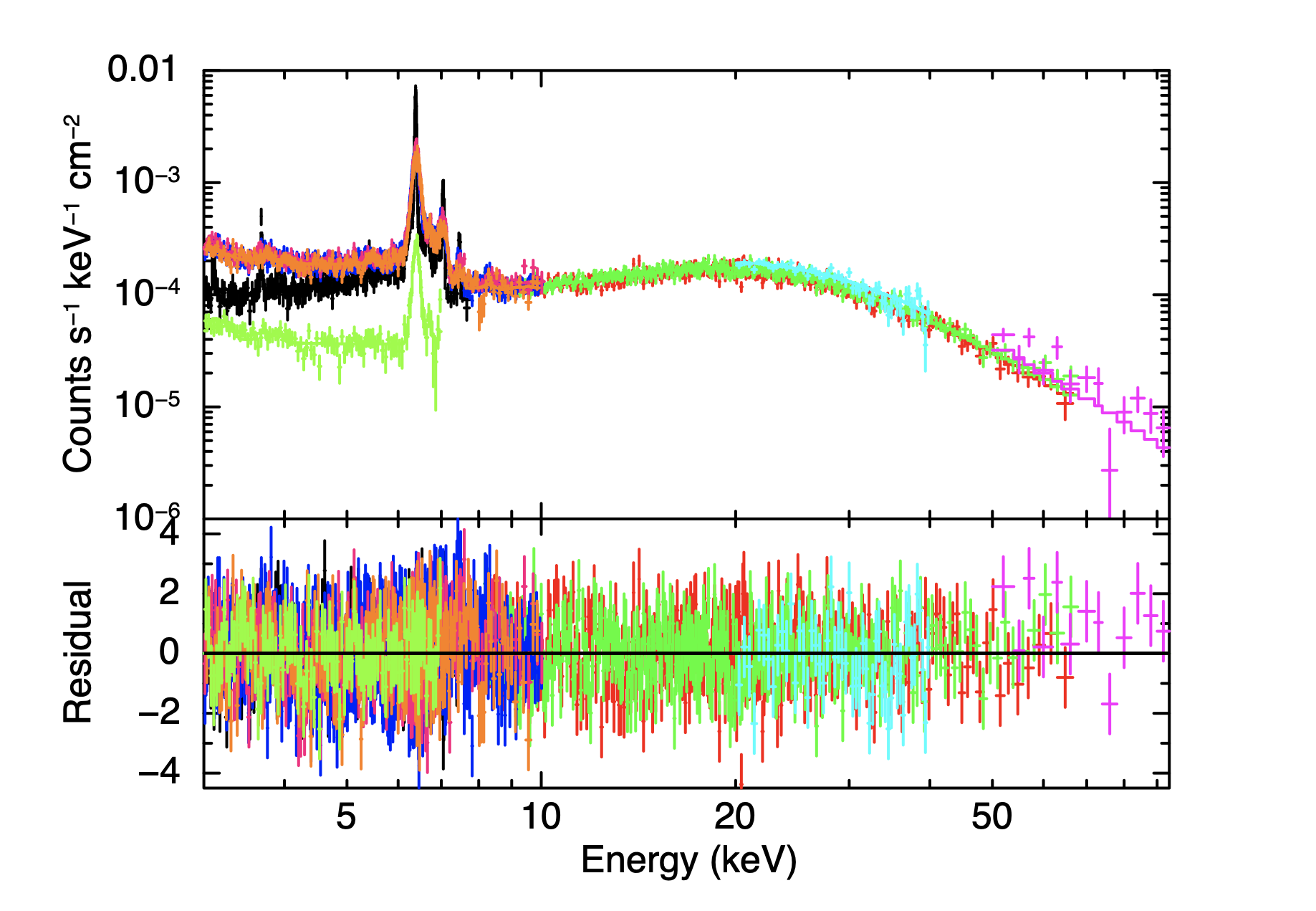}{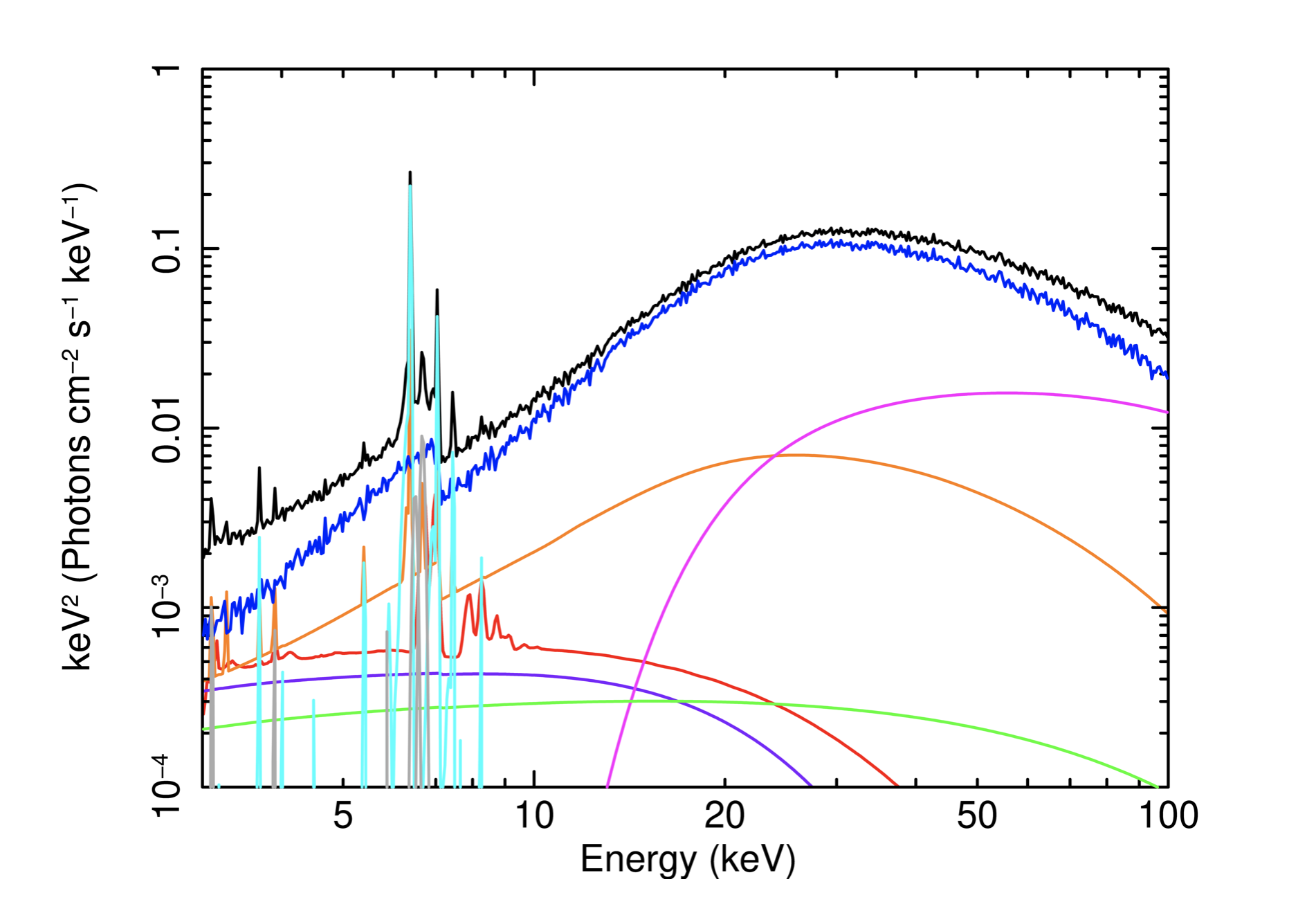}
\caption{(Left) The folded spectra of the Circinus galaxy. Black:
Chandra/HETG. Blue: Suzaku/XIS-FI.  Dark red:
Suzaku/XIS-BI. Light blue; Suzaku/PIN.  Magenta:
Suzaku/GSO. Orange: XMM-Newton/PN. Red:
NuSTAR/FPMA. Green: NuSTAR/FPMB. Light green:
Chandra/ACIS. (Right) Black: total. Magenta: direct
component. Green: scattered component. Blue: reflection continuum of
the updated XCLUMPY model (xclumpyv\_R.fits). Light Blue:
fluorescence lines of the updated XCLUMPY model (xclumpyvd\_L.fits),
Orange: diffuse contamination. Purple: CGX1. Red: CGX2. Gray: added
emission lines. The noises noticeable 
in the reflection continuum model are due to the statistic fluctuation in the
simulated spectra, which have negligible effects on the fitting. \label{figure:broad}}
\end{figure*}

We successfully reproduce the broadband (3--100 keV) spectra of the
Circinus galaxy with the above model. The best-fit parameters are
summarized in Table~\ref{table:broadbest}. Figure~\ref{figure:broad}
shows the folded X-ray spectra and best-fit models.
We obtain the main torus parameters 
$N_\mathrm{H}^\mathrm{Equ}=2.16^{+0.24}_{-0.17}\times 10^{25}\ \mathrm{cm^{-2}}$,
$\sigma=10.3^{+0.8}_{-0.3}\ \mathrm{degrees}$, and $i=78.3^{+0.5}_{-0.9}\ \mathrm{degrees}$. 
We also find that the supersolar abundance
($Z=1.52^{+0.05}_{-0.06}$) significantly improve the fit, mainly to reproduce 
the Compton-shoulder fraction (see Appendix~\ref{appendix:metal}).
The cross-normalization factors of XMM-Newton/pn and Suzaku/XIS
are consistent with \citet{2017AJ....153....2M} but those of 
NuSTAR/FPMs are not.\footnote{We infer that, this is because
\citet{2017AJ....153....2M} calibrated
NuSTAR/FPMs in the 1--5 keV and 3--7 keV bands, whereas
our cross-normalization factors 
are determined by using only the 8--10 keV band.}
The best-fit value of \textsf{const2} (the relative normalization factor of line
components to the continuum in XCLUMPY), $1.08^{+0.06}_{-0.05}$, is slightly over 1.0.
Since the torus parameters, such as $N_{\mathrm{H}}^{\mathrm{Equ}}$,
affect the equivalent
width of the Fe K$\alpha$ line (see Figure 4 in
\citealt{2019ApJ...877...95T}), this may be explained if 
the actual distribution of
clumps in the elevation direction is not perfectly represented by a 
single gaussian model as assumed (e.g., more material exists 
very close to the equatorial plane). Nevertheless, the result that
\textsf{const2} is close to unity supports that the assumed geometry 
is a good approximation. The bolometric luminosity is
estimated to be $L_\mathrm{bol}=1.27^{+0.09}_{-0.13}\times 10^{44}\
\mathrm{erg\ s^{-1}}$ by assuming the relation of X-ray and bolometric
luminosities of $L_\mathrm{bol}=20\times L_\mathrm{2-10}$
\citep{2009MNRAS.392.1124V}.\footnote{The error is estimated by fixing
the photon index at the best-fit value.}  We compare these results with
previous studies in Section~\ref{discussion}.

\begin{deluxetable*}{lCCCc}[htb]
 \tablecaption{Best Fit Parameters of the Broadband Spectra \label{table:broadbest}}
 \tablewidth{0pt}
 \tablehead{
 \colhead{Region} & \colhead{No.} & \colhead{Parameter} & \colhead{Best Fit Value} & \colhead{Units}
 }
 \startdata
 AGN & (1) & N_\mathrm{H}^\mathrm{Equ} & 21.6^{+2.4}_{-1.7} & $10^{24}\ \mathrm{cm}^{-2}$ \\
 & (2) & N_\mathrm{H}^\mathrm{LOS} & 5.95^{+0.76}_{-0.60} & $10^{24}\ \mathrm{cm}^{-2}$ \\
 & (3) & Z & 1.52^{+0.05}_{-0.06} & solar value \\
 & (4) & \sigma & 10.3^{+0.8}_{-0.3a} & degree \\
 & (5) & i & 78.3^{+0.5}_{-0.9} & degree \\
 & (6) & \Gamma & 1.68^{+0.05}_{-0.09} &  \\
 & (7) & E_\mathrm{cut} & 48.0^{+5.0}_{-5.4} & keV \\
 & (8) & N_\mathrm{Line} & 1.08^{+0.06}_{-0.05} &  \\
 & (9) & \log(r_\mathrm{in}/r_\mathrm{g}) & 4.84^{+0.08}_{-0.08} &  \\
 & (10) & N_\mathrm{Dir} & 0.80^{+0.11}_{-0.18} & $\mathrm{photons\ keV^{-1}\ cm^{-2}\ s^{-1}}$ \\
 & (11) & f_\mathrm{scat} & 2.2^{+1.8}_{-1.1}\times 10^{-2} & \% \\
 & (12) & L_{2-10} & 6.34^{+0.43}_{-0.64} & $10^{42}\ \mathrm{erg s^{-1}}$ \\
 \hline
 diffuse & (13) & Z & 0.96^{+0.12}_{-0.11} & solar value \\
 & (14) & N_\mathrm{pexmon} & 6.75^{+0.90}_{-1.22}\times 10^{-3} & $\mathrm{photons\ keV^{-1}\ cm^{-2}\ s^{-1}}$ \\
 & (15) & N_\mathrm{zcutoffpl} & 2.14^{+0.47}_{-0.47}\times 10^{-4} & $\mathrm{photons\ keV^{-1}\ cm^{-2}\ s^{-1}}$ \\
 \hline
 & (16) & C_{\mathrm{FI}} & 0.85^{+0.02}_{-0.02} &  \\
 & (17) & C_{\mathrm{BI}} & 0.90^{+0.03}_{-0.03} &  \\
 & (18) & C_{\mathrm{pn}} & 0.81^{+0.03}_{-0.03} &  \\
 & (19) & C_{\mathrm{FPMA}} & 0.80^{+0.03}_{-0.03} &  \\
 & (20) & C_{\mathrm{FPMB}} & 0.83^{+0.03}_{-0.03} &  \\
 & (21) & T_{\mathrm{FI}} & 1.88^{+0.30}_{-0.30} &  \\
 & (22) & T_{\mathrm{pn}} & 2.29^{+0.39}_{-0.39} &  \\
 & (23) & z_\mathrm{HETG} & 1.64^{+0.10}_{-0.09}\times 10^{-3} &  \\
 & (24) & z_\mathrm{FI} & 0.70^{+0.22}_{-0.22}\times 10^{-3} &  \\
 & (25) & z_\mathrm{BI} & -0.33^{+0.39}_{-0.41}\times 10^{-3} &  \\
 & (26) & z_\mathrm{pn} & -1.41^{+0.46}_{-0.47}\times 10^{-3} &  \\
 & (27) & z_\mathrm{ACIS} & -0.72^{+0.88}_{-1.01}\times 10^{-3} &  \\
 & (28) & F_{2-10} & 1.39\times 10^{-11} & $\mathrm{erg\ cm^{-2}\ s^{-1}}$ \\
 & (29) & F_{10-100} & 2.62\times 10^{-10} & $\mathrm{erg\ cm^{-2}\ s^{-1}}$ \\
 &  & \chi^2/\mathrm{dof} & 2275/1991 & 
 \enddata
 \tablecomments{
 (1) Hydrogen column density along the equatorial plane. (2) Hydrogen 
 column density along the line of sight. (3) Metal abundance of
 the torus. (4) Torus angular width. (5) Inclination angle. (6)
 Photon index. (7) Cutoff energy. (8) Relative normalization of
 the emission lines to the reflection continuum. (9) Torus inner
 radius in units of gravitational radius. (10) Normalization of
 the direct component. (11) Scattering fraction in percent.
 (12) Intrinsic AGN luminosity in the 2--10 keV band.
 (13) Metal abundance of the diffuse emission.
 Iron abundance was set to this value.
 (14) Normalization of the reflection component in the diffuse
 emission. (15) Normalization of the scattered component in the
 diffuse emission. (16)--(20) Cross-normalization
 factors. (21)--(22) Time-variablity factors. (23)--(27) Redshift of
 each spectrum. Those of NuSTAR and Suzaku/HXD
 are fixed at 0.0014. (28)--(29) Obscured AGN flux in the 2--10 keV and 10--100 keV band.
 } 
 \tablenotetext{a}{The parameter reaches a limit of its allowed range.}
\end{deluxetable*}

\begin{deluxetable*}{lCCCC}
 \tablecaption{Emission Lines Added in the Model \label{table:lines}}
 \tablewidth{0pt} \tablehead{ \colhead{Region} & \colhead{Energy} &
 \colhead{Sigma} & \colhead{Norm.} & \colhead{ID} \\ & \colhead{(keV)} &
 \colhead{(eV)} & \colhead{($\mathrm{photons\ cm^{-2}\ s^{-1}}$)} & }
 \colnumbers \startdata AGN & 3.11 & 5.8 & 3.9^{+1.4}_{-1.3}\times
 10^{-6} & \mathrm{S_{\ XVI}\ K\beta} \\ & 3.90 & 4.8 &
 1.68^{+0.87}_{-0.87}\times 10^{-6} & \mathrm{Ar_{\ XVIII}} \\ & 5.88
 & 1.0 & 0.89^{+1.18}_{-0.89\tablenotemark{a}}\times 10^{-6} &
 \mathrm{Cr_{\ XXIV}} \\ & 6.50 & 50.0 & 1.41^{+0.48}_{-0.51}\times
 10^{-5} & \mathrm{Fe_{\ XX}} \\ & 6.68^{+0.01}_{-0.01} &
 50.6^{+17.5}_{-12.9} & 2.86^{+0.46}_{-0.40}\times 10^{-5} &
 \mathrm{Fe_{\ XXV}} \\ \hline diffuse & 3.11 & 5.8 &
 2.8^{+1.2}_{-1.2}\times 10^{-6} & \mathrm{S_{\ XVI}\ K\beta} \\ &
 3.27 & 1.2 & 2.16^{+1.02}_{-1.02}\times 10^{-6} & \mathrm{Ar_{\
 XVIII}} \\ & 3.69 & 1.0 & 1.44^{+0.89}_{-0.94}\times 10^{-6} &
 \mathrm{Ca_{\ II-XIV}+Ar_{\ XVII}} \\ & 3.90 & 4.8 &
 1.57^{+0.85}_{-0.88}\times 10^{-6} & \mathrm{Ar_{\ XVIII}} \\ & 5.41
 & 0.1 & 1.49^{+0.99}_{-1.01}\times 10^{-6} & \mathrm{Cr_{\ I}} \\ &
 6.68\mathrm{(linked)} & 10.0 & 4.9^{+2.4}_{-2.4}\times 10^{-6} &
 \mathrm{Fe_{\ XXV}} \\ \enddata \tablecomments{ Column (1):
 region. Column (2): line energy. These are fixed at the values
 reported in \citet{2014ApJ...791...81A} except for $\mathrm{Fe_{\
 XXV}}$ line. It is allowed to vary within 50 eV centered at 6.66 keV
 in the broadband fitting. The energy of diffuse $\mathrm{Fe_{\
 XXV}}$ line is linked to that of the AGN component. Column (3): line
 width. These are fixed at the values reported in
 \citet{2014ApJ...791...81A} except for AGN $\mathrm{Fe_{\ XXV}}$
 line. The width of AGN 6.66 keV line is left free in the broadband
 fitting. Column (4): normalization of the lines. These were also
 left free in the broadband fitting. Lines whose best-fit
 normalizations are less than $1.0\times 10^{-10}$ were not
 included. Column (5): line ID.
 }
 \tablenotetext{a}{The parameter reaches a limit of its allowed range.}
\end{deluxetable*}

\section{Detailed Analysis of Fe K$\alpha$ Line} \label{feline}

\subsection{Imaging Analysis} \label{subsection:imaging}

\citet{2013MNRAS.436.2500M} analyzed Chandra
imaging data of the Circinus galaxy and detected a spatial 
extent of the Fe K$\alpha$ line emitting region. The spatial extent has
line-broadening effects in grating spectra, 
which we have to correct for in order to accurately measure the intrinsic 
spectral line width.\footnote{\texttt{https://cxc.harvard.edu/newsletters/news\_24/}}
In this subsection, we confirm the spatial extent of the
Fe K$\alpha$ line emitting region by comparing the radial profile obtained from a
Chandra image with that obtained from simulations. We analyze the 0th
order image of ObsID = 4771, one of the longest exposure data with
HETGs, which is little affected by pile-up thanks to the grating mode.

\begin{figure*}[htb]
 \plottwo{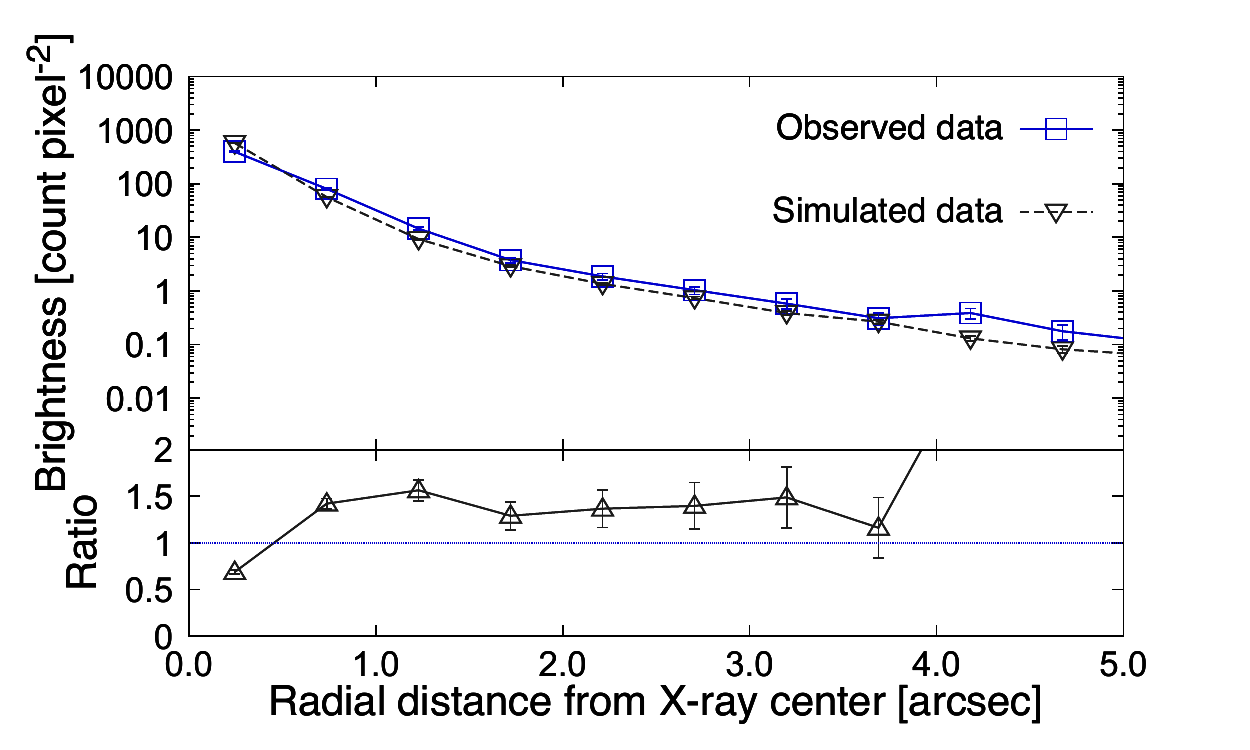}{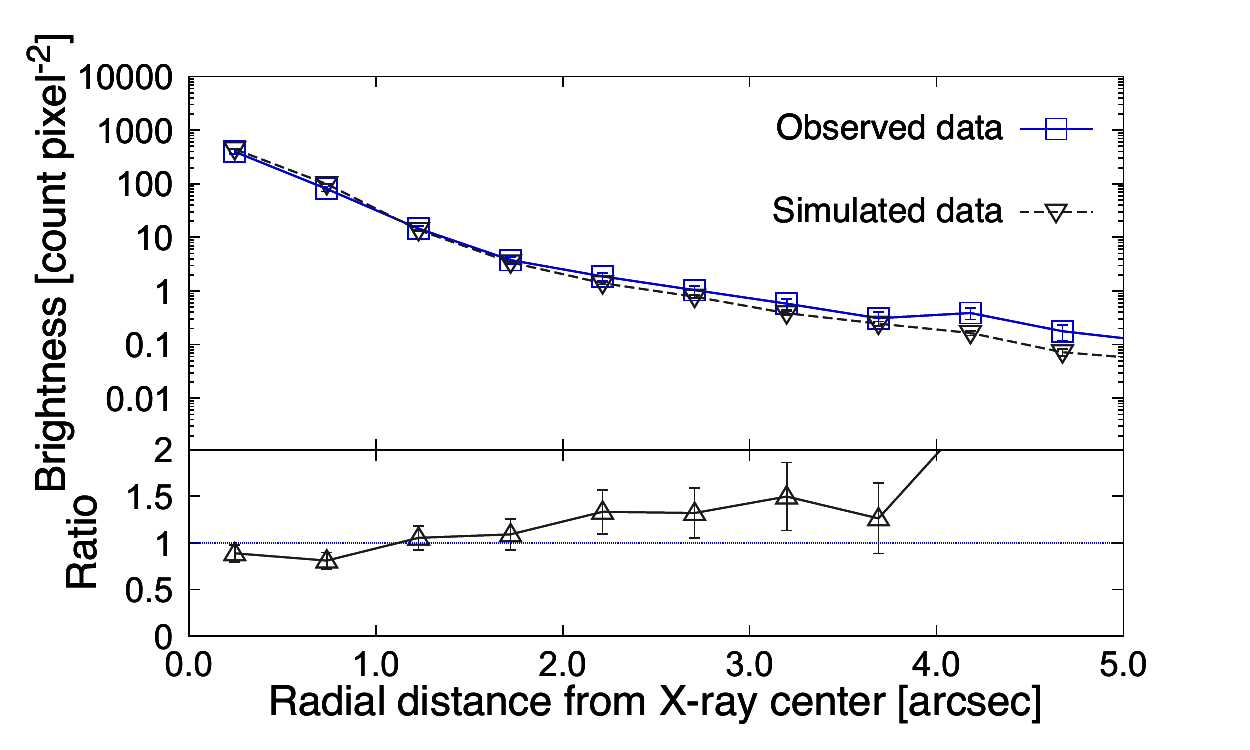}
 \caption{Comparisons of the radial profile in the 6--7 keV band
 between the observed data and simulation.
 (left) AspectBlur=0.00\arcsec\ (right)
 AspectBlur=0.25\arcsec \label{figure:rfile}}
\end{figure*}

We perform simulations with the MARX code \citep{2012SPIE.8443E..1AD},
which produces simulated event files based on a Monte-Carlo method.
The input spectrum is determined from our broadband spectral
fitting in Section~\ref{results}. 
Since the AspectBlur parameter in MARX is still under calibration, we perform simulations
with two different settings, AspectBlur=0.0\arcsec\ and 0.25\arcsec, referring to
\citet{2020ApJ...902...49F} and the latest PSF calibration by the MARX
team\footnote{\texttt{https://cxc.cfa.harvard.edu/ciao/why/aspectblur.html}}, respectively.
Figure~\ref{figure:rfile} compares the simulated and observed
source radial profiles. This confirms that the Circinus galaxy has a
spatial extent in the 6.0--7.0 keV band regardless of the adopted value of AspectBlur.
Note that the extent is not circularly symmetric but has a complex
structure \citep{2013MNRAS.436.2500M}. Hence, we cannot refer to this plot to
estimate the extent along the dispersion direction of the HEG data in a given observation.

\subsection{Spectral Analysis} \label{subsection:spectral}

Since the width of the line-broadening effect due to
a spatial extent is inversely proportional to the grating order
(Appendix~\ref{appendix:HETG}), in principle, we can separate the effect
by simultaneously analyzing multiple spectra of different grating orders.
Thus, we perform a simultaneous fitting to the Chandra/HETG first,
second, and third order spectra in the 6.3--6.5 keV band. Considering
the small photon statistics, we bin each spectrum to contain at least
one count per bin, and use Cash statistics (C-stat;
\citealt{1979ApJ...228..939C}) to determine the best-fit parameters.  We
use basically the same model as model1 (Equation~\ref{equation:model})
but take into account the line broadening effect by the spatial extent,
which is proportional to the inverse of the grating order (see
Appendix~\ref{appendix:HETG}).  The model is represented as follows in
the XSPEC terminology:

\begin{eqnarray}
 \label{equation:fe}
 \mathrm{model4} &=& \textsf{const3*phabs} \nonumber\\
 &*& (\textsf{zvphabs*cabs*zcutoffpl + const1*zcutoffpl} \nonumber\\
 &+& \textsf{atable\{xclumpyv\_R.fits\}} \nonumber\\
 &+& \textsf{(1$-$const5)*atable\{xclumpyvd\_L.fits\}} \nonumber\\
 &+& \textsf{const5*gsmooth*atable\{xclumpyvd\_L.fits\}} \nonumber\\
 &+& \textsf{zgausses})
\end{eqnarray}
\begin{enumerate}
 \item The \textsf{const5} represents the fraction of the extended
 emission in the total intensity ($f_{\mathrm{diff}}$),
 which is left as a free parameter.
 \item The 4th and 5th terms correspond to the fluorescence lines
 from the point-like and extended regions. The \textsf{gsmooth}
 model represents the spatial broadening effect. Its width is
 left as a free parameter
 ($\sigma^{\mathrm{diff}}_{\mathrm{E}}$).
 The torus inner radius in the 4th term is left free, whereas that in
 the 5th one is fixed at $\log(r_\mathrm{in}/r_\mathrm{g}) = 7.0$ (i.e., effectively no
 intrinsic spectral broadening). 
 \item The cross-normalization factors (\textsf{const3}) of the second
 and third order spectra are left as free parameters, whereas
 that of the first order one is fixed at unity. The 
 normalization of the lines (\textsf{xclumpyvd\_L.fits}) and the torus inner
 radius ($r_\mathrm{in}/r_\mathrm{g}$) are set free. 
 The other parameters are fixed to the values in
 Tables~\ref{table:broadbest} and \ref{table:lines}.
\end{enumerate}

\begin{figure*}[htbp]
 \plottwo{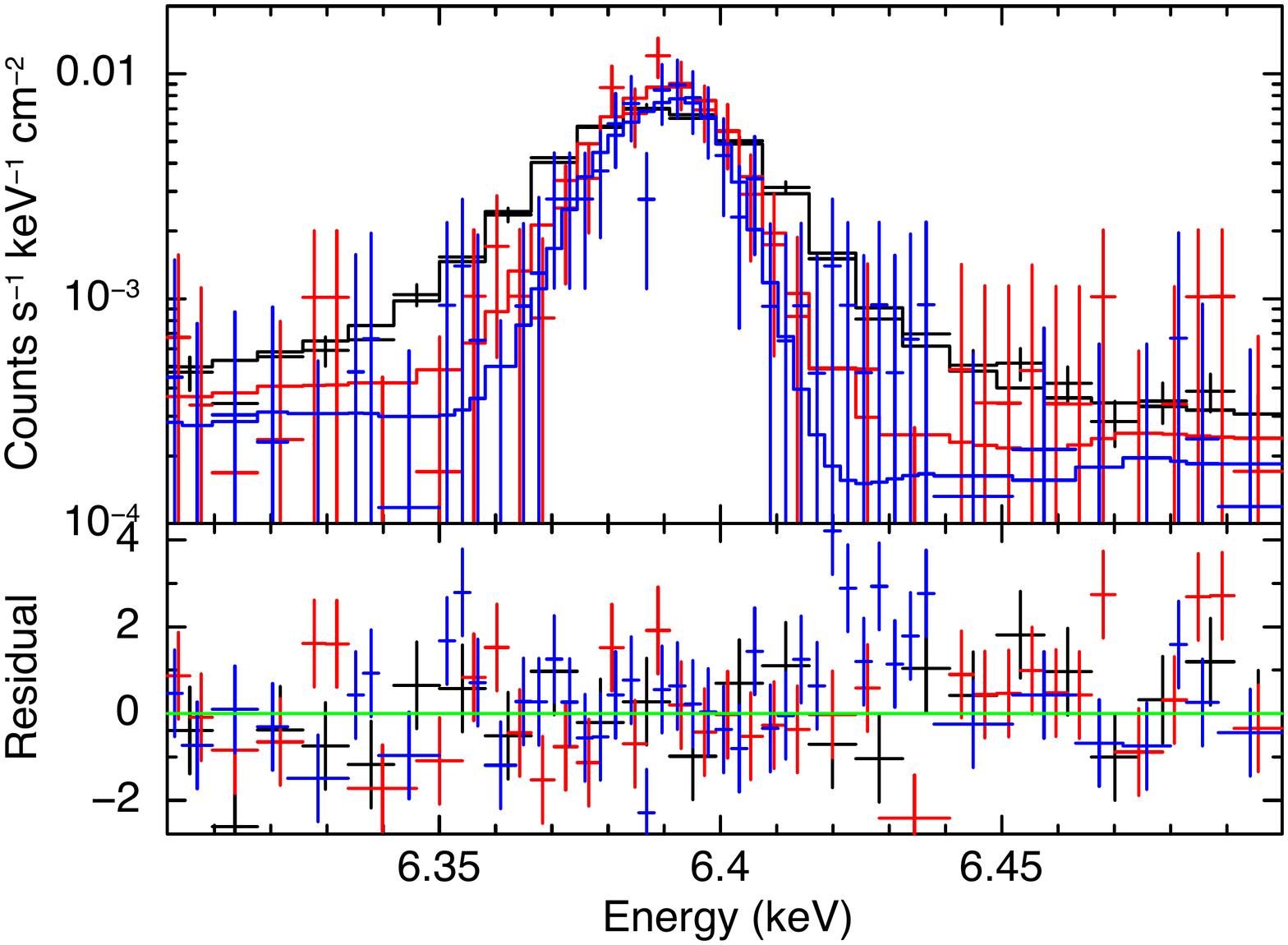}{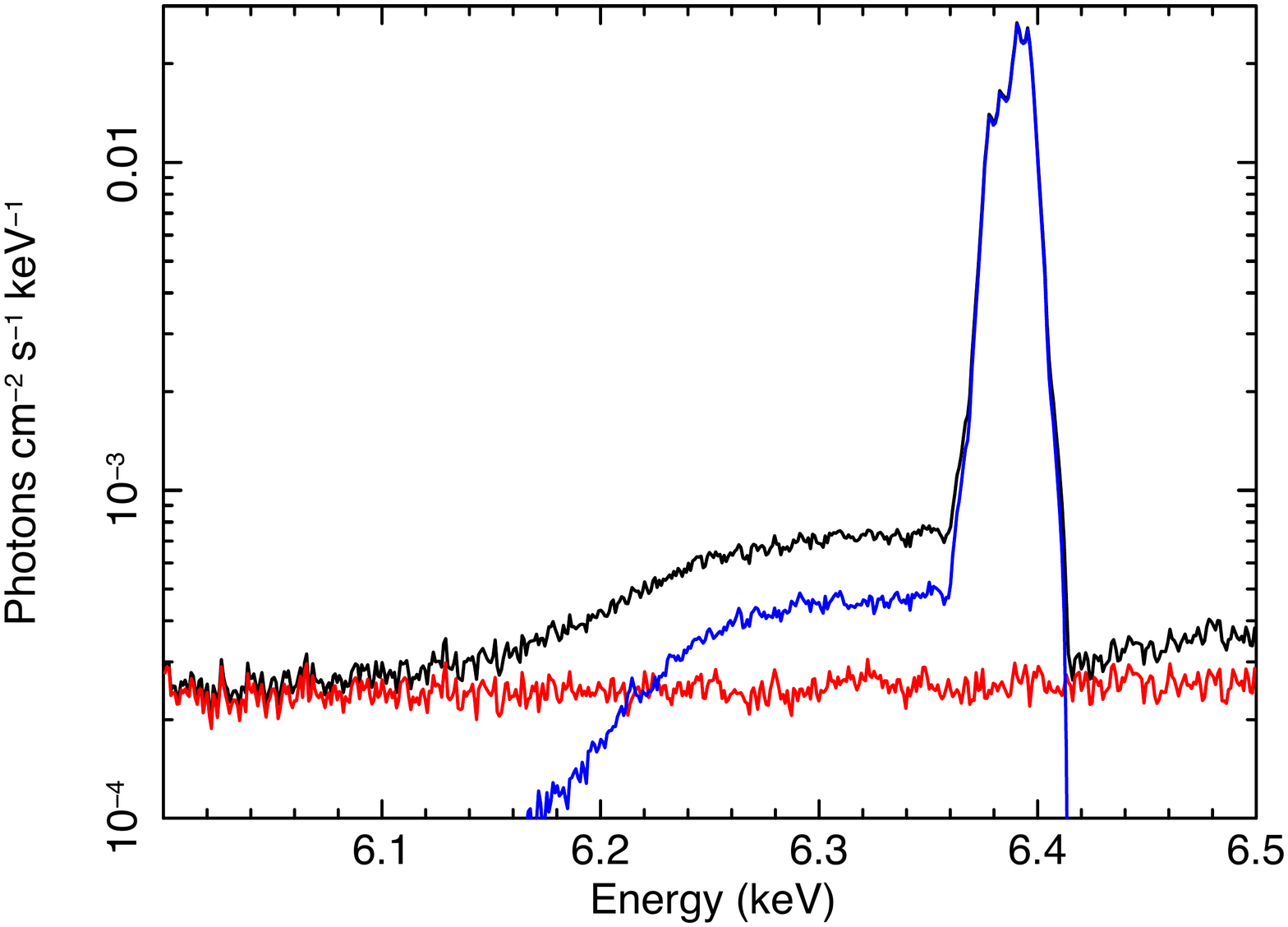}
 \caption{(Left) The folded spectra and fitting residuals
 around the Fe K$\alpha$ line. Black: Chandra/HETG first order spectrum. Red: second order. Blue: third order. (Right) The best-fit models. The black, red, and
 blue lines represent the total component, the continuum, and the
 Fe K$\alpha$ line, respectively.\label{figure:Fewidth}}
\end{figure*}

The best-fit parameters are summarized in
Table~\ref{table:Fewidth}. Figure~\ref{figure:Fewidth} shows the folded
X-ray spectra and the best-fit models.
We estimate the inner radius of the torus
to be
$\log\left(r_\mathrm{in}/r_\mathrm{g}\right)=5.28^{+0.42}_{-0.23}$. 
Since 
the mass of the SMBH is $(1.7\pm 0.3)\times 10^6~M_{\odot}$
\citep{2003ApJ...590..162G}, it corresponds to 
$r_\mathrm{in}=1.6^{+1.5}_{-0.9}\times 10^{-2}\ \mathrm{pc}$. 
At the same time, the spectral analysis also constrains the 
spatial
extent of Fe K$\alpha$ in the Circinus galaxy
averaged over the dispersion directions of the whole HEG
data used in our analysis: the fraction of the extended emission is $0.19^{+0.16}_{-0.09}$ and the angular width is
$\sigma^{\mathrm{diff}}_{\mathrm{angl}} = 0.66^{+0.33}_{-0.21}\
\mathrm{arcsec}$.

\begin{deluxetable*}{CCCCCCCCC}
 \tablecaption{Best Fit Parameters of Fe K$\alpha$ Line \label{table:Fewidth}}
 \tablewidth{0pt}
 \tablehead{
 \colhead{$C_{2\mathrm{nd}}$} & \colhead{$C_{3\mathrm{rd}}$} & \colhead{$z_{2\mathrm{nd}}$} & \colhead{$z_{3\mathrm{rd}}$} & \colhead{$f_{\mathrm{diff}}$} & \colhead{$\sigma^{\mathrm{diff}}_{\mathrm{E}}$ (eV)} & \colhead{$\log(r_\mathrm{in}/r_\mathrm{g})$} & \colhead{$\mathrm{Norm.}$} & \colhead{C/dof} \\
 (1) & (2) & (3) & (4) & (5) & (6) & (7) & (8) & 
 }
 \startdata
 0.83^{+0.11}_{-0.10} & 0.64^{+0.09}_{-0.08} & 1.57^{+0.25}_{-0.25}\times 10^{-3} & 1.53^{+0.25}_{-0.26}\times 10^{-3} & 0.19^{+0.16}_{-0.09} & 24.8^{+12.4}_{-7.7} & 5.28^{+0.42}_{-0.23} & 0.91^{+0.04}_{-0.04} & 98.8/100
 \enddata
 \tablecomments{
 Columns (1) and (2): Cross-normalization factors for the second and third
 order spectra, respectively. Columns (3) and (4): Redshifts for the
 second and third order spectra, respectively. Column (5): Fraction
 of the extended emission. Column (6): Width of the spatial
 extent. Column (7): Torus inner radius in units of gravitational
 radius. Column (8): Normalization of xclumpyvd\_L.fits.
 }
\end{deluxetable*}

\section{Discussion} \label{discussion}

We have updated the XCLUMPY model by incorporating two new features:
variable metal abundance and Doppler effects due to a Keplerian motion
of each clump. Using this model, we have analyzed currently the best
available broadband (3--100 keV) X-ray spectra of the Circinus galaxy
observed with Suzaku, XMM-Newton, NuSTAR and Chandra,
including deep Chandra/HETG data of 0.62 Ms. These rich datasets,
together with the state-of-art reflection model from an AGN torus, have
enabled us to best determine the shape of the broadband spectrum and Fe
K$\alpha$ line profile of the Circinus galaxy. Below, we discuss our
results focusing on the torus properties.

\subsection{Broadband Spectra}

We confirm the previous results based on broadband X-ray spectroscopy
(e.g., \citealt{2014ApJ...791...81A}; \citealt{2019ApJ...877...95T})
that the AGN torus in the Circinus galaxy is Compton-thick
($N_\mathrm{H}^\mathrm{Equ}=2.16^{+0.24}_{-0.17}\times 10^{25}\
\mathrm{cm^{-2}}$; 
$N_\mathrm{H}^\mathrm{LOS}=5.95^{+0.76}_{-0.60}\times 10^{24}\
\mathrm{cm^{-2}}$), geometrically thin ($\sigma=10.3^{+0.8}_{-0.3}\
\mathrm{degrees}$), and viewed edge-on ($i=78.3^{+0.5}_{-0.9}\ \mathrm{degrees}$). Our results
supercede the earlier work by \citet{2019ApJ...877...95T}, who first
applied XCLUMPY to the broadband spectrum of the Circinus galaxy
observed with Suzaku, XMM-Newton, and NuSTAR. They obtained
$N_\mathrm{H}^\mathrm{Equ}=9.08^{+0.14}_{-0.08}\times 10^{24}\
\mathrm{cm^{-2}}$ 
($N_\mathrm{H}^\mathrm{LOS}=4.86^{+0.07}_{-0.04}\times 10^{24}\
\mathrm{cm}^{-2}$), $\sigma=14.7^{+0.5}_{-0.4}\ \mathrm{degrees}$,
$i=78.3^{+0.2}_{-0.2}\ \mathrm{degrees}$. Our inclination angle is consistent with
the \citet{2019ApJ...877...95T} value, whereas the hydrogen column
density along the equatorial plane and torus angular width are
somewhat larger and smaller than theirs, respectively. 
We infer that the differences arise because
\citet{2019ApJ...877...95T} assumed the solar abundance and did not
separately treat the diffuse emission, which contaminates the AGN
spectrum in a soft band. In \citet{2019ApJ...877...95T} these effects
worked to overestimate the photon index of the intrinsic power-law
component ($\Gamma=1.80^{+0.01}_{-0.03}$ compared with our result of
$\Gamma=1.68^{+0.05}_{-0.09}$) and hence required a larger solid 
angle of the reflector (i.e., a larger torus angular width) to
reproduce the observed reflection spectrum. Since the column density
along the line of sight ($N_\mathrm{H}^\mathrm{LOS}$) is well
constrained by the spectrum, the large torus angular width may have
led them to underestimate that along the equatorial plane
($N_\mathrm{H}^\mathrm{Equ}$).

\citet{2014ApJ...791...81A} analyzed the XMM-Newton
and NuSTAR data of the Circinus galaxy 
observed in 2013 with the MYTorus and Torus models, and obtained
$N_{\mathrm{H}}^{\mathrm{Equ}}=\ $6.6--10$\times 10^{24}\ \mathrm{cm}^{-2}$,
$\Gamma=\ $2.2--2.4.
Our hydrogen column density along the equatorial plane
($N_{\mathrm{H}}^{\mathrm{Equ}}=\ 21.6^{+2.4}_{-1.7}\times 10^{24}\ \mathrm{cm}^{-2}$)
and photon index ($\Gamma=\ 1.68^{+0.05}_{-0.09}$) are
larger and smaller than their results, respectively.
As mentioned in 
\citet{2019ApJ...877...95T}, 
we infer that 
because the XCLUMPY model contains a larger flux of the unabsorbed
reflection component, it leads to a flatter intrinsic spectrum.
Also, the differences in the adopted torus geometry inevitably affect
the estimate of $N_{\mathrm{H}}^{\mathrm{Equ}}$. We refer the reader to
\citet{2019ApJ...877...95T} for detailed comparison of spectral analysis
results between the XCLUMPY model and other torus models.

Our result that the metal abundance is super-solar
($Z=1.52^{+0.05}_{-0.06}$) is consistent with
\citet{2003MNRAS.343L...1M}, who estimated the iron abundance to be
1.2--1.7 solar from the Fe K$\alpha$ edge. As mentioned in
Section~\ref{section:intro}, \citet{2018ApJ...867...80H} derived $Z
\approx 1.75$ from the Compton shoulder of the Fe K$\alpha$ line
utilizing a torus model developed by \citet{2016ApJ...818..164F}. Our
result is slightly smaller than their result, which was based on a more
geometrically-thick torus model.

According to our best-fit parameters of XCLUMPY, the expected number of
clumps along the line of sight is $\sim$3.
The hydrogen number density of each clump ($n_{\mathrm{H}}$) can be
calculated with Equation~\ref{equation:los} (see also Equation 5 in
\citealt{2019ApJ...877...95T}). In our case, 
$n_{\mathrm{H}} = 8.2\times 10^8\ \mathrm{cm}^{-3}$, which is almost
consistent with those estimated by \citet{2014MNRAS.439.1403M} for
nearby Seyfert galaxies.

\begin{equation}\label{equation:los}
   n_{\mathrm{H}} = \frac{3N_{\mathrm{H}}^{\mathrm{Equ}}}{40R_{\mathrm{clump}}}
\end{equation}

\subsection{Torus Inner Radius} 

We have performed a simultaneous spectral analysis of the
Chandra/HETG first, second, and third order spectra
of the Circinus galaxy, where we
take into account the spatial extent of the Fe K$\alpha$ line emitting
region. We obtain $\log(r_\mathrm{in}/r_\mathrm{g}) =
5.28^{+0.42}_{-0.23}$, 
which corresponds to a Keplerian velocity of
$690^{+200}_{-270}~\mathrm{km\ s^{-1}}$
at the inner edge of the torus, 
and to $r_\mathrm{in}= 1.6^{+1.6}_{-0.9}\times 10^{-2}$ pc for an SMBH
mass of $(1.7\pm 0.3)\times 10^6\ M_{\odot}$ (Section~\ref{subsection:spectral}).
Our inner radius is 2.5 times larger than that reported in
\citet{2011ApJ...738..147S}. This is most probably because they analyzed
the first order spectrum by ignoring the spatial extent and fitted the
Fe K$\alpha$ line with a single Gaussian model, both led them to
overestimate the true line width.
By contrast, the torus inner radius obtained here is
consistent with the location of 
the Si K$\alpha$ line emitting region estimated by
\citet{2016MNRAS.459L.105L} ($0.03^{+0.06}_{-0.015}$ pc), which was
later interpreted by \citet{2019MNRAS.490.4344L}
to mainly originate from the polar outflows.
We note that our torus inner radius 
is smaller than the maser-disk radius ($r=0.11\pm0.02\ \mathrm{pc}$, \citealt{2003ApJ...590..162G}).

The innermost radius of the dusty region in a torus (dusty torus) can be
estimated from the dust sublimation radius ($R_\mathrm{sub}$), the limit
where dust grains can exist against radiative heating from the central
engine.
\citet{2008ApJ...685..147N,2008ApJ...685..160N} derived the formula of
the sublimation radius of a clumpy torus as:
\begin{equation}
 R_\mathrm{sub} = 0.4\left(\frac{L_\mathrm{bol}}{10^{45}~
 \mathrm{erg\ s^{-1}}}\right)^{0.5}\left(\frac{1500~\mathrm{K}}
 {T_\mathrm{sub}}\right)^{2.6}\ \mathrm{pc}, 
\end{equation}
where $L$ is the bolometric luminosity of the AGN and $T_\mathrm{sub}$
is the sublimation temperature.
Adopting $T_\mathrm{sub} = $ 1500 K and the bolometric
luminosity derived from the broadband spectral analysis (Section~\ref{subsection:broad}),
we obtain that of the Circinus galaxy to be $R_\mathrm{sub}\sim 0.14\
\mathrm{pc}$. \citet{2007A&A...476..713K} found that the torus inner
radius in an AGN derived from the near-infrared reverberation mapping was
${\sim}3$ times smaller than the sublimation radius calculated as
above; the factor 3 difference can be explained by considering the
anisotropy of emission from the accretion disk \citep{2010ApJ...724L.183K}. 
If we apply this relation, the inner radius of the dusty torus in the
Circinus galaxy is estimated as $\approx$0.05 pc.

The torus inner radius we have derived from the Fe K$\alpha$ line width
is still significantly smaller than this value. Since the near-infrared
reverberation observation only measures the location of dust, this
result indicates that there is a significant amount of dust-free gas
in the inner side of the dusty torus.
Our result basically supports the arguments by
\citet{2015ApJ...802...98M} and \citet{2015ApJ...812..113G} that a
significant fraction of the Fe K$\alpha$ line originates from a region
closer to the SMBH than the dusty torus. However, the torus inner radius
we have derived is larger than these previous estimates, strongly
suggesting that the region is well outside the broad line region. We
note that our result is currently limited by the uncertainty of the
spatial extent in analyzing the grating spectra. Future non-dispersive
high resolution spectroscopy, like that by the X-Ray Imaging and
Spectroscopy Mission (XRISM), will push forward our understanding on the
whole structure of an AGN including the torus and its inner region.

\section{Conclusion}

\begin{enumerate}

 \item We update the XCLUMPY model developed by \citet{2019ApJ...877...95T}, an X-ray
 spectral model from a clumpy torus in an AGN, by incorporating
 line broadening effect due to the Keplerian rotation of the
 torus with variable metal abundance. This model enables us to
 constrain the inner radius of the torus from the fluorescence
 line profiles.

 \item We apply the model to currently the best available broadband
 X-ray spectra (3--100 keV) of the Circinus galaxy. We confirm
 that the Circinus galaxy has a geometry thin and Compton-thick
 torus with super-solar ($Z\approx1.5$) metal abundance.

 \item From the simultaneous analysis of the 
 Chandra/HETG first, second, and third order spectra,
 we derive the torus inner radius emitting
 Fe K$\alpha$ to be
 $\log(r_\mathrm{in}/r_\mathrm{g})=5.28^{+0.42}_{-0.23}$
 or $1.6^{+1.6}_{-0.9}\times 10^{-2}\ \mathrm{pc}$, which is about 3 times smaller than 
 that estimated from the dust sublimation
 radius. This suggests that the inner side of the torus is
 composed of dust-free gas.

\end{enumerate}

This work has been financially supported by the Grant-in-Aid for
Scientific 20H01946 (Y.U.) and for JSPS Research Fellowships
20J00119 (A.T.) and 19J22216 (S.Y.). This research has made use of data and software
provided by the high energy astrophysics science archive research center
(HEASARC) and the Chandra X-ray Center (CXC). We have also
utilized the NASA/IPAC Extragalactic Database (NED).

%

\vspace{5mm}
\facilities{Suzaku, XMM-Newton, NuSTAR, Chandra}


\software{HEAsoft v6.27 \citep{2014ascl.soft08004N}, MONACO \citep{2011ApJ...740..103O, 2016MNRAS.462.2366O}, SAS v18.0.0 \citep{2004ASPC..314..759G}, CIAO v4.12, MARX \citep{2012SPIE.8443E..1AD}, XSPEC \citep{1996ASPC..101...17A}
       }



\appendix

\section{Effect of Spatial Extent in HETG spectra} \label{appendix:HETG}

\citet{2016MNRAS.463L.108L} reported that the width of Fe K$\alpha$ line
is often larger in the Chandra/HETG first order spectrum than in the
second and third order spectra. It was pointed out that the discrepancy can be explained by the
spatial extent of the target\footnote{\texttt{https://cxc.harvard.edu/newsletters/news\_24/}}; 
the total broadening in physical coordinates $x$ on the detector is given by

\begin{equation}
 \label{equation:HETG_cal}
 \sigma_x^2 = \sigma_i^2 + \left(\frac{Rm\lambda}{pc}\right)^2\sigma_v^2+\left(F\sigma_{\theta}\right)^2
\end{equation}
where $\sigma_i$ is the instrumental broadening, 
$\sigma_v$ the
intrinsic line width (Doppler broadening), $\sigma_\theta$ the
spatial extent, $R$ the Rowland distance of the HETGS, $m$ the
grating order, $\lambda$ the wavelength of the line, $p$ the
grating period, and $F$ the focal length of the High Resolution Mirror
Assembly (HRMA).
(In Equation~\ref{equation:HETG_cal}, we assume that all the broadening
profile can be modelled by Gaussians.) In the standard processing, the
response matrix is created by assuming a point source, and therefore
the observed line width is given by
\begin{equation}
 {\sigma_v'}^2 = \sigma_v^2 + \left(\frac{pc}{Rm\lambda}\right)^2\left(F\sigma_\theta\right)^2
 = \sigma_v^2 + \left(\frac{1750}{m}\sigma_{\theta}\right)^2,
\end{equation}
where ${\sigma_v'}^2$ and $\sigma_v^2$ are given in 
units of km s$^{-1}$ and $\sigma_{\theta}$ in units of arcsec.
This shows that the spatial broadening effect is inversely proportional
to the grating order.
In fact, if we ignored the effect of the spatial extent 
of Fe K$\alpha$ line emitting region in the Circinus galaxy, 
we would obtain $r_{\mathrm{in}}=(5.6\pm 1.4)\times 10^{-3}\ \mathrm{pc}$ and
$r_{\mathrm{in}}=(1.4\pm 0.8)\times 10^{-2}\ \mathrm{pc}$ from the first and 
second/third order spectra, respectively, consistent with 
the trend reported by \citet{2016MNRAS.463L.108L}.

\section{Dependence of Torus Inner Radius on Radial Density Profile}

We evaluate the dependence of the torus inner radius
derived from the spectral analysis on the assumed radial profile index
of clump number density ($q$).
\citet{2015ApJ...803...57I} analyzed the infrared spectral energy distribution
of the Circinus galaxy with the CLUMPY model and constrained 
$q=0.6\pm 0.3$. Following the result, we
make three models with $q=$0.3, 0.6, and 0.9, and apply them 
to the 6.3--6.5 keV HETG spectra. The other model parameters
are the same as those reported in Section~\ref{subsection:spectral}.

\begin{deluxetable}{cC}
 \tablecaption{Torus inner radius in various radial density profiles. \label{table:radial_density}}
 \tablewidth{0pt}
 \tablehead{
 \colhead{\hspace*{1.0cm}$q$}\hspace*{1.0cm} & \colhead{\hspace*{1.0cm}$\log (r_{\mathrm{in}}/r_{\mathrm{g}})$}\hspace*{1.0cm}
 }
 \colnumbers
 \startdata
 0.3 & 5.19^{+0.31}_{-0.27} \\
 0.6 & 5.27^{+0.36}_{-0.23} \\
 0.9 & 5.30^{+0.38}_{-0.21} \\
 \enddata
 \tablecomments{
 Column (1): index of radial density profile. Column (2): torus inner radius in units of gravitational radius.
 }
\end{deluxetable}

Table~\ref{table:radial_density} summarizes the result. We
find that a model with a smaller value of $q$ results in a smaller 
torus inner radius. 
However, we confirm that the difference is small and does 
not change our conclusion.

\begin{deluxetable}{C|CCCC}
 \tablecaption{Best-fit torus parameters in two different metal abundances. \label{table:metal}}
 \tablewidth{0pt}
 \tablehead{
 \colhead{Z}\vline & \colhead{$N_{\mathrm{H}}^{\mathrm{Equ}}$} & $\sigma$ & $i$ & $\chi^2/{\mathrm{dof}}$ \\
 (1) & (2) & (3) & (4) & 
 }
 \startdata
 1.0 & 30.5_{-1.9}^{+1.5\tablenotemark{a}} & 11.5_{-0.9}^{+1.0} & 78.8_{-0.7}^{+0.6} & 2341/1992 \\
 1.52^{+0.05}_{-0.06} & 21.6^{+2.4}_{-1.7} & 10.3^{+0.8}_{-0.3} & 78.3^{+0.5}_{-0.9} & 2275/1992 \\
 \enddata
 \tablecomments{
 Column (1): metal abundance of the torus in units of solar abundance.
 Column (2): hydrogen column density along the equatorial plane
 in units of $10^{24}\ \mathrm{cm}^{-2}$.
 Column (3): torus angular width in units of degrees.
 Column (4): inclination angle in units of degrees.
 }
 \tablenotetext{a}{The parameter reaches a limit of its allowed range.}
\end{deluxetable}

\begin{figure}[]
 \plotone{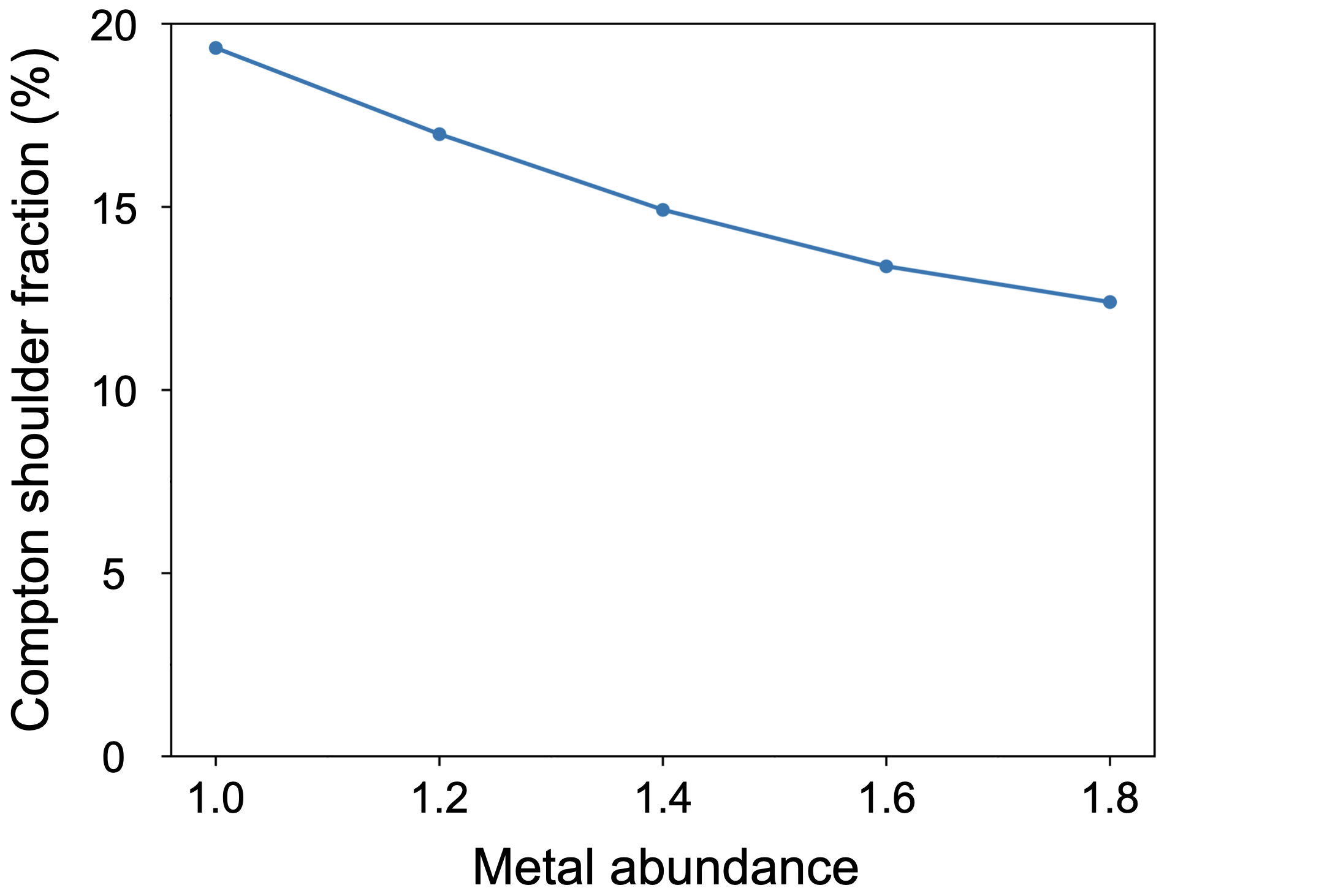}
 \caption{Dependence of the Compton-shoulder fraction on the metal
abundance. We adopt the basic torus
 parameters: $N_{\mathrm{H}}^{\mathrm{Equ}}=2.0\times 10^{25}\ \mathrm{cm}^{-2},\ \sigma=10\degr,\ i=78\degr$.
 \label{figure:cs_frac}}
\end{figure}

\section{Dependence on Metal Abundance} \label{appendix:metal}

Table~\ref{table:metal} summarizes the results of the
broadband spectral fitting in two different settings: $Z=1.0$ (fixed)
and $Z=1.52$ (free, best-fit). We see that the supersolar abundance of
the torus significantly improves the fit. Among the torus parameters, 
the hydrogen column density along the equatorial plane shows the 
largest difference. We interpret that, since the photoelectric
absorption is well constrained by the spectra, which is dominated by
metals, higher metalicity leads to a smaller column density of
hydrogen. The main driving factor that determines the metal abundance
is the fraction of Compton shoulder in the Fe K$\alpha$ line.
Figure~\ref{figure:cs_frac} shows the dependence of the Compton-shoulder
fraction on the metal abundance while the other torus parameters are
kept the same. As seen, the Compton-shoulder fraction become smaller
with increasing metal abundance.

\section{Fe K$\alpha$ Line Emitting Region in XCLUMPY} \label{appendix:feline}

\begin{figure*}[htbp]
 \epsscale{0.9}
 \plottwo{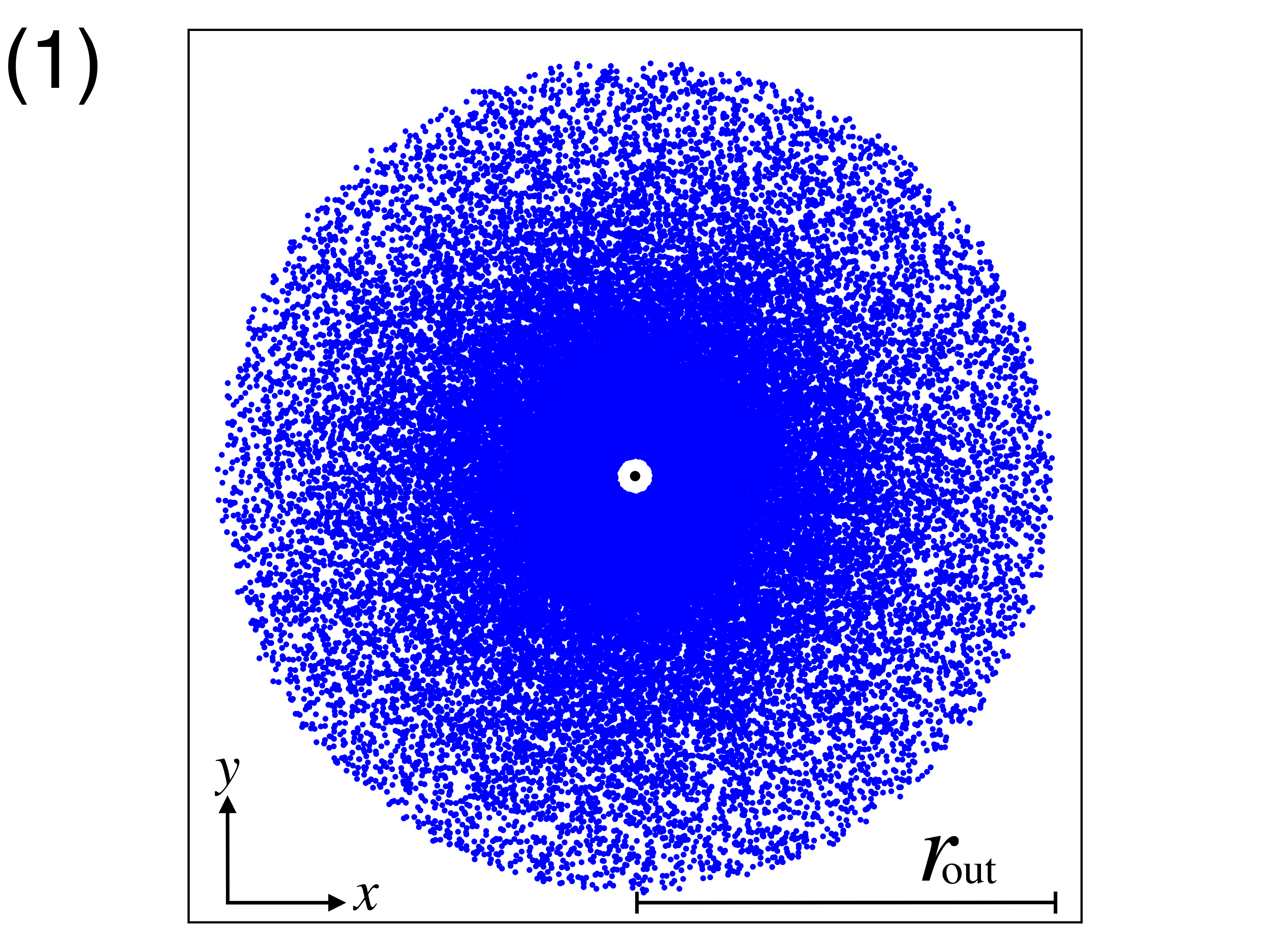}{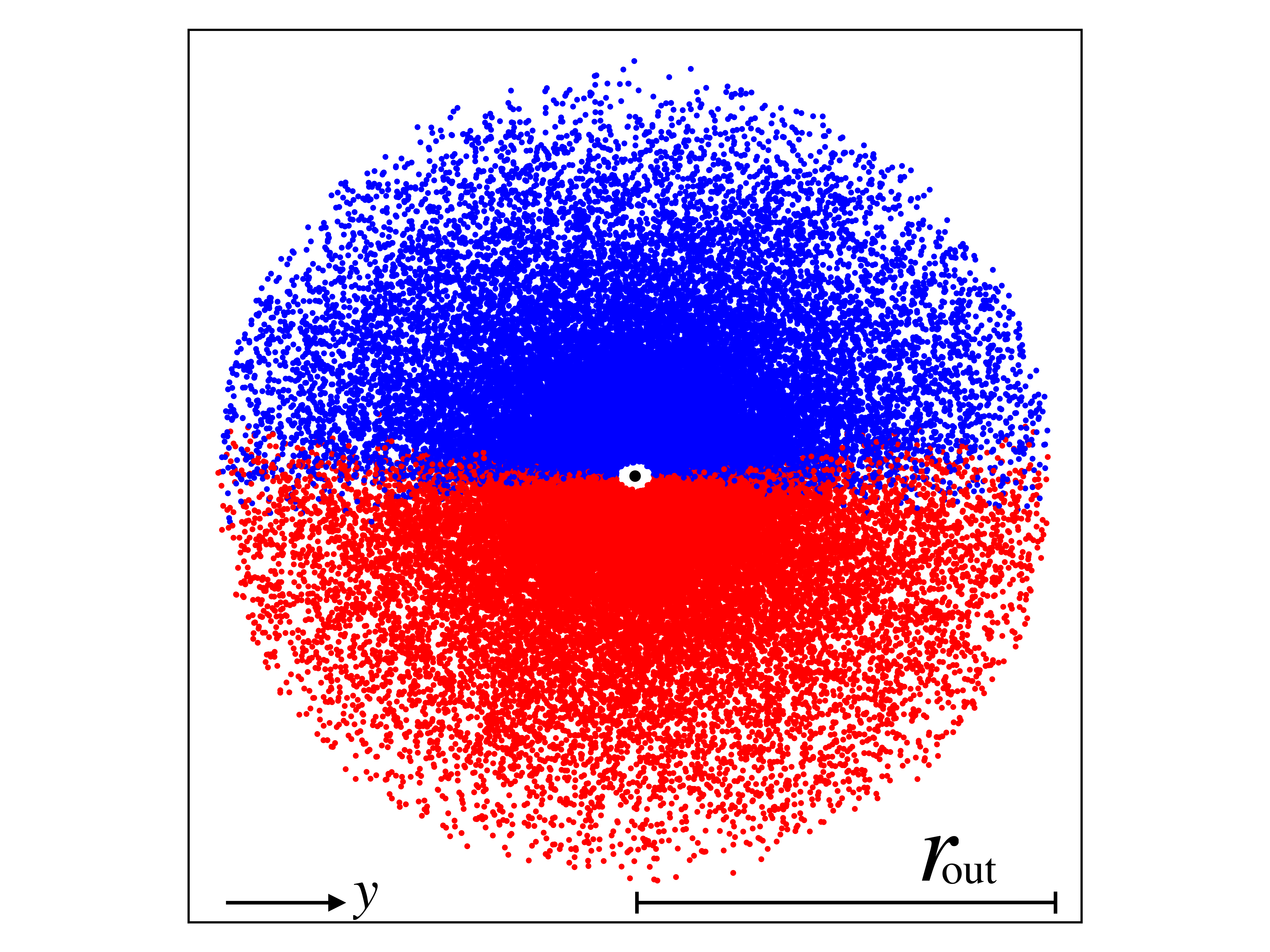}
 \plottwo{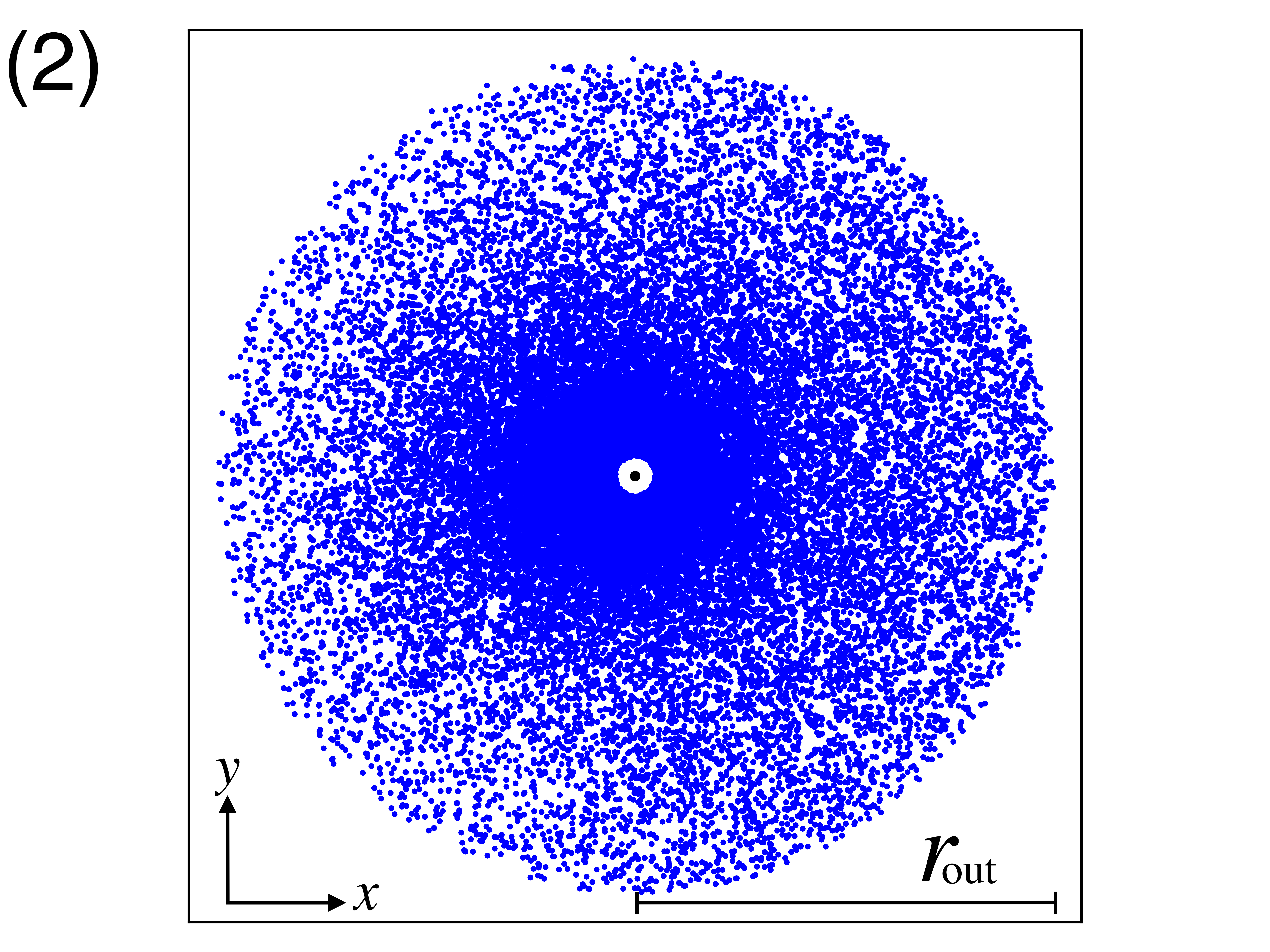}{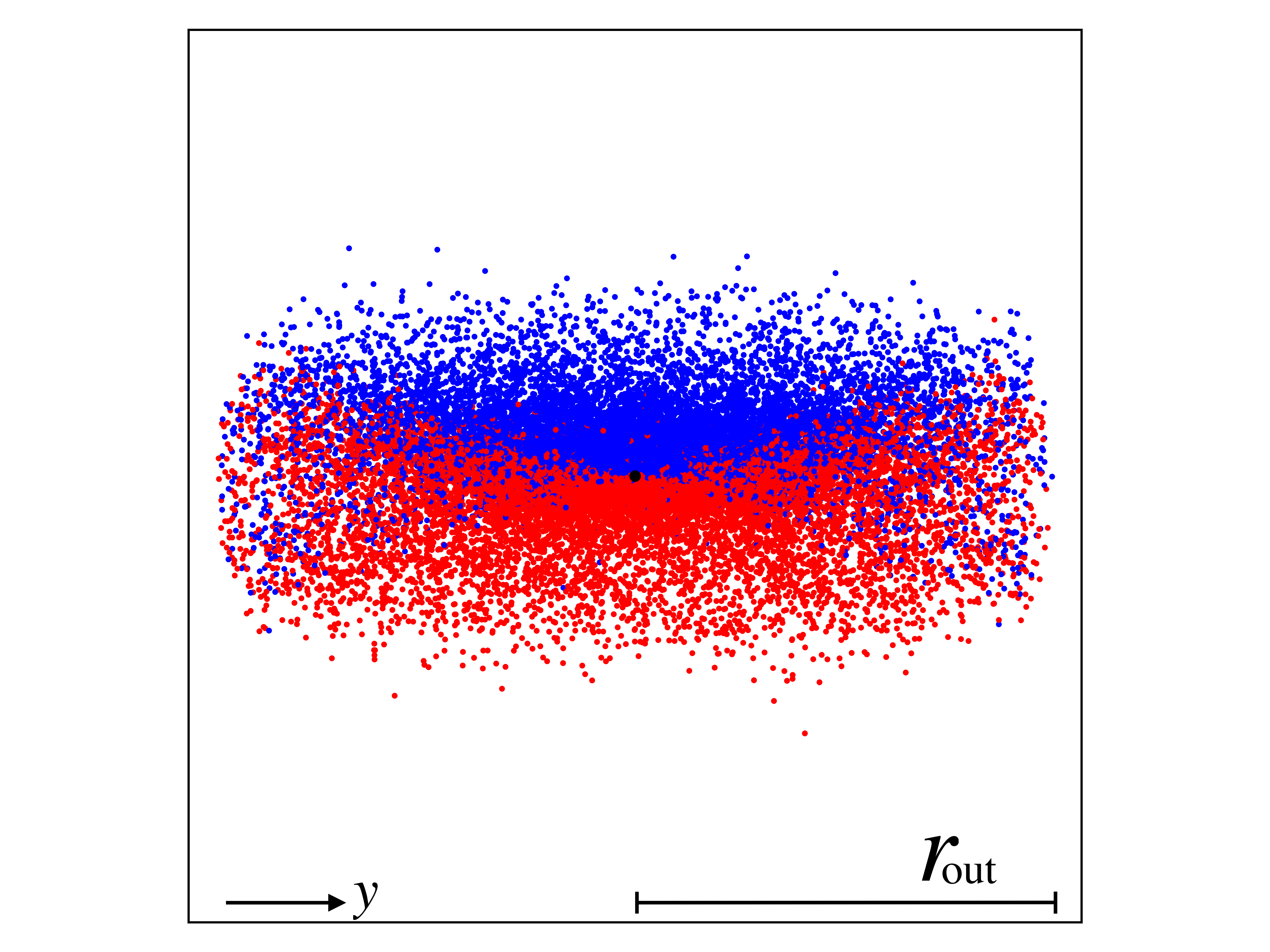}
 \plottwo{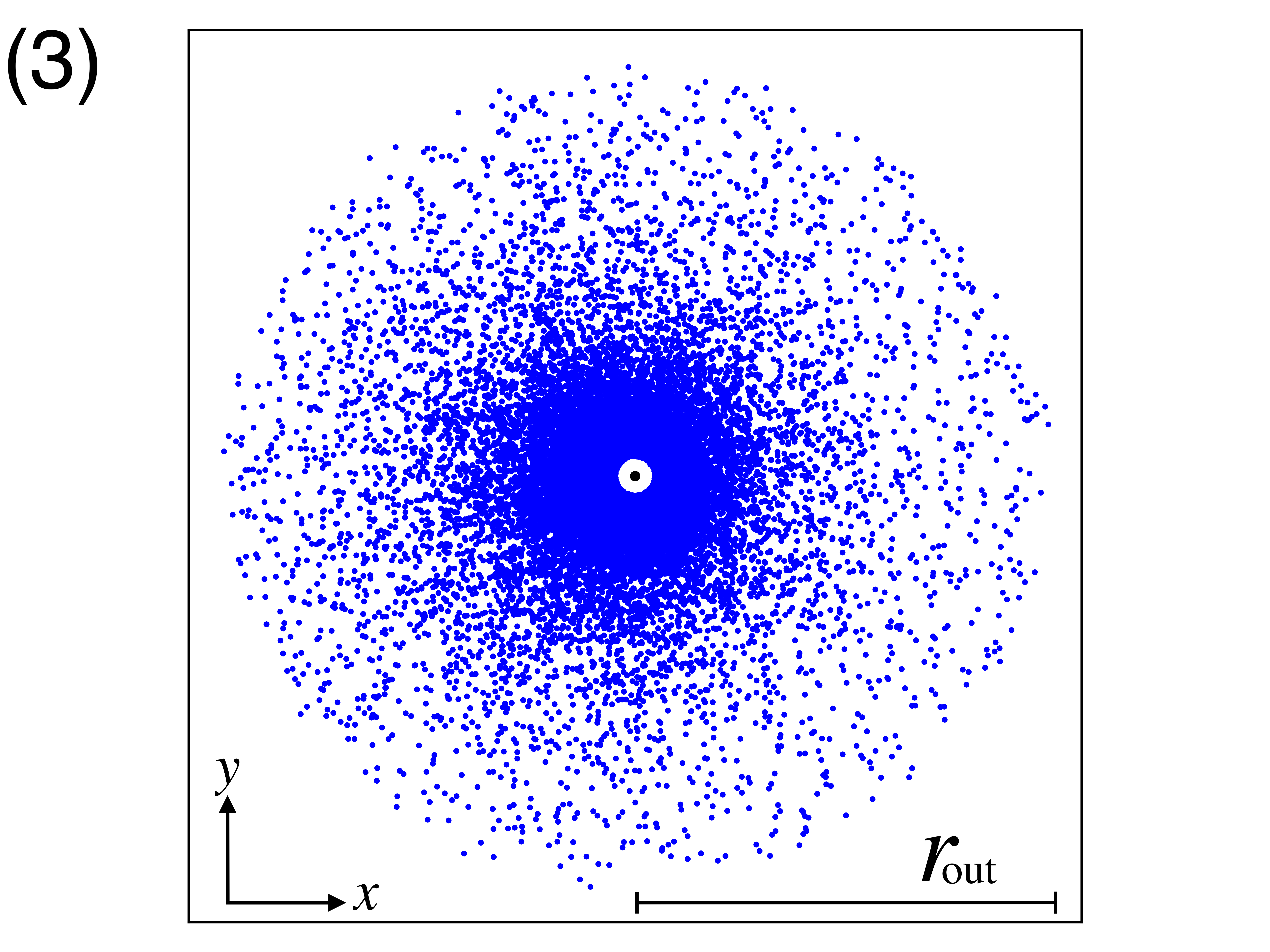}{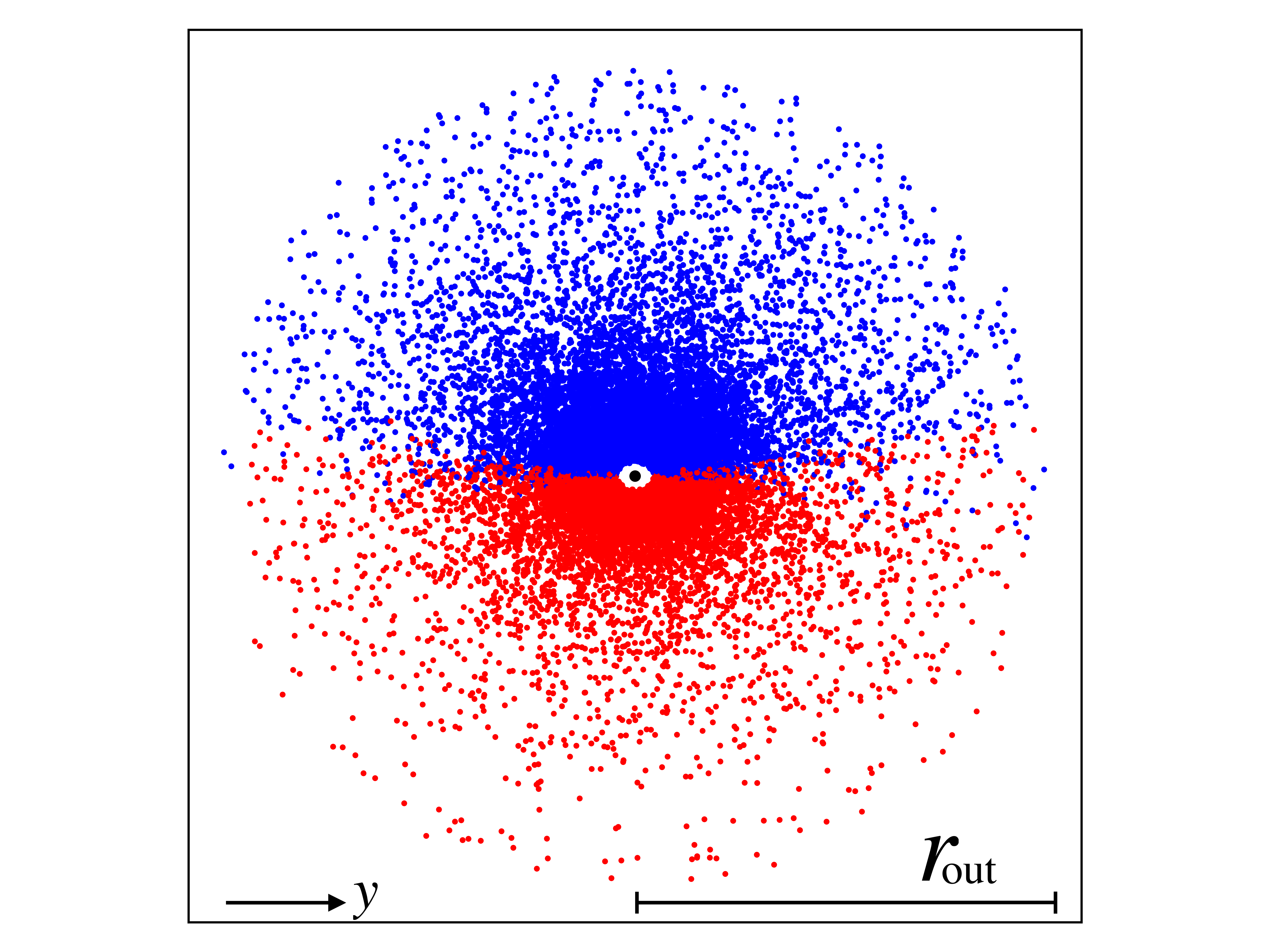}
 \plottwo{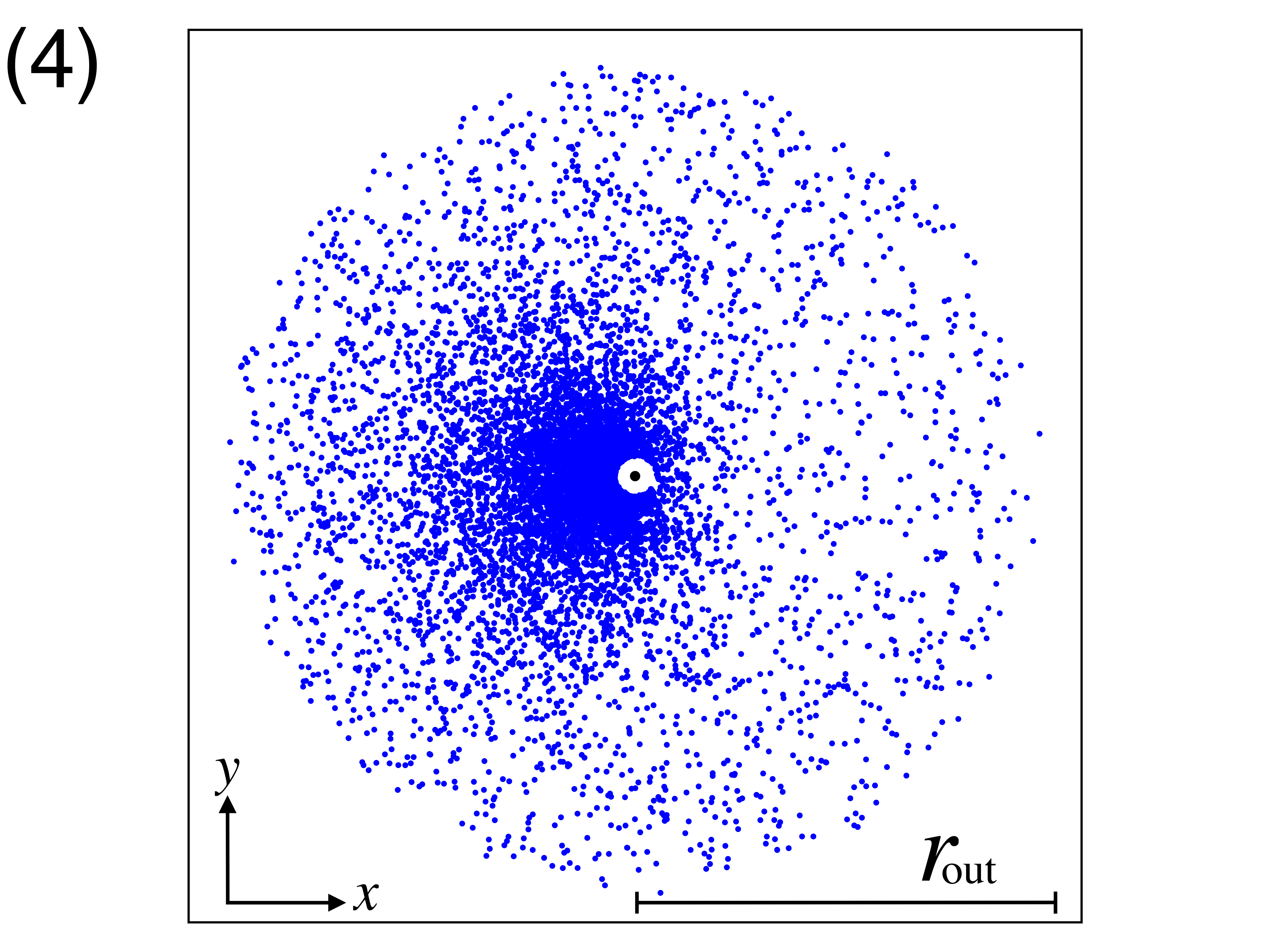}{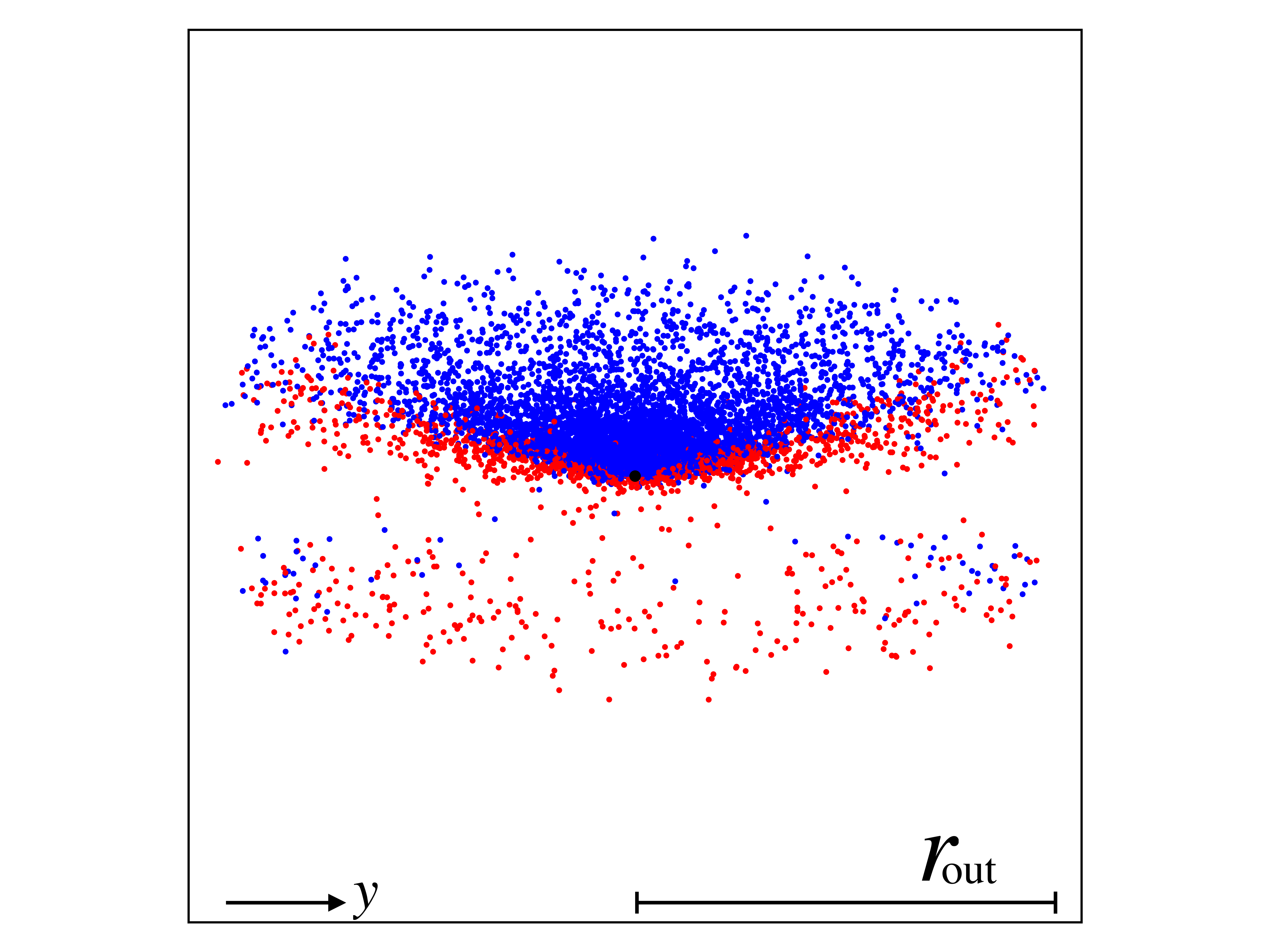}
 \caption{Maps of the Fe K$\alpha$ line emitting region.
 The red and blue points represent the positions where observed Fe K$\alpha$ 
 photons (except for Compton shoulder) are originated. The central black points
 represent the SMBH. 
 We set default torus parameters: $Z=1.5,\ \sigma=10\degr$. 
 The other parameters are set to 
(1) $N_\mathrm{H}^\mathrm{Equ}=1.0\times 10^{24}\
 \mathrm{cm}^{-2},\ i=30\degr$, 
(2) $N_\mathrm{H}^\mathrm{Equ}=1.0\times 10^{24}\
 \mathrm{cm}^{-2},\ i=78\degr$, 
(3) $N_\mathrm{H}^\mathrm{Equ}=2.0\times 10^{25}\
 \mathrm{cm}^{-2},\ i=30\degr$, 
and (4) $N_\mathrm{H}^\mathrm{Equ}=2.0\times 10^{25}\
 \mathrm{cm}^{-2},\ i=78\degr$,
from the top to the bottom.
 (Left): The positions projected onto the equatorial plane.
 An observer is viewing from the right direction.
 (Right): The positions viewed from the observer. 
 The red and blue color represent the front side and the back side, respectively.
 \label{figure:asymmetry}}
\end{figure*}

\begin{figure*}[htbp]
 \epsscale{0.9}
 \plottwo{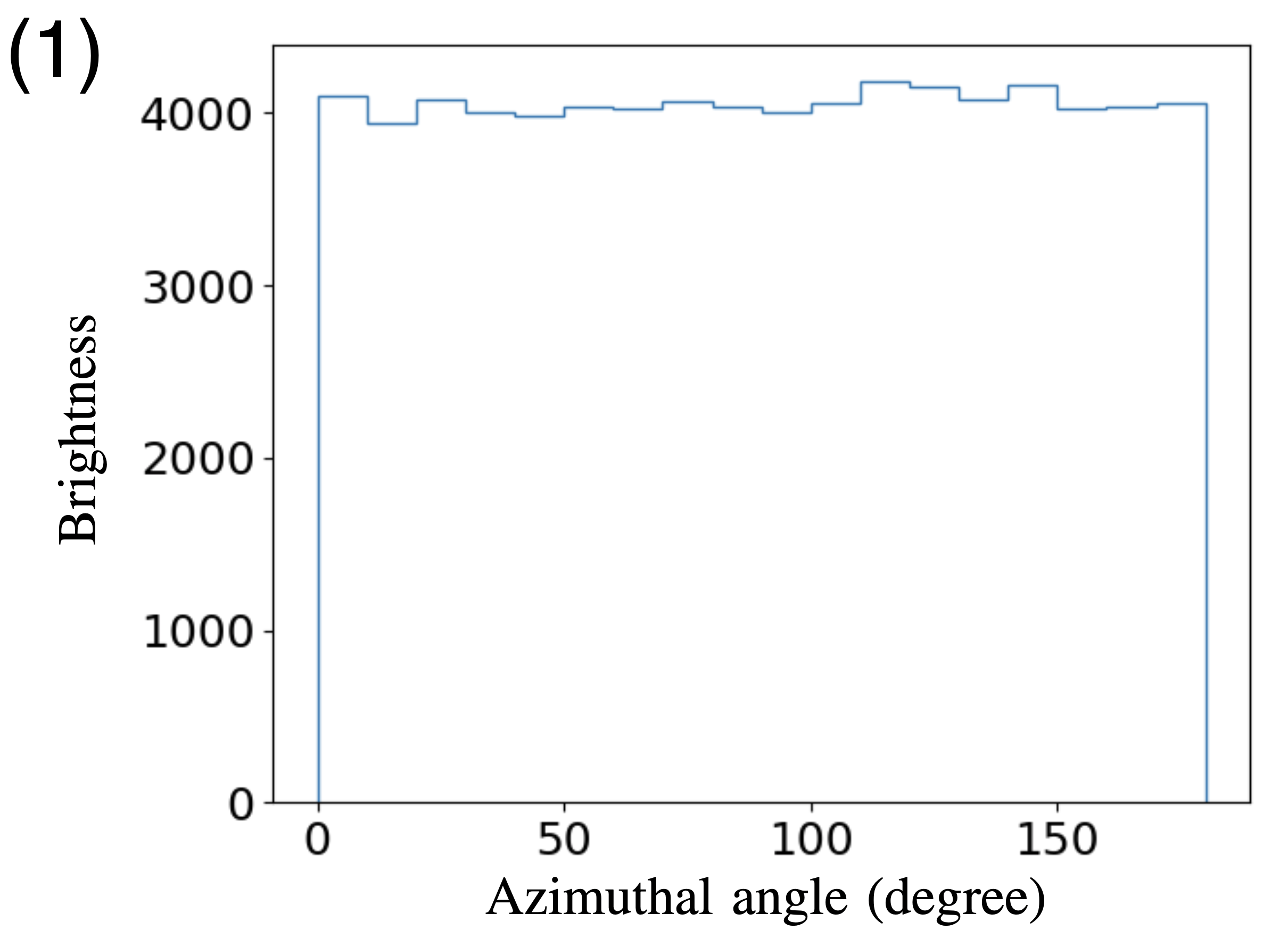}{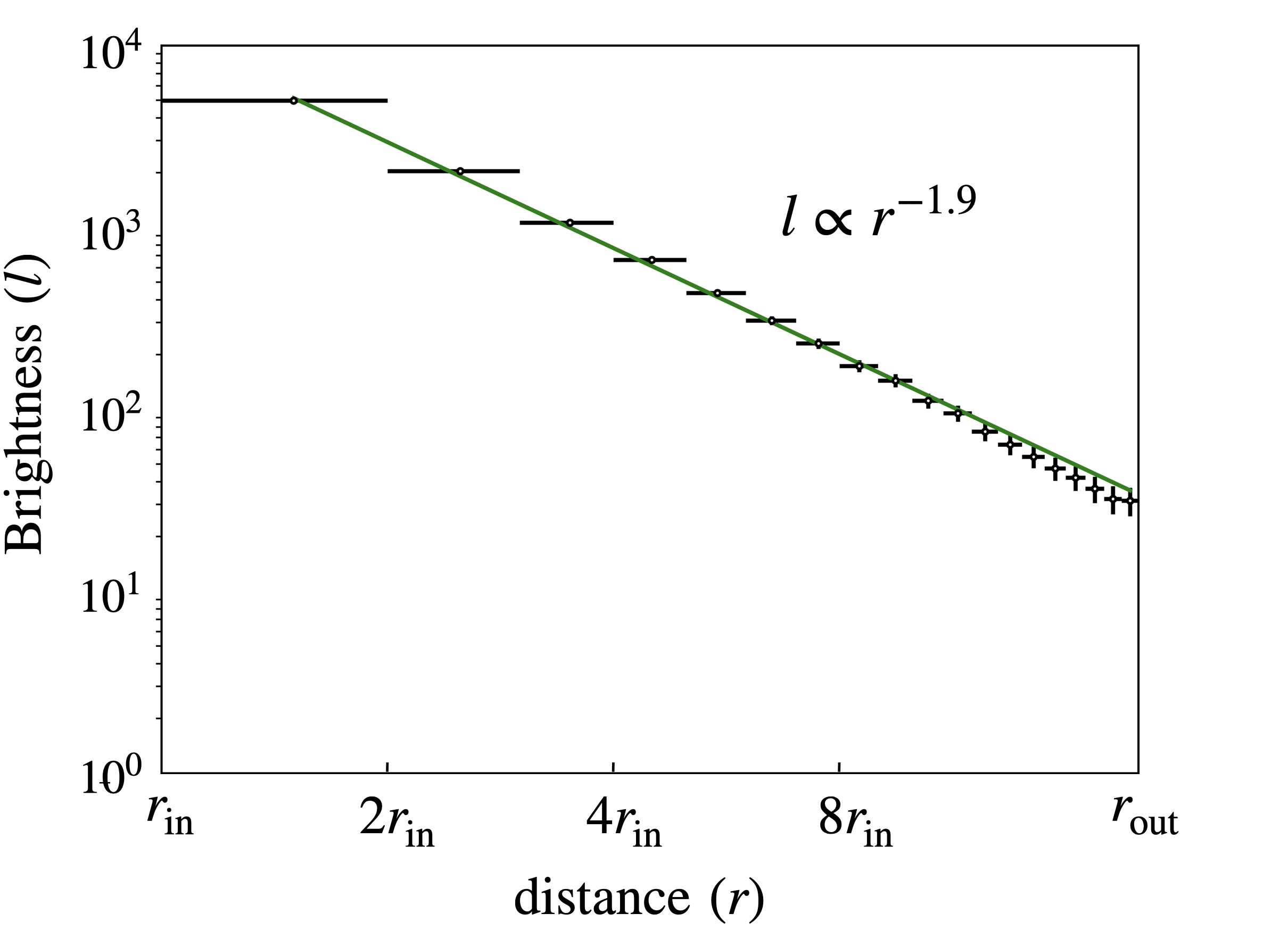}
 \plottwo{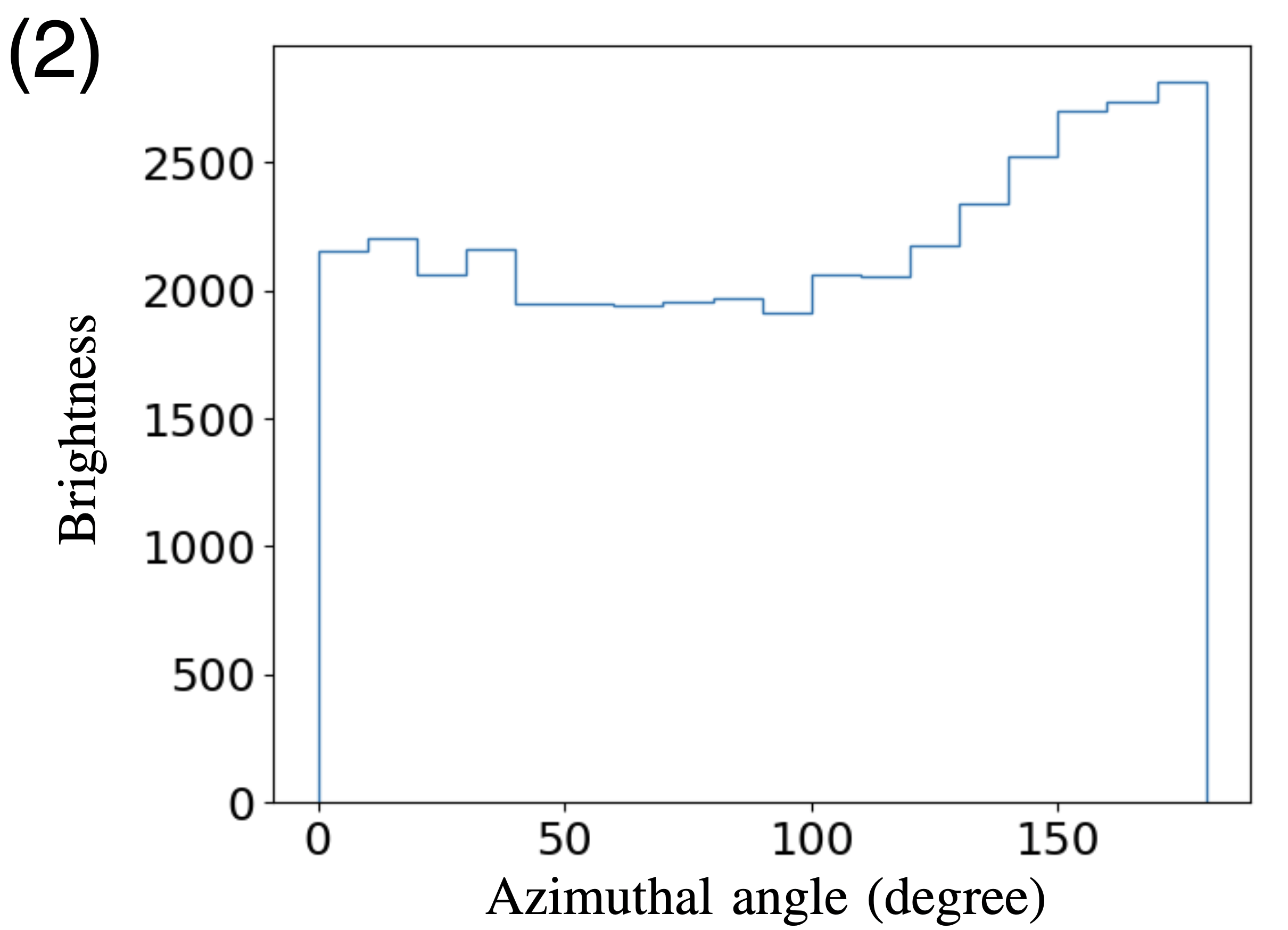}{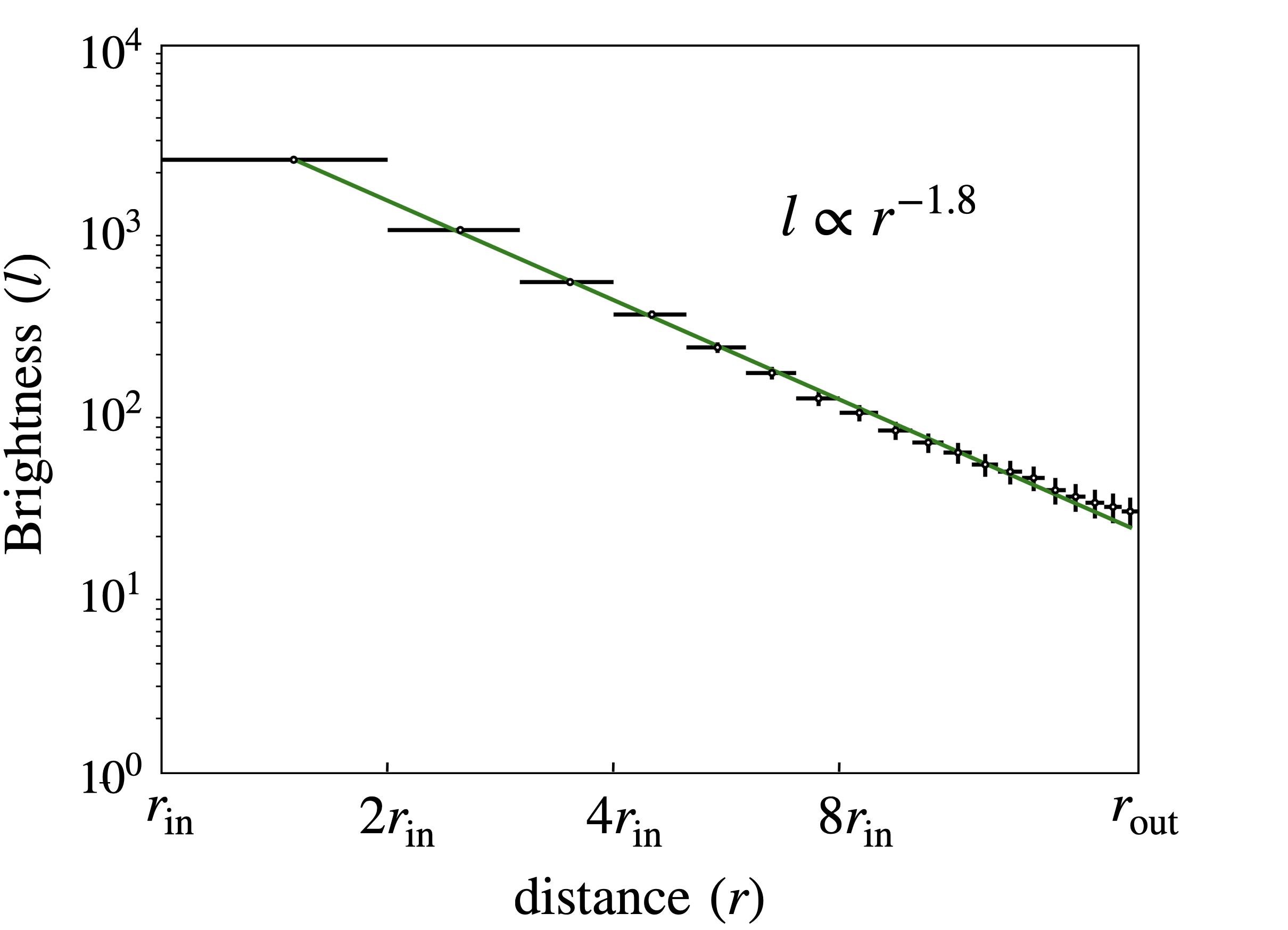}
 \plottwo{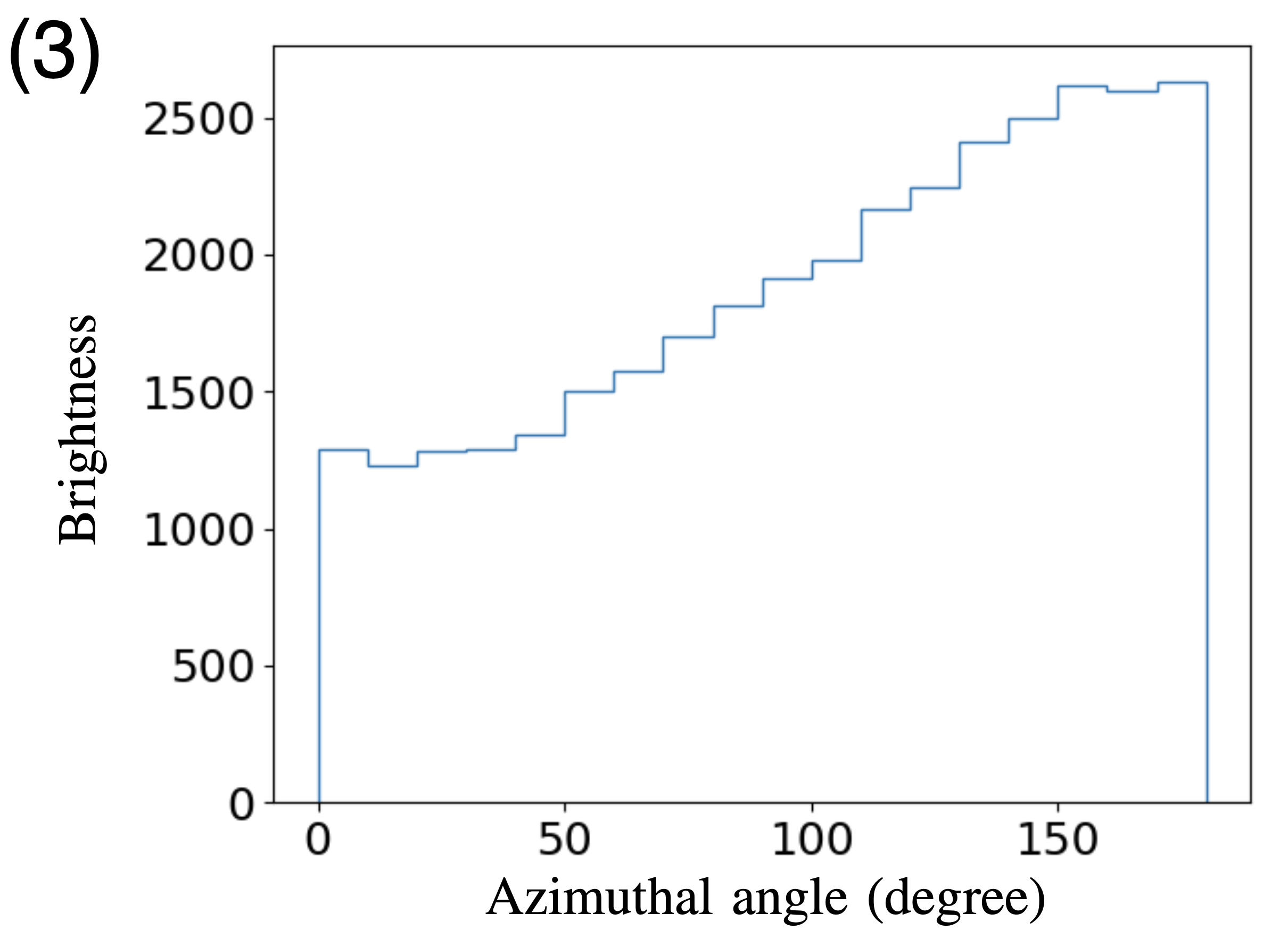}{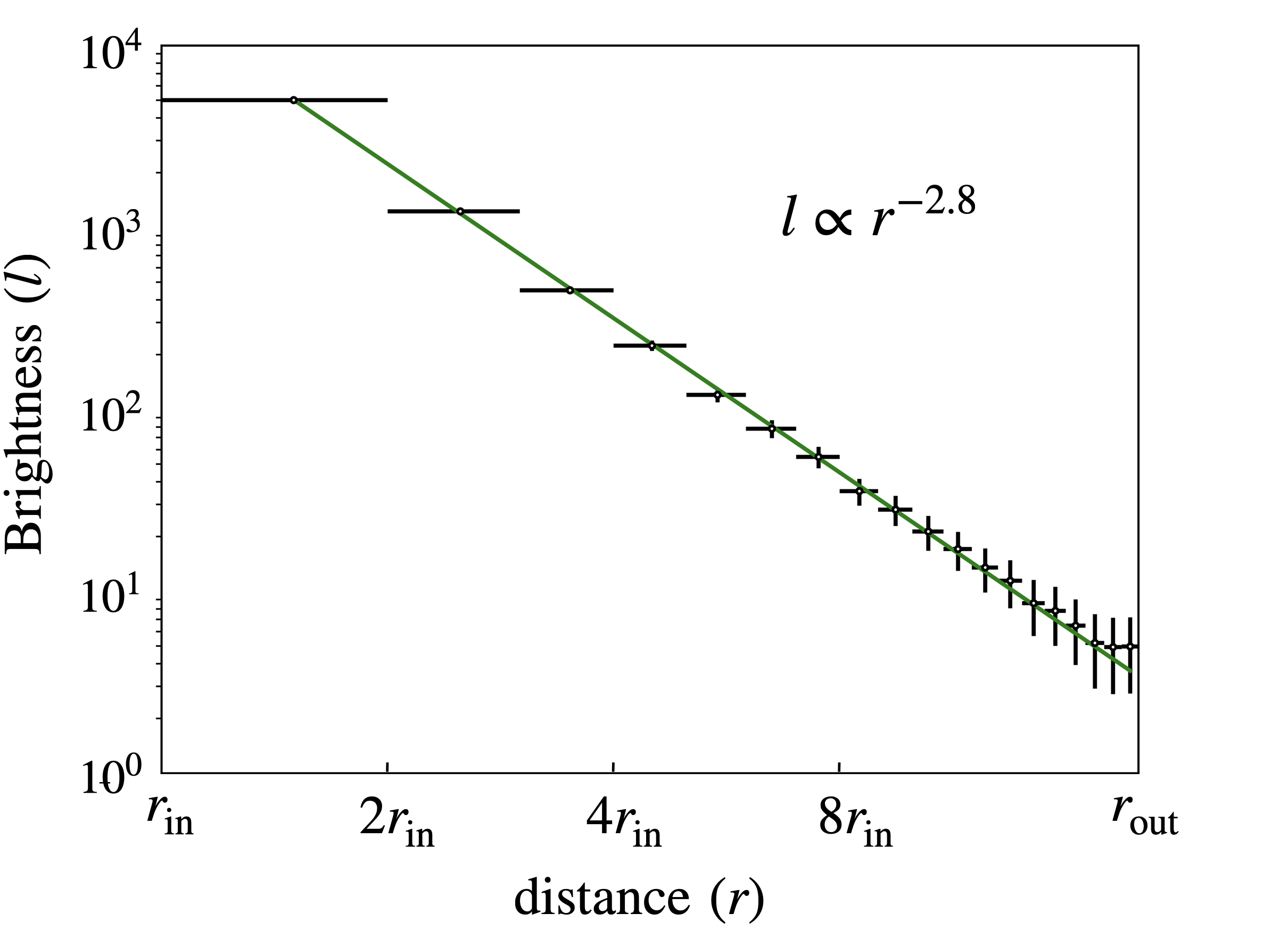}
 \plottwo{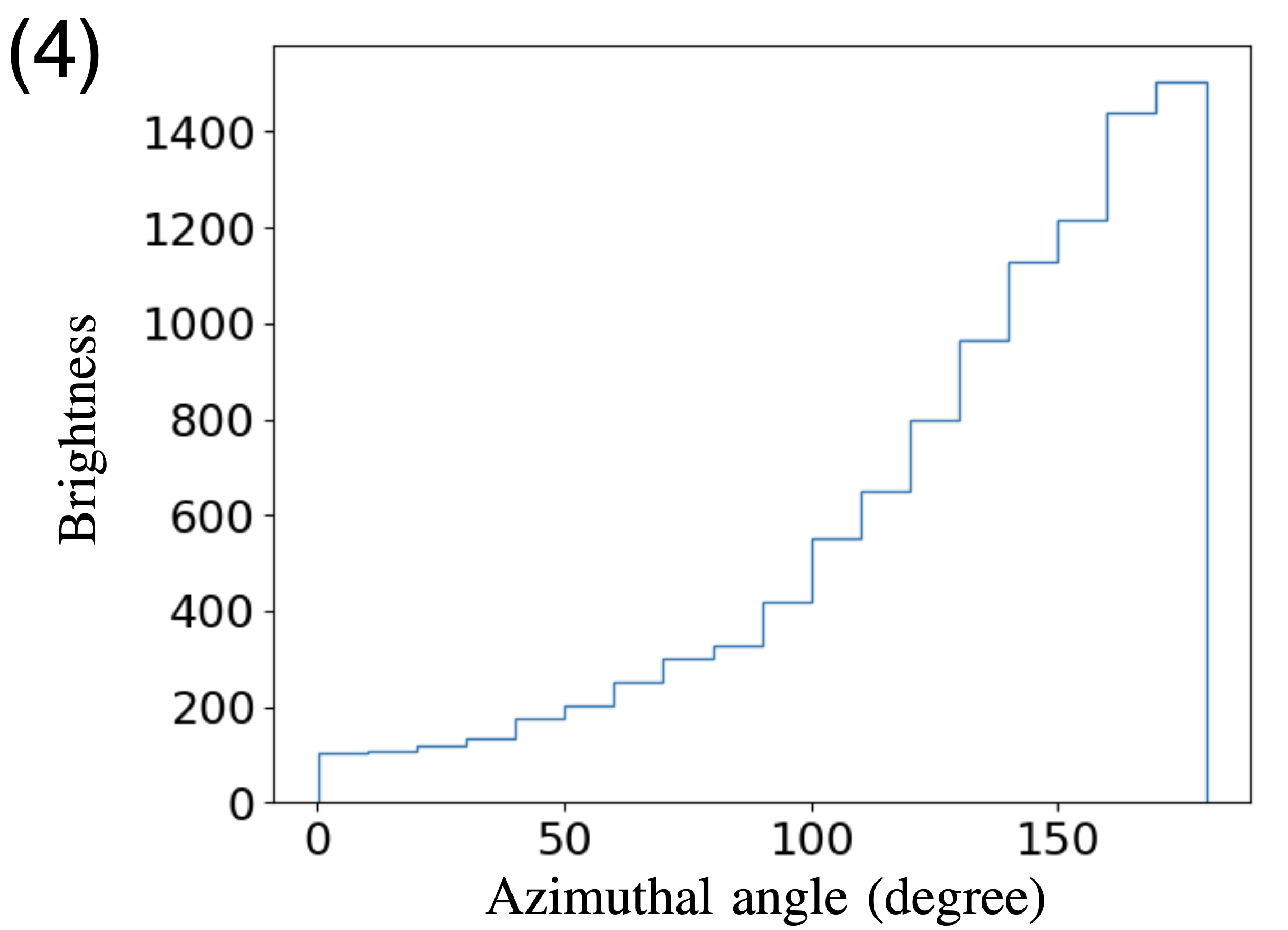}{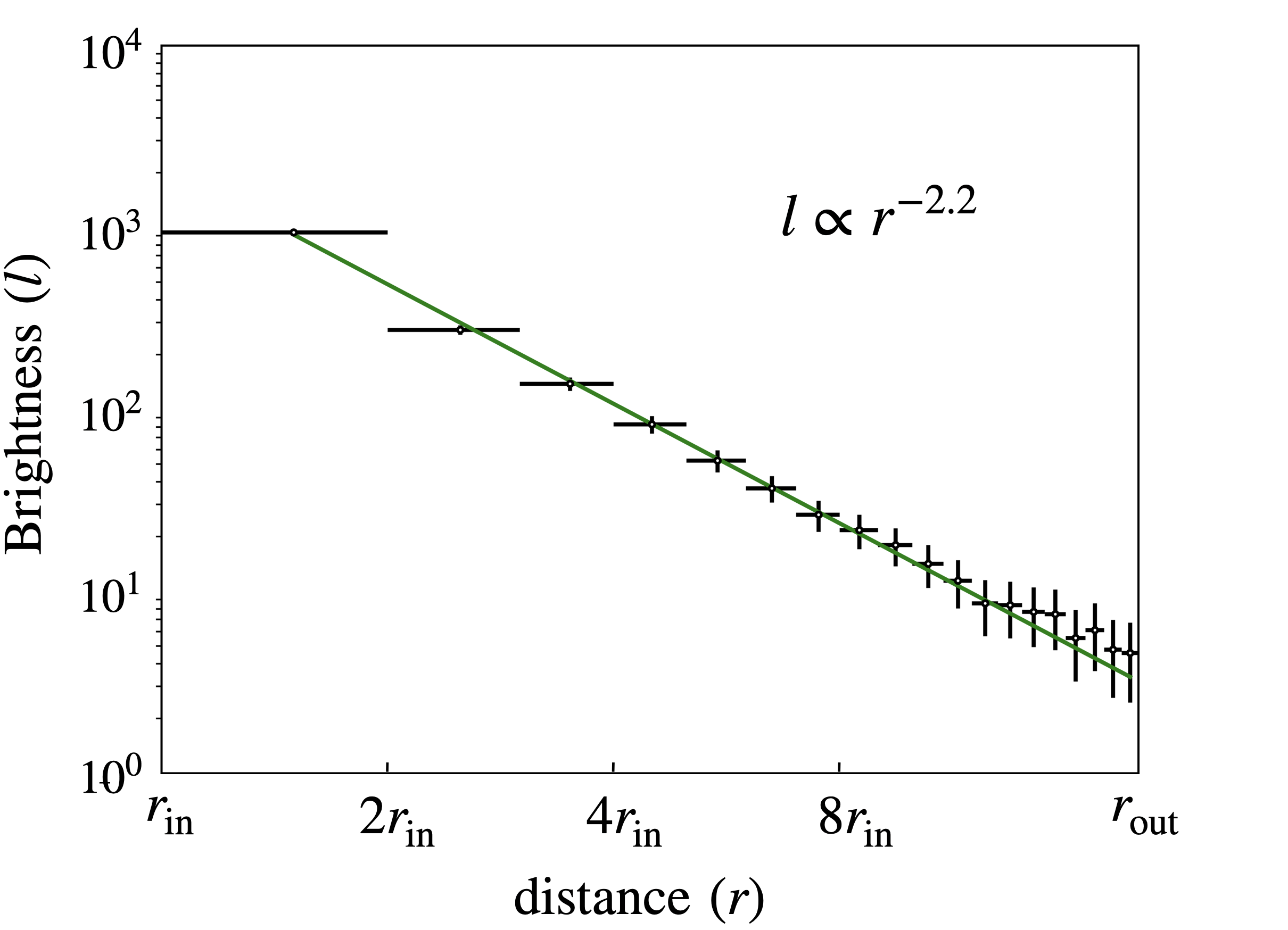}
 \caption{
 (Left): Azimuthal distribution 
 of the Fe K$\alpha$ emitting positions projected onto the equatorial plane. 
 The vertical axis is the number of photons (in an arbitrary unit),
 and the horizontal axis is the
 azimuthal angle with respect to the line of sight in degree.
 (Right): The radial profile. 
 The vertical axis is the surface brightness ($l$) in an arbitrary unit,
 and the horizontal axis is the
 distance from the SMBH $r$. 
 The green lines represent the best-fit power-law models. 
 The torus parameters are the same as in Figure~\ref{figure:asymmetry}.
 \label{figure:histogram}}
\end{figure*}

\begin{figure*}[htbp]
 \plottwo{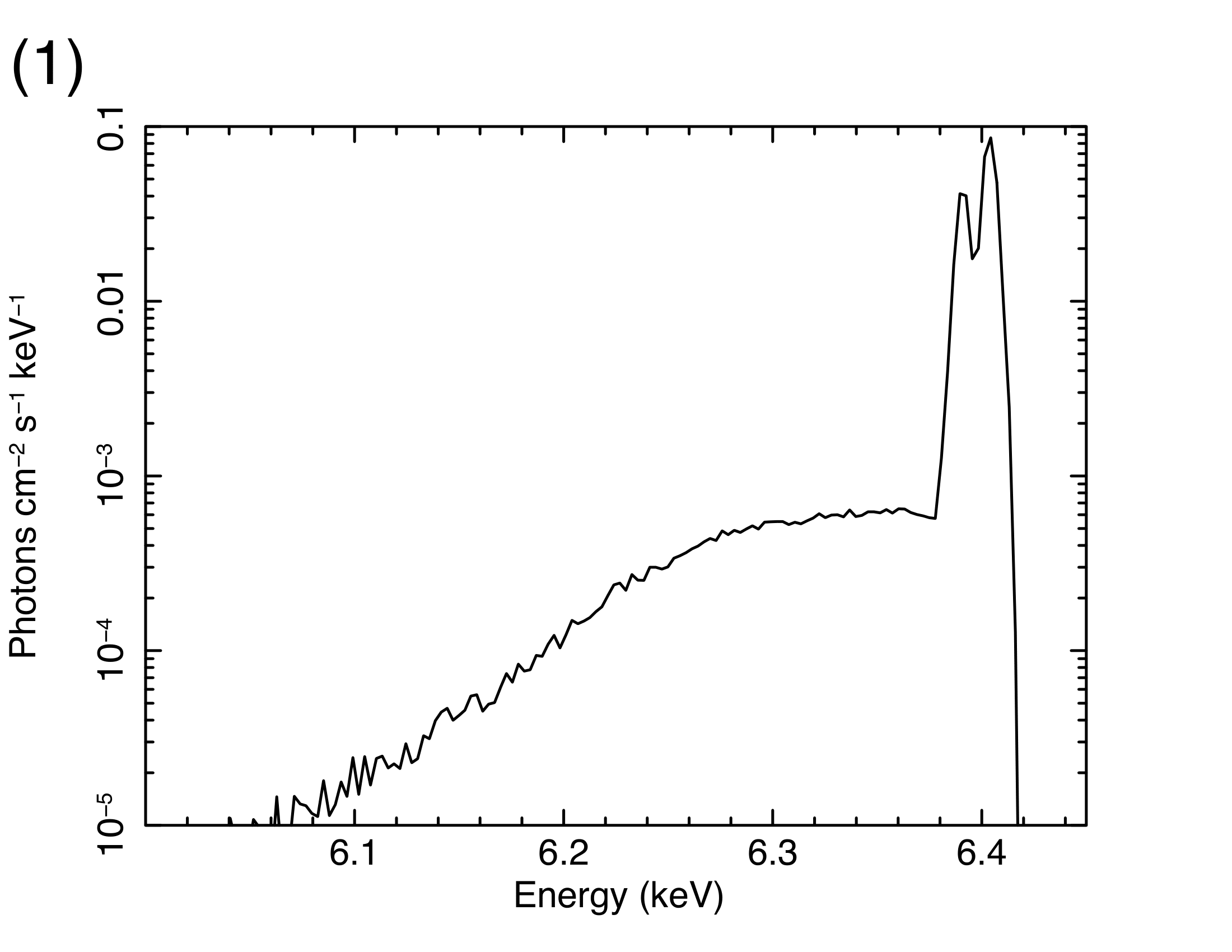}{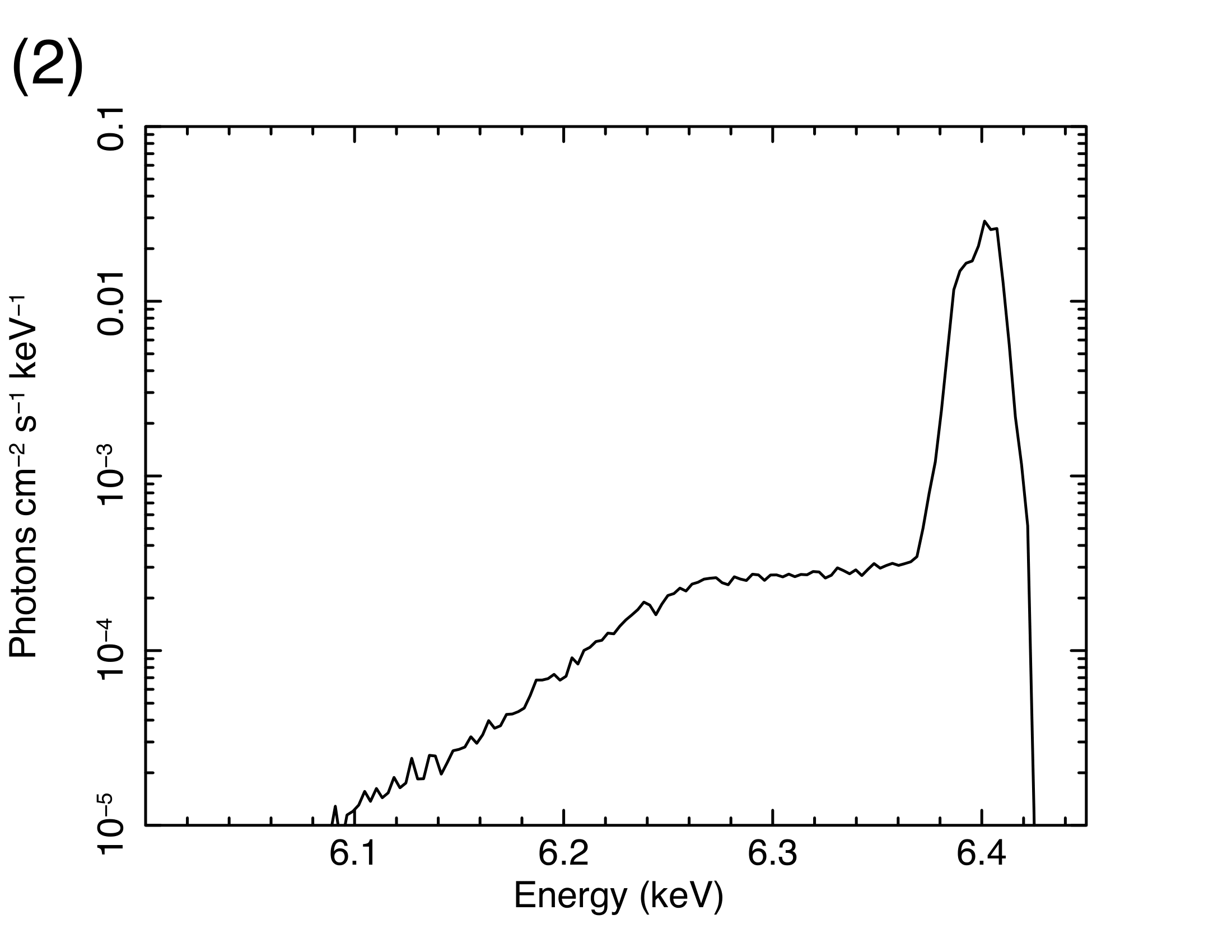}
 \plottwo{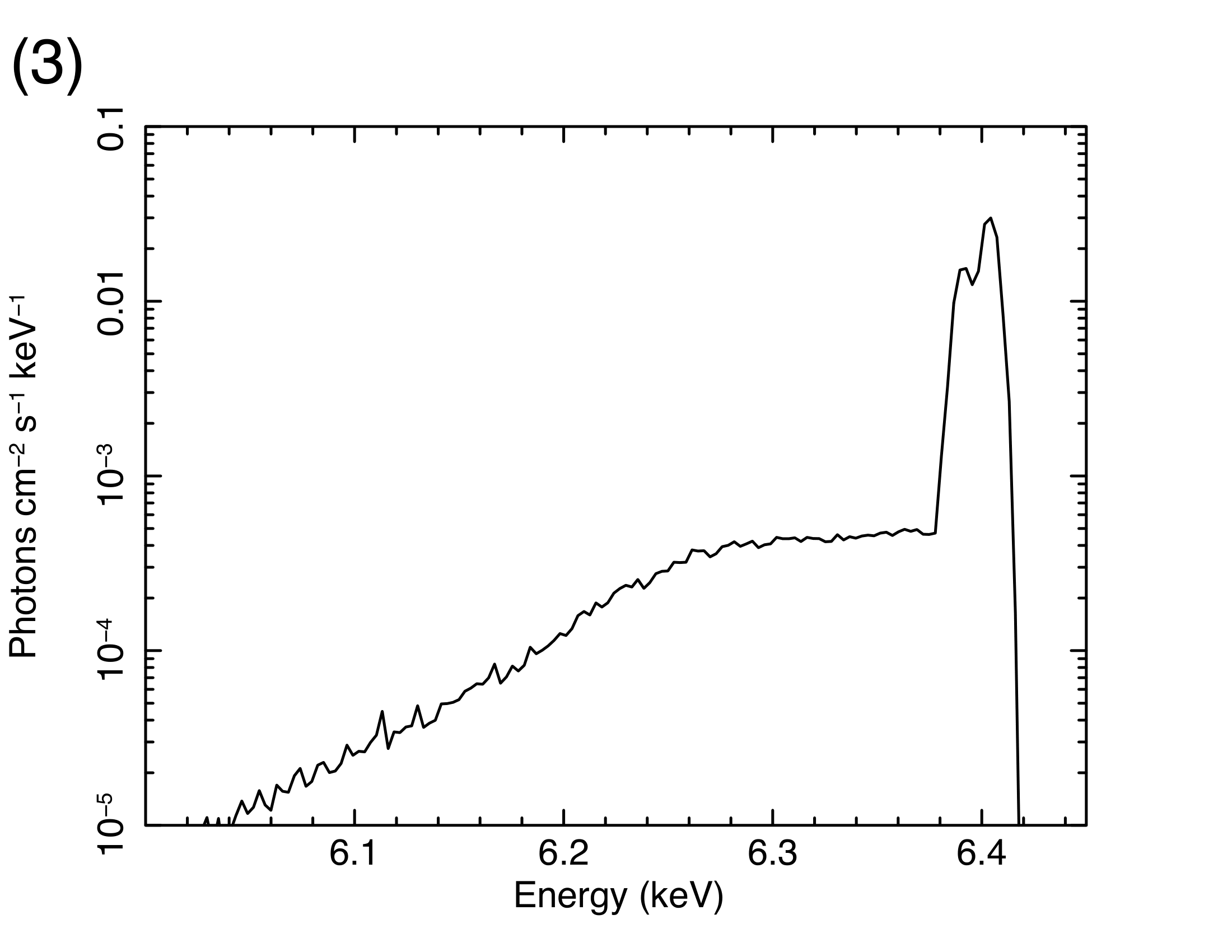}{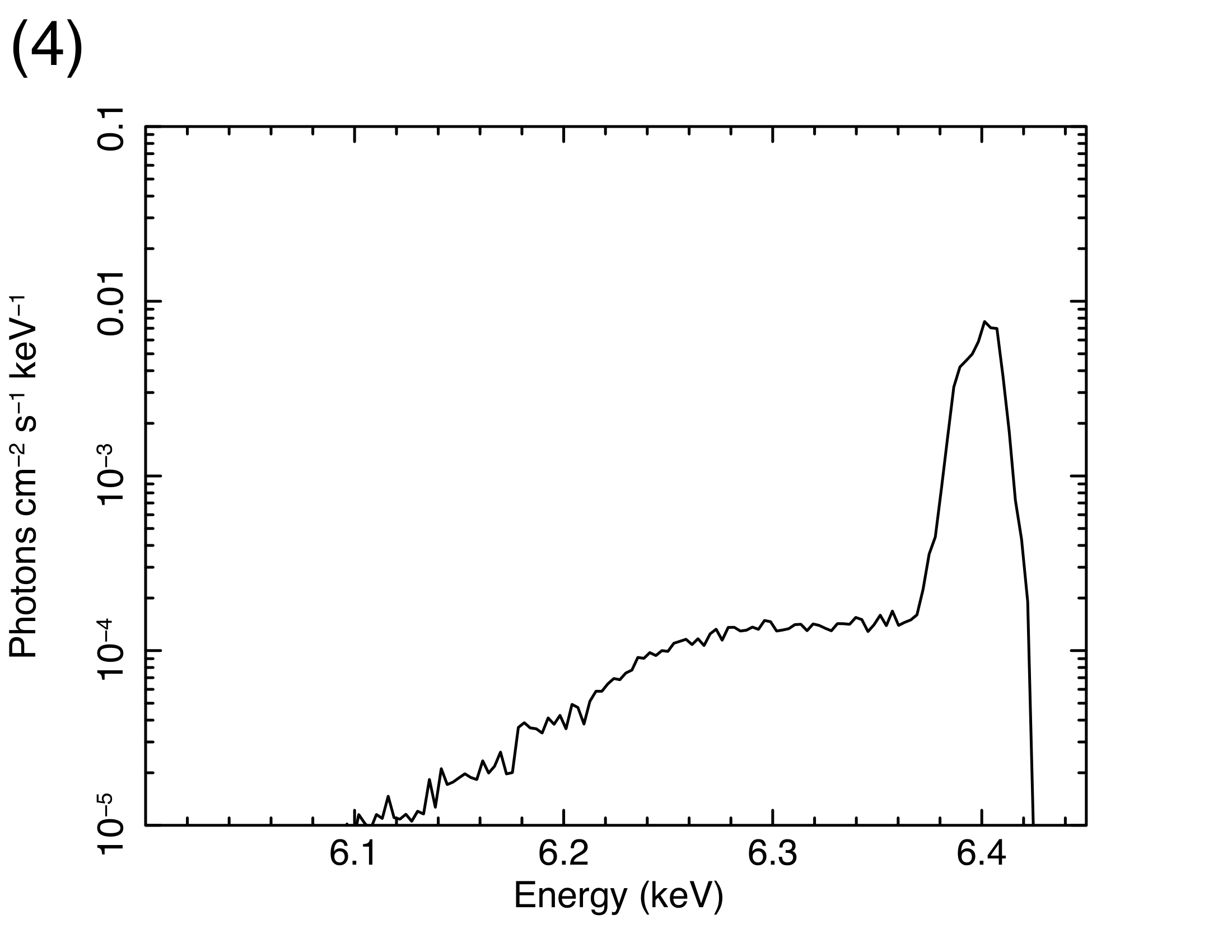}
 \caption{Fe K$\alpha$ line profiles for
 the torus parameters in Figure~\ref{figure:asymmetry}. 
 We assume $\log(r_\mathrm{in}/r_\mathrm{g})=5.28$.
 \label{figure:plmodel}}
\end{figure*}

The ray-tracing simulations performed to make the XCLUMPY table model
enable us to check the positions where the fluorescence lines are produced.
The left and right panels of 
Figure~\ref{figure:asymmetry} plot the
last interaction points of the observed Fe K$\alpha$
lines (except for the Compton shoulder) projected onto the equatorial plane
(i.e., viewed from the polar axis direction) 
and onto a plane perpendicular to the line of sight
(i.e., viewed from the observer), respectively, 
for 4 sets of the torus parameters ((1)--(4)).
The left and right panels of 
Figure~\ref{figure:histogram} display their azimuthal 
distribution 
and the radial profile of emissivity (photon numbers per unit area),  
respectively.
For each parameter set, we fit the radial profile with a power-law,
whose best fit result is overplotted. 
As noticed, when a Compton-thick torus is viewed nearly edge on,
photons emitted from the far-side torus become dominant.
Figure~\ref{figure:plmodel} shows the intrinsic Fe K$\alpha$ line profiles.

\bibliography{citation}{}

\begin{thebibliography}{}
\expandafter\ifx\csname natexlab\endcsname\relax\def\natexlab#1{#1}\fi
\providecommand{\url}[1]{\href{#1}{#1}}
\providecommand{\dodoi}[1]{doi:~\href{http://doi.org/#1}{\nolinkurl{#1}}}
\providecommand{\doeprint}[1]{\href{http://ascl.net/#1}{\nolinkurl{http://ascl.net/#1}}}
\providecommand{\doarXiv}[1]{\href{https://arxiv.org/abs/#1}{\nolinkurl{https://arxiv.org/abs/#1}}}

\bibitem[{{Anders} \& {Grevesse}(1989)}]{1989GeCoA..53..197A}
{Anders}, E., \& {Grevesse}, N. 1989, \gca, 53, 197,
  \dodoi{10.1016/0016-7037(89)90286-X}

\bibitem[{{Ar{\'e}valo} {et~al.}(2014){Ar{\'e}valo}, {Bauer}, {Puccetti},
  {Walton}, {Koss}, {Boggs}, {Brandt}, {Brightman}, {Christensen}, {Comastri},
  {Craig}, {Fuerst}, {Gandhi}, {Grefenstette}, {Hailey}, {Harrison}, {Luo},
  {Madejski}, {Madsen}, {Marinucci}, {Matt}, {Saez}, {Stern}, {Stuhlinger},
  {Treister}, {Urry}, \& {Zhang}}]{2014ApJ...791...81A}
{Ar{\'e}valo}, P., {Bauer}, F.~E., {Puccetti}, S., {et~al.} 2014, \apj, 791,
  81, \dodoi{10.1088/0004-637X/791/2/81}

\bibitem[{{Arnaud}(1996)}]{1996ASPC..101...17A}
{Arnaud}, K.~A. 1996, in Astronomical Society of the Pacific Conference Series,
  Vol. 101, Astronomical Data Analysis Software and Systems V, ed. G.~H.
  {Jacoby} \& J.~{Barnes}, 17

\bibitem[{{Balokovi{\'c}} {et~al.}(2018){Balokovi{\'c}}, {Brightman},
  {Harrison}, {Comastri}, {Ricci}, {Buchner}, {Gandhi}, {Farrah}, \&
  {Stern}}]{2018ApJ...854...42B}
{Balokovi{\'c}}, M., {Brightman}, M., {Harrison}, F.~A., {et~al.} 2018, \apj,
  854, 42, \dodoi{10.3847/1538-4357/aaa7eb}

\bibitem[{{Bianchi} {et~al.}(2002){Bianchi}, {Matt}, {Fiore}, {Fabian},
  {Iwasawa}, \& {Nicastro}}]{2002A&A...396..793B}
{Bianchi}, S., {Matt}, G., {Fiore}, F., {et~al.} 2002, \aap, 396, 793,
  \dodoi{10.1051/0004-6361:20021414}

\bibitem[{{Brightman} \& {Nandra}(2011)}]{2011MNRAS.413.1206B}
{Brightman}, M., \& {Nandra}, K. 2011, \mnras, 413, 1206,
  \dodoi{10.1111/j.1365-2966.2011.18207.x}

\bibitem[{{Brightman} {et~al.}(2015){Brightman}, {Balokovi{\'c}}, {Stern},
  {Ar{\'e}valo}, {Ballantyne}, {Bauer}, {Boggs}, {Craig}, {Christensen},
  {Comastri}, {Fuerst}, {Gandhi}, {Hailey}, {Harrison}, {Hickox}, {Koss},
  {LaMassa}, {Puccetti}, {Rivers}, {Vasudevan}, {Walton}, \&
  {Zhang}}]{2015ApJ...805...41B}
{Brightman}, M., {Balokovi{\'c}}, M., {Stern}, D., {et~al.} 2015, \apj, 805,
  41, \dodoi{10.1088/0004-637X/805/1/41}

\bibitem[{{Buchner} {et~al.}(2019){Buchner}, {Brightman}, {Nandra}, {Nikutta},
  \& {Bauer}}]{2019A&A...629A..16B}
{Buchner}, J., {Brightman}, M., {Nandra}, K., {Nikutta}, R., \& {Bauer}, F.~E.
  2019, \aap, 629, A16, \dodoi{10.1051/0004-6361/201834771}

\bibitem[{{Canizares} {et~al.}(2005){Canizares}, {Davis}, {Dewey}, {Flanagan},
  {Galton}, {Huenemoerder}, {Ishibashi}, {Markert}, {Marshall}, {McGuirk},
  {Schattenburg}, {Schulz}, {Smith}, \& {Wise}}]{2005PASP..117.1144C}
{Canizares}, C.~R., {Davis}, J.~E., {Dewey}, D., {et~al.} 2005, \pasp, 117,
  1144, \dodoi{10.1086/432898}

\bibitem[{{Cash}(1979)}]{1979ApJ...228..939C}
{Cash}, W. 1979, \apj, 228, 939, \dodoi{10.1086/156922}

\bibitem[{{Davis} {et~al.}(2012){Davis}, {Bautz}, {Dewey}, {Heilmann}, {Houck},
  {Huenemoerder}, {Marshall}, {Nowak}, {Schattenburg}, {Schulz}, \&
  {Smith}}]{2012SPIE.8443E..1AD}
{Davis}, J.~E., {Bautz}, M.~W., {Dewey}, D., {et~al.} 2012, in Society of
  Photo-Optical Instrumentation Engineers (SPIE) Conference Series, Vol. 8443,
  Space Telescopes and Instrumentation 2012: Ultraviolet to Gamma Ray, ed.
  T.~{Takahashi}, S.~S. {Murray}, \& J.-W.~A. {den Herder}, 84431A,
  \dodoi{10.1117/12.926937}

\bibitem[{{Fabbiano} {et~al.}(2020){Fabbiano}, {Paggi}, {Karovska}, {Elvis},
  {Nardini}, \& {Wang}}]{2020ApJ...902...49F}
{Fabbiano}, G., {Paggi}, A., {Karovska}, M., {et~al.} 2020, \apj, 902, 49,
  \dodoi{10.3847/1538-4357/abb5ad}

\bibitem[{{Freeman} {et~al.}(1977){Freeman}, {Karlsson}, {Lynga}, {Burrell},
  {van Woerden}, {Goss}, \& {Mebold}}]{1977A&A....55..445F}
{Freeman}, K.~C., {Karlsson}, B., {Lynga}, G., {et~al.} 1977, \aap, 55, 445

\bibitem[{{Fukazawa} {et~al.}(2009){Fukazawa}, {Mizuno}, {Watanabe}, {Kokubun},
  {Takahashi}, {Kawano}, {Nishino}, {Sasada}, {Shirai}, {Takahashi}, {Umeki},
  {Yamasaki}, {Yasuda}, {Bamba}, {Ohno}, {Takahashi}, {Ushio}, {Enoto},
  {Kitaguchi}, {Makishima}, {Nakazawa}, {Uehara}, {Yamada}, {Yuasa}, {Isobe},
  {Kawaharada}, {Tanaka}, {Tashiro}, {Terada}, \&
  {Yamaoka}}]{2009PASJ...61S..17F}
{Fukazawa}, Y., {Mizuno}, T., {Watanabe}, S., {et~al.} 2009, \pasj, 61, S17,
  \dodoi{10.1093/pasj/61.sp1.S17}

\bibitem[{{Fukazawa} {et~al.}(2011){Fukazawa}, {Hiragi}, {Mizuno}, {Nishino},
  {Hayashi}, {Yamasaki}, {Shirai}, {Takahashi}, \&
  {Ohno}}]{2011ApJ...727...19F}
{Fukazawa}, Y., {Hiragi}, K., {Mizuno}, M., {et~al.} 2011, \apj, 727, 19,
  \dodoi{10.1088/0004-637X/727/1/19}

\bibitem[{{Furui} {et~al.}(2016){Furui}, {Fukazawa}, {Odaka}, {Kawaguchi},
  {Ohno}, \& {Hayashi}}]{2016ApJ...818..164F}
{Furui}, S., {Fukazawa}, Y., {Odaka}, H., {et~al.} 2016, \apj, 818, 164,
  \dodoi{10.3847/0004-637X/818/2/164}

\bibitem[{{Gabriel} {et~al.}(2004){Gabriel}, {Denby}, {Fyfe}, {Hoar}, {Ibarra},
  {Ojero}, {Osborne}, {Saxton}, {Lammers}, \& {Vacanti}}]{2004ASPC..314..759G}
{Gabriel}, C., {Denby}, M., {Fyfe}, D.~J., {et~al.} 2004, in Astronomical
  Society of the Pacific Conference Series, Vol. 314, Astronomical Data
  Analysis Software and Systems (ADASS) XIII, ed. F.~{Ochsenbein}, M.~G.
  {Allen}, \& D.~{Egret}, 759

\bibitem[{{Gandhi} {et~al.}(2015){Gandhi}, {H{\"o}nig}, \&
  {Kishimoto}}]{2015ApJ...812..113G}
{Gandhi}, P., {H{\"o}nig}, S.~F., \& {Kishimoto}, M. 2015, \apj, 812, 113,
  \dodoi{10.1088/0004-637X/812/2/113}

\bibitem[{{Garmire} {et~al.}(2003){Garmire}, {Bautz}, {Ford}, {Nousek}, \&
  {Ricker}}]{2003SPIE.4851...28G}
{Garmire}, G.~P., {Bautz}, M.~W., {Ford}, P.~G., {Nousek}, J.~A., \& {Ricker},
  George~R., J. 2003, in Society of Photo-Optical Instrumentation Engineers
  (SPIE) Conference Series, Vol. 4851, X-Ray and Gamma-Ray Telescopes and
  Instruments for Astronomy., ed. J.~E. {Truemper} \& H.~D. {Tananbaum},
  28--44, \dodoi{10.1117/12.461599}

\bibitem[{{Greenhill} {et~al.}(2003){Greenhill}, {Booth}, {Ellingsen},
  {Herrnstein}, {Jauncey}, {McCulloch}, {Moran}, {Norris}, {Reynolds}, \&
  {Tzioumis}}]{2003ApJ...590..162G}
{Greenhill}, L.~J., {Booth}, R.~S., {Ellingsen}, S.~P., {et~al.} 2003, \apj,
  590, 162, \dodoi{10.1086/374862}

\bibitem[{{Harrison} {et~al.}(2013){Harrison}, {Craig}, {Christensen},
  {Hailey}, {Zhang}, {Boggs}, {Stern}, {Cook}, {Forster}, {Giommi},
  {Grefenstette}, {Kim}, {Kitaguchi}, {Koglin}, {Madsen}, {Mao}, {Miyasaka},
  {Mori}, {Perri}, {Pivovaroff}, {Puccetti}, {Rana}, {Westergaard}, {Willis},
  {Zoglauer}, {An}, {Bachetti}, {Barri{\`e}re}, {Bellm}, {Bhalerao},
  {Brejnholt}, {Fuerst}, {Liebe}, {Markwardt}, {Nynka}, {Vogel}, {Walton},
  {Wik}, {Alexander}, {Cominsky}, {Hornschemeier}, {Hornstrup}, {Kaspi},
  {Madejski}, {Matt}, {Molendi}, {Smith}, {Tomsick}, {Ajello}, {Ballantyne},
  {Balokovi{\'c}}, {Barret}, {Bauer}, {Blandford}, {Brandt}, {Brenneman},
  {Chiang}, {Chakrabarty}, {Chenevez}, {Comastri}, {Dufour}, {Elvis}, {Fabian},
  {Farrah}, {Fryer}, {Gotthelf}, {Grindlay}, {Helfand}, {Krivonos}, {Meier},
  {Miller}, {Natalucci}, {Ogle}, {Ofek}, {Ptak}, {Reynolds}, {Rigby},
  {Tagliaferri}, {Thorsett}, {Treister}, \& {Urry}}]{2013ApJ...770..103H}
{Harrison}, F.~A., {Craig}, W.~W., {Christensen}, F.~E., {et~al.} 2013, \apj,
  770, 103, \dodoi{10.1088/0004-637X/770/2/103}

\bibitem[{{Hikitani} {et~al.}(2018){Hikitani}, {Ohno}, {Fukazawa}, {Kawaguchi},
  \& {Odaka}}]{2018ApJ...867...80H}
{Hikitani}, M., {Ohno}, M., {Fukazawa}, Y., {Kawaguchi}, T., \& {Odaka}, H.
  2018, \apj, 867, 80, \dodoi{10.3847/1538-4357/aae1fe}

\bibitem[{{Ichikawa} {et~al.}(2015){Ichikawa}, {Packham}, {Ramos Almeida},
  {Asensio Ramos}, {Alonso-Herrero}, {Gonz{\'a}lez-Mart{\'\i}n},
  {Lopez-Rodriguez}, {Ueda}, {D{\'\i}az-Santos}, {Elitzur}, {H{\"o}nig},
  {Imanishi}, {Levenson}, {Mason}, {Perlman}, \& {Alsip}}]{2015ApJ...803...57I}
{Ichikawa}, K., {Packham}, C., {Ramos Almeida}, C., {et~al.} 2015, \apj, 803,
  57, \dodoi{10.1088/0004-637X/803/2/57}

\bibitem[{{Ikeda} {et~al.}(2009){Ikeda}, {Awaki}, \&
  {Terashima}}]{2009ApJ...692..608I}
{Ikeda}, S., {Awaki}, H., \& {Terashima}, Y. 2009, \apj, 692, 608,
  \dodoi{10.1088/0004-637X/692/1/608}

\bibitem[{{Jansen} {et~al.}(2001){Jansen}, {Lumb}, {Altieri}, {Clavel}, {Ehle},
  {Erd}, {Gabriel}, {Guainazzi}, {Gondoin}, {Much}, {Munoz}, {Santos},
  {Schartel}, {Texier}, \& {Vacanti}}]{2001A&A...365L...1J}
{Jansen}, F., {Lumb}, D., {Altieri}, B., {et~al.} 2001, \aap, 365, L1,
  \dodoi{10.1051/0004-6361:20000036}

\bibitem[{{Kawaguchi} \& {Mori}(2010)}]{2010ApJ...724L.183K}
{Kawaguchi}, T., \& {Mori}, M. 2010, \apjl, 724, L183,
  \dodoi{10.1088/2041-8205/724/2/L183}

\bibitem[{{Kawamuro} {et~al.}(2019){Kawamuro}, {Izumi}, \&
  {Imanishi}}]{2019PASJ...71...68K}
{Kawamuro}, T., {Izumi}, T., \& {Imanishi}, M. 2019, \pasj, 71, 68,
  \dodoi{10.1093/pasj/psz045}

\bibitem[{{Kawamuro} {et~al.}(2016){Kawamuro}, {Ueda}, {Tazaki}, {Ricci}, \&
  {Terashima}}]{2016ApJS..225...14K}
{Kawamuro}, T., {Ueda}, Y., {Tazaki}, F., {Ricci}, C., \& {Terashima}, Y. 2016,
  \apjs, 225, 14, \dodoi{10.3847/0067-0049/225/1/14}

\bibitem[{{Kishimoto} {et~al.}(2007){Kishimoto}, {H{\"o}nig}, {Beckert}, \&
  {Weigelt}}]{2007A&A...476..713K}
{Kishimoto}, M., {H{\"o}nig}, S.~F., {Beckert}, T., \& {Weigelt}, G. 2007,
  \aap, 476, 713, \dodoi{10.1051/0004-6361:20077911}

\bibitem[{{Kokubun} {et~al.}(2007){Kokubun}, {Makishima}, {Takahashi},
  {Murakami}, {Tashiro}, {Fukazawa}, {Kamae}, {Madejski}, {Nakazawa},
  {Yamaoka}, {Terada}, {Yonetoku}, {Watanabe}, {Tamagawa}, {Mizuno}, {Kubota},
  {Isobe}, {Takahashi}, {Sato}, {Takahashi}, {Hong}, {Kawaharada}, {Kawano},
  {Mitani}, {Murashima}, {Suzuki}, {Abe}, {Miyawaki}, {Ohno}, {Tanaka},
  {Yanagida}, {Itoh}, {Ohnuki}, {Tamura}, {Endo}, {Hirakuri}, {Hiruta},
  {Kitaguchi}, {Kishishita}, {Sugita}, {Takahashi}, {Takeda}, {Enoto},
  {Hirasawa}, {Katsuta}, {Matsumura}, {Onda}, {Sato}, {Ushio}, {Ishikawa},
  {Murase}, {Odaka}, {Suzuki}, {Yaji}, {Yamada}, {Yamasaki}, {Yuasa}, \& {Hxd
  Team}}]{2007PASJ...59S..53K}
{Kokubun}, M., {Makishima}, K., {Takahashi}, T., {et~al.} 2007, \pasj, 59, 53,
  \dodoi{10.1093/pasj/59.sp1.S53}

\bibitem[{{Kormendy} \& {Ho}(2013)}]{2013ARA&A..51..511K}
{Kormendy}, J., \& {Ho}, L.~C. 2013, \araa, 51, 511,
  \dodoi{10.1146/annurev-astro-082708-101811}

\bibitem[{{Koshida} {et~al.}(2009){Koshida}, {Yoshii}, {Kobayashi}, {Minezaki},
  {Sakata}, {Sugawara}, {Enya}, {Suganuma}, {Tomita}, {Aoki}, \&
  {Peterson}}]{2009ApJ...700L.109K}
{Koshida}, S., {Yoshii}, Y., {Kobayashi}, Y., {et~al.} 2009, \apjl, 700, L109,
  \dodoi{10.1088/0004-637X/700/2/L109}

\bibitem[{{Koyama} {et~al.}(2007){Koyama}, {Tsunemi}, {Dotani}, {Bautz},
  {Hayashida}, {Tsuru}, {Matsumoto}, {Ogawara}, {Ricker}, {Doty}, {Kissel},
  {Foster}, {Nakajima}, {Yamaguchi}, {Mori}, {Sakano}, {Hamaguchi},
  {Nishiuchi}, {Miyata}, {Torii}, {Namiki}, {Katsuda}, {Matsuura}, {Miyauchi},
  {Anabuki}, {Tawa}, {Ozaki}, {Murakami}, {Maeda}, {Ichikawa}, {Prigozhin},
  {Boughan}, {Lamarr}, {Miller}, {Burke}, {Gregory}, {Pillsbury}, {Bamba},
  {Hiraga}, {Senda}, {Katayama}, {Kitamoto}, {Tsujimoto}, {Kohmura}, {Tsuboi},
  \& {Awaki}}]{2007PASJ...59S..23K}
{Koyama}, K., {Tsunemi}, H., {Dotani}, T., {et~al.} 2007, \pasj, 59, 23,
  \dodoi{10.1093/pasj/59.sp1.S23}

\bibitem[{{Liu}(2016{\natexlab{a}})}]{2016MNRAS.463L.108L}
{Liu}, J. 2016{\natexlab{a}}, \mnras, 463, L108, \dodoi{10.1093/mnrasl/slw164}

\bibitem[{{Liu}(2016{\natexlab{b}})}]{2016MNRAS.459L.105L}
---. 2016{\natexlab{b}}, \mnras, 459, L105, \dodoi{10.1093/mnrasl/slw048}

\bibitem[{{Liu} {et~al.}(2019){Liu}, {H{\"o}nig}, {Ricci}, \&
  {Paltani}}]{2019MNRAS.490.4344L}
{Liu}, J., {H{\"o}nig}, S.~F., {Ricci}, C., \& {Paltani}, S. 2019, \mnras, 490,
  4344, \dodoi{10.1093/mnras/stz2908}

\bibitem[{{Madsen} {et~al.}(2017){Madsen}, {Beardmore}, {Forster}, {Guainazzi},
  {Marshall}, {Miller}, {Page}, \& {Stuhlinger}}]{2017AJ....153....2M}
{Madsen}, K.~K., {Beardmore}, A.~P., {Forster}, K., {et~al.} 2017, \aj, 153, 2,
  \dodoi{10.3847/1538-3881/153/1/2}

\bibitem[{{Magdziarz} \& {Zdziarski}(1995)}]{1995MNRAS.273..837M}
{Magdziarz}, P., \& {Zdziarski}, A.~A. 1995, \mnras, 273, 837,
  \dodoi{10.1093/mnras/273.3.837}

\bibitem[{{Marinucci} {et~al.}(2013){Marinucci}, {Miniutti}, {Bianchi}, {Matt},
  \& {Risaliti}}]{2013MNRAS.436.2500M}
{Marinucci}, A., {Miniutti}, G., {Bianchi}, S., {Matt}, G., \& {Risaliti}, G.
  2013, \mnras, 436, 2500, \dodoi{10.1093/mnras/stt1759}

\bibitem[{{Markowitz} {et~al.}(2014){Markowitz}, {Krumpe}, \&
  {Nikutta}}]{2014MNRAS.439.1403M}
{Markowitz}, A.~G., {Krumpe}, M., \& {Nikutta}, R. 2014, \mnras, 439, 1403,
  \dodoi{10.1093/mnras/stt2492}

\bibitem[{{Massaro} {et~al.}(2006){Massaro}, {Bianchi}, {Matt}, {D'Onofrio}, \&
  {Nicastro}}]{2006A&A...455..153M}
{Massaro}, F., {Bianchi}, S., {Matt}, G., {D'Onofrio}, E., \& {Nicastro}, F.
  2006, \aap, 455, 153, \dodoi{10.1051/0004-6361:20054772}

\bibitem[{{Minezaki} \& {Matsushita}(2015)}]{2015ApJ...802...98M}
{Minezaki}, T., \& {Matsushita}, K. 2015, \apj, 802, 98,
  \dodoi{10.1088/0004-637X/802/2/98}

\bibitem[{{Mitsuda} {et~al.}(2007){Mitsuda}, {Bautz}, {Inoue}, {Kelley},
  {Koyama}, {Kunieda}, {Makishima}, {Ogawara}, {Petre}, {Takahashi}, {Tsunemi},
  {White}, {Anabuki}, {Angelini}, {Arnaud}, {Awaki}, {Bamba}, {Boyce}, {Brown},
  {Chan}, {Cottam}, {Dotani}, {Doty}, {Ebisawa}, {Ezoe}, {Fabian}, {Figueroa},
  {Fujimoto}, {Fukazawa}, {Furusho}, {Furuzawa}, {Gendreau}, {Griffiths},
  {Haba}, {Hamaguchi}, {Harrus}, {Hasinger}, {Hatsukade}, {Hayashida}, {Henry},
  {Hiraga}, {Holt}, {Hornschemeier}, {Hughes}, {Hwang}, {Ishida}, {Ishisaki},
  {Isobe}, {Itoh}, {Iyomoto}, {Kahn}, {Kamae}, {Katagiri}, {Kataoka},
  {Katayama}, {Kawai}, {Kilbourne}, {Kinugasa}, {Kissel}, {Kitamoto}, {Kohama},
  {Kohmura}, {Kokubun}, {Kotani}, {Kotoku}, {Kubota}, {Madejski}, {Maeda},
  {Makino}, {Markowitz}, {Matsumoto}, {Matsumoto}, {Matsuoka}, {Matsushita},
  {McCammon}, {Mihara}, {Misaki}, {Miyata}, {Mizuno}, {Mori}, {Mori}, {Morii},
  {Moseley}, {Mukai}, {Murakami}, {Murakami}, {Mushotzky}, {Nagase}, {Namiki},
  {Negoro}, {Nakazawa}, {Nousek}, {Okajima}, {Ogasaka}, {Ohashi}, {Oshima},
  {Ota}, {Ozaki}, {Ozawa}, {Parmar}, {Pence}, {Porter}, {Reeves}, {Ricker},
  {Sakurai}, {Sanders}, {Senda}, {Serlemitsos}, {Shibata}, {Soong}, {Smith},
  {Suzuki}, {Szymkowiak}, {Takahashi}, {Tamagawa}, {Tamura}, {Tamura},
  {Tanaka}, {Tashiro}, {Tawara}, {Terada}, {Terashima}, {Tomida}, {Torii},
  {Tsuboi}, {Tsujimoto}, {Tsuru}, {Turner}, {Ueda}, {Ueno}, {Ueno}, {Uno},
  {Urata}, {Watanabe}, {Yamamoto}, {Yamaoka}, {Yamasaki}, {Yamashita},
  {Yamauchi}, {Yamauchi}, {Yaqoob}, {Yonetoku}, \&
  {Yoshida}}]{2007PASJ...59S...1M}
{Mitsuda}, K., {Bautz}, M., {Inoue}, H., {et~al.} 2007, \pasj, 59, S1,
  \dodoi{10.1093/pasj/59.sp1.S1}

\bibitem[{{Molendi} {et~al.}(2003){Molendi}, {Bianchi}, \&
  {Matt}}]{2003MNRAS.343L...1M}
{Molendi}, S., {Bianchi}, S., \& {Matt}, G. 2003, \mnras, 343, L1,
  \dodoi{10.1046/j.1365-8711.2003.06783.x}

\bibitem[{{Murphy} \& {Yaqoob}(2009)}]{2009MNRAS.397.1549M}
{Murphy}, K.~D., \& {Yaqoob}, T. 2009, \mnras, 397, 1549,
  \dodoi{10.1111/j.1365-2966.2009.15025.x}

\bibitem[{{Nasa High Energy Astrophysics Science Archive Research Center
  (Heasarc)}(2014)}]{2014ascl.soft08004N}
{Nasa High Energy Astrophysics Science Archive Research Center (Heasarc)}.
  2014, {HEAsoft: Unified Release of FTOOLS and XANADU}.
\newblock \doeprint{1408.004}

\bibitem[{{Nenkova} {et~al.}(2008{\natexlab{a}}){Nenkova}, {Sirocky},
  {Ivezi{\'c}}, \& {Elitzur}}]{2008ApJ...685..147N}
{Nenkova}, M., {Sirocky}, M.~M., {Ivezi{\'c}}, {\v{Z}}., \& {Elitzur}, M.
  2008{\natexlab{a}}, \apj, 685, 147, \dodoi{10.1086/590482}

\bibitem[{{Nenkova} {et~al.}(2008{\natexlab{b}}){Nenkova}, {Sirocky},
  {Nikutta}, {Ivezi{\'c}}, \& {Elitzur}}]{2008ApJ...685..160N}
{Nenkova}, M., {Sirocky}, M.~M., {Nikutta}, R., {Ivezi{\'c}}, {\v{Z}}., \&
  {Elitzur}, M. 2008{\natexlab{b}}, \apj, 685, 160, \dodoi{10.1086/590483}

\bibitem[{{Odaka} {et~al.}(2011){Odaka}, {Aharonian}, {Watanabe}, {Tanaka},
  {Khangulyan}, \& {Takahashi}}]{2011ApJ...740..103O}
{Odaka}, H., {Aharonian}, F., {Watanabe}, S., {et~al.} 2011, \apj, 740, 103,
  \dodoi{10.1088/0004-637X/740/2/103}

\bibitem[{{Odaka} {et~al.}(2016){Odaka}, {Yoneda}, {Takahashi}, \&
  {Fabian}}]{2016MNRAS.462.2366O}
{Odaka}, H., {Yoneda}, H., {Takahashi}, T., \& {Fabian}, A. 2016, \mnras, 462,
  2366, \dodoi{10.1093/mnras/stw1764}

\bibitem[{{Ogawa} {et~al.}(2021){Ogawa}, {Ueda}, {Tanimoto}, \&
  {Yamada}}]{2021ApJ...906...84O}
{Ogawa}, S., {Ueda}, Y., {Tanimoto}, A., \& {Yamada}, S. 2021, \apj, 906, 84,
  \dodoi{10.3847/1538-4357/abccce}

\bibitem[{{Qiu} {et~al.}(2019){Qiu}, {Soria}, {Wang}, {Wiktorowicz}, {Liu},
  {Bai}, {Bogomazov}, {Di Stefano}, {Walton}, \& {Xu}}]{2019ApJ...877...57Q}
{Qiu}, Y., {Soria}, R., {Wang}, S., {et~al.} 2019, \apj, 877, 57,
  \dodoi{10.3847/1538-4357/ab16e7}

\bibitem[{{Ramos Almeida} \& {Ricci}(2017)}]{2017NatAs...1..679R}
{Ramos Almeida}, C., \& {Ricci}, C. 2017, Nature Astronomy, 1, 679,
  \dodoi{10.1038/s41550-017-0232-z}

\bibitem[{{Reynolds} {et~al.}(2014){Reynolds}, {Ueda}, {Awaki}, {Gallo},
  {Gandhi}, {Haba}, {Kawamuro}, {LaMassa}, {Lohfink}, {Ricci}, {Tazaki}, \&
  {Zoghbi}}]{2014arXiv1412.1177R}
{Reynolds}, C., {Ueda}, Y., {Awaki}, H., {et~al.} 2014, arXiv e-prints,
  arXiv:1412.1177.
\newblock \doarXiv{1412.1177}

\bibitem[{{Ricci} {et~al.}(2017){Ricci}, {Trakhtenbrot}, {Koss}, {Ueda},
  {Schawinski}, {Oh}, {Lamperti}, {Mushotzky}, {Treister}, {Ho}, {Weigel},
  {Bauer}, {Paltani}, {Fabian}, {Xie}, \& {Gehrels}}]{2017Natur.549..488R}
{Ricci}, C., {Trakhtenbrot}, B., {Koss}, M.~J., {et~al.} 2017, \nat, 549, 488,
  \dodoi{10.1038/nature23906}

\bibitem[{{Sambruna} {et~al.}(2001){Sambruna}, {Netzer}, {Kaspi}, {Brandt},
  {Chartas}, {Garmire}, {Nousek}, \& {Weaver}}]{2001ApJ...546L..13S}
{Sambruna}, R.~M., {Netzer}, H., {Kaspi}, S., {et~al.} 2001, \apjl, 546, L13,
  \dodoi{10.1086/318068}

\bibitem[{{Shidatsu} {et~al.}(2017){Shidatsu}, {Ueda}, \&
  {Fabrika}}]{2017ApJ...839...46S}
{Shidatsu}, M., {Ueda}, Y., \& {Fabrika}, S. 2017, \apj, 839, 46,
  \dodoi{10.3847/1538-4357/aa67e7}

\bibitem[{{Shu} {et~al.}(2010){Shu}, {Yaqoob}, \& {Wang}}]{2010ApJS..187..581S}
{Shu}, X.~W., {Yaqoob}, T., \& {Wang}, J.~X. 2010, \apjs, 187, 581,
  \dodoi{10.1088/0067-0049/187/2/581}

\bibitem[{{Shu} {et~al.}(2011){Shu}, {Yaqoob}, \& {Wang}}]{2011ApJ...738..147S}
---. 2011, \apj, 738, 147, \dodoi{10.1088/0004-637X/738/2/147}

\bibitem[{{Str{\"u}der} {et~al.}(2001){Str{\"u}der}, {Briel}, {Dennerl},
  {Hartmann}, {Kendziorra}, {Meidinger}, {Pfeffermann}, {Reppin}, {Aschenbach},
  {Bornemann}, {Br{\"a}uninger}, {Burkert}, {Elender}, {Freyberg}, {Haberl},
  {Hartner}, {Heuschmann}, {Hippmann}, {Kastelic}, {Kemmer}, {Kettenring},
  {Kink}, {Krause}, {M{\"u}ller}, {Oppitz}, {Pietsch}, {Popp}, {Predehl},
  {Read}, {Stephan}, {St{\"o}tter}, {Tr{\"u}mper}, {Holl}, {Kemmer}, {Soltau},
  {St{\"o}tter}, {Weber}, {Weichert}, {von Zanthier}, {Carathanassis}, {Lutz},
  {Richter}, {Solc}, {B{\"o}ttcher}, {Kuster}, {Staubert}, {Abbey}, {Holland},
  {Turner}, {Balasini}, {Bignami}, {La Palombara}, {Villa}, {Buttler},
  {Gianini}, {Lain{\'e}}, {Lumb}, \& {Dhez}}]{2001A&A...365L..18S}
{Str{\"u}der}, L., {Briel}, U., {Dennerl}, K., {et~al.} 2001, \aap, 365, L18,
  \dodoi{10.1051/0004-6361:20000066}

\bibitem[{{Takahashi} {et~al.}(2007){Takahashi}, {Abe}, {Endo}, {Endo}, {Ezoe},
  {Fukazawa}, {Hamaya}, {Hirakuri}, {Hong}, {Horii}, {Inoue}, {Isobe}, {Itoh},
  {Iyomoto}, {Kamae}, {Kasama}, {Kataoka}, {Kato}, {Kawaharada}, {Kawano},
  {Kawashima}, {Kawasoe}, {Kishishita}, {Kitaguchi}, {Kobayashi}, {Kokubun},
  {Kotoku}, {Kouda}, {Kubota}, {Kuroda}, {Madejski}, {Makishima}, {Masukawa},
  {Matsumoto}, {Mitani}, {Miyawaki}, {Mizuno}, {Mori}, {Mori}, {Murashima},
  {Murakami}, {Nakazawa}, {Niko}, {Nomachi}, {Okada}, {Ohno}, {Oonuki}, {Ota},
  {Ozawa}, {Sato}, {Shinoda}, {Sugiho}, {Suzuki}, {Taguchi}, {Takahashi},
  {Takahashi}, {Takeda}, {Tamura}, {Tamura}, {Tanaka}, {Tanihata}, {Tashiro},
  {Terada}, {Tominaga}, {Uchiyama}, {Watanabe}, {Yamaoka}, {Yanagida}, \&
  {Yonetoku}}]{2007PASJ...59S..35T}
{Takahashi}, T., {Abe}, K., {Endo}, M., {et~al.} 2007, \pasj, 59, 35,
  \dodoi{10.1093/pasj/59.sp1.S35}

\bibitem[{{Tanimoto} {et~al.}(2018){Tanimoto}, {Ueda}, {Kawamuro}, {Ricci},
  {Awaki}, \& {Terashima}}]{2018ApJ...853..146T}
{Tanimoto}, A., {Ueda}, Y., {Kawamuro}, T., {et~al.} 2018, \apj, 853, 146,
  \dodoi{10.3847/1538-4357/aaa47c}

\bibitem[{{Tanimoto} {et~al.}(2019){Tanimoto}, {Ueda}, {Odaka}, {Kawaguchi},
  {Fukazawa}, \& {Kawamuro}}]{2019ApJ...877...95T}
{Tanimoto}, A., {Ueda}, Y., {Odaka}, H., {et~al.} 2019, \apj, 877, 95,
  \dodoi{10.3847/1538-4357/ab1b20}

\bibitem[{{Tanimoto} {et~al.}(2020){Tanimoto}, {Ueda}, {Odaka}, {Ogawa},
  {Yamada}, {Kawaguchi}, \& {Ichikawa}}]{2020ApJ...897....2T}
---. 2020, \apj, 897, 2, \dodoi{10.3847/1538-4357/ab96bc}

\bibitem[{{Turner} {et~al.}(2001){Turner}, {Abbey}, {Arnaud}, {Balasini},
  {Barbera}, {Belsole}, {Bennie}, {Bernard}, {Bignami}, {Boer}, {Briel},
  {Butler}, {Cara}, {Chabaud}, {Cole}, {Collura}, {Conte}, {Cros}, {Denby},
  {Dhez}, {Di Coco}, {Dowson}, {Ferrando}, {Ghizzardi}, {Gianotti}, {Goodall},
  {Gretton}, {Griffiths}, {Hainaut}, {Hochedez}, {Holland}, {Jourdain},
  {Kendziorra}, {Lagostina}, {Laine}, {La Palombara}, {Lortholary}, {Lumb},
  {Marty}, {Molendi}, {Pigot}, {Poindron}, {Pounds}, {Reeves}, {Reppin},
  {Rothenflug}, {Salvetat}, {Sauvageot}, {Schmitt}, {Sembay}, {Short},
  {Spragg}, {Stephen}, {Str{\"u}der}, {Tiengo}, {Trifoglio}, {Tr{\"u}mper},
  {Vercellone}, {Vigroux}, {Villa}, {Ward}, {Whitehead}, \&
  {Zonca}}]{2001A&A...365L..27T}
{Turner}, M.~J.~L., {Abbey}, A., {Arnaud}, M., {et~al.} 2001, \aap, 365, L27,
  \dodoi{10.1051/0004-6361:20000087}

\bibitem[{{Vasudevan} \& {Fabian}(2009)}]{2009MNRAS.392.1124V}
{Vasudevan}, R.~V., \& {Fabian}, A.~C. 2009, \mnras, 392, 1124,
  \dodoi{10.1111/j.1365-2966.2008.14108.x}

\bibitem[{{Walton} {et~al.}(2014){Walton}, {Harrison}, {Grefenstette},
  {Miller}, {Bachetti}, {Barret}, {Boggs}, {Christensen}, {Craig}, {Fabian},
  {Fuerst}, {Hailey}, {Madsen}, {Parker}, {Ptak}, {Rana}, {Stern}, {Webb}, \&
  {Zhang}}]{2014ApJ...793...21W}
{Walton}, D.~J., {Harrison}, F.~A., {Grefenstette}, B.~W., {et~al.} 2014, \apj,
  793, 21, \dodoi{10.1088/0004-637X/793/1/21}

\bibitem[{{Weisskopf} {et~al.}(2002){Weisskopf}, {Brinkman}, {Canizares},
  {Garmire}, {Murray}, \& {Van Speybroeck}}]{2002PASP..114....1W}
{Weisskopf}, M.~C., {Brinkman}, B., {Canizares}, C., {et~al.} 2002, \pasp, 114,
  1, \dodoi{10.1086/338108}

\bibitem[{{Willingale} {et~al.}(2013){Willingale}, {Starling}, {Beardmore},
  {Tanvir}, \& {O'Brien}}]{2013MNRAS.431..394W}
{Willingale}, R., {Starling}, R.~L.~C., {Beardmore}, A.~P., {Tanvir}, N.~R., \&
  {O'Brien}, P.~T. 2013, \mnras, 431, 394, \dodoi{10.1093/mnras/stt175}

\bibitem[{{Yamada} {et~al.}(2020){Yamada}, {Ueda}, {Tanimoto}, {Oda},
  {Imanishi}, {Toba}, \& {Ricci}}]{2020ApJ...897..107Y}
{Yamada}, S., {Ueda}, Y., {Tanimoto}, A., {et~al.} 2020, \apj, 897, 107,
  \dodoi{10.3847/1538-4357/ab94b1}

\bibitem[{{Yang} {et~al.}(2009){Yang}, {Wilson}, {Matt}, {Terashima}, \&
  {Greenhill}}]{2009ApJ...691..131Y}
{Yang}, Y., {Wilson}, A.~S., {Matt}, G., {Terashima}, Y., \& {Greenhill}, L.~J.
  2009, \apj, 691, 131, \dodoi{10.1088/0004-637X/691/1/131}

\end{thebibliography}
\bibliographystyle{aasjournal}



\end{document}